\documentclass[a4paper,12pt]{article}

\usepackage{epsfig,graphicx,amsmath,amssymb}
\usepackage{url}
\usepackage{hyperref}
\usepackage{subcaption}
\newcommand{\ee}{e^{+} e^{-}}
\def\amu{a_\mu}

\def \ee {e^+e^-}

\newcommand{\btm}{\bibitem}
\newcommand{\mpi}{M_\pi}
\def \piee {\pi^0\to e^-e^+}
\newcommand{\fplus}{f^{\eta\pi}_+}
\newcommand{\fzero}{f^{\eta\pi}_0}
\newcommand{\kzero}{{K^0}}
\newcommand{\kplus}{{K^+}}
\newcommand{\piplus}{{\pi^+} }
\newcommand{\pizero}{{\pi^0} }
\newcommand{\mpid}{m_\pi^2}
\newcommand{\mkd}{m_K^2}
\def \Kl {\mathrm{K_L}}
\def \Ks {\mathrm{K_S}}
\begin{document}

\thispagestyle{empty}

$\phantom{.}$

\hfill

\begin{center}
{\Large {\bf MesonNet 2013 International Workshop\\
Mini-proceedings} \\
\vspace{0.75cm}}

\vspace{1cm}

{\large June 17--19, 2013 in Prague, Czech Republic}

\vspace{2cm}

{\it Editors:}
K.~Kampf (Prague), A.~Kupsc (Uppsala) and P.~Masjuan (Mainz)

\vspace{2.5cm}

ABSTRACT

\end{center}

\vspace{0.3cm}

\noindent
The mini-proceedings of the MesonNet 2013 International Workshop held in Prague from June 17$^{\rm th}$ to 19$^{\rm th}$, 2013, are presented. MesonNet  is a research network within EU  HadronPhysics3  project
(1/2012 -- 12/2014).

\medskip\noindent
The web page of the conference, which contains all talks, can be found at
\begin{center}
\url{http://ipnp.mff.cuni.cz/mesonnet13} 
\end{center}

\vspace{0.5cm}

\noindent
We acknowledge the support of the EU HadronPhysics3 project and thank Charles University in Prague for its hospitality.

\vspace{5mm}
\noindent
This work is a part of the activity of the MesonNet:
\begin{center}
[\url{https://sites.google.com/site/mesonnetwork/}]
\end{center}

\newpage

{$\phantom{=}$}

\vspace{0.5cm}

\tableofcontents

\newpage

\section{Introduction to the Workshop}

\addtocontents{toc}{\hspace{1cm}{\sl K.~Kampf, A.~Kupsc}\par}

\vspace{5mm}

\noindent
K.~Kampf$^1$ and A.~Kupsc$^2$

\vspace{5mm}

\noindent
$^1$Institute of Particle and Nuclear Physics, Faculty of Mathematics
and Physics, Charles University in Prague, Czech Republic\\
$^2$Department of Physics and Astronomy, Uppsala University, Sweden\\

\vspace{5mm}

MesonNet  is  a  research  network within  EU  HadronPhysics3  project
(1/2012 -- 12/2014).  The main objective is the  coordination of light
meson studies  at different European  accelerator research facilities:
COSY (J\"ulich),  DAPHNE (Frascati), ELSA (Bonn),  GSI (Darmstadt) and
MAMI (Mainz).

The network  includes also EU researchers carrying  out experiments at
VEPP-2000  (BINP),  CEBAF (JLAB)  and  heavy flavor-factories  (Babar,
Belle  II, BESIII  experiments).   The scope  are processes  involving
lightest  neutral mesons: $\pi^0$,  $\eta$, $\omega$,  $\eta'$, $\phi$
and the  lightest scalar  resonances. The emphasis  is on  meson decay
studies  but we  include  also meson  production  processes and  meson
baryon interactions.  The majority  of the participants of the network
are experimentalists  while close collaboration with  theory groups is
essential for  the planning of  experiments and the  interpretation of
the data.

We have  identified some  specific research topics  for which  a close
collaboration  between  experiment  and  theory leads  to  significant
progress.   The  first  project  is  related to  the  studies  of  the
isospin-violating $\eta$ meson decay into three pions occurring due to
the  light  quark  mass  difference  $m_d-m_u$.   The  decay  has  the
potential to provide  one of the best constraints  for the light quark
mass  ratios  and is  a  sensitive test  of  convergence  of the  CHPT
expansion. The immediate aim is  to resolve issues in the experimental
and theoretical description of  the decays.  Hadronic decays of $\eta$
and $\eta'$ mesons are sensitive tools for investigations of $\pi\pi$ and $\pi\eta$
interactions.   The  second  joined   project  aims  at  a  systematic
determination of the transition form factors of the $\pi^0$ and $\eta$
mesons  with focus  on the  case involving  two virtual  photons.  The
knowledge of the form factors is important for the calculations of the
Standard  Model contributions to  $g-2$ of  the muon  and to  the rare
$\pi^0$  and $\eta$ decays  into a  lepton-antilepton pair.   The muon
$g-2$  and  the  branching  ratio  for  $\pi^0\to  e^+e^-$  decay  are
currently among the few observables  where hints of  a deviation  from the
Standard Model  predictions are reported.  A topical meeting  on light
mesons transition  form factors was  organized in June 2012  in Krakow
\cite{Czerwinski:2012ry1}.

The present workshop gives an overview of activities at the halfway of
the  project.  MesonNet  is a  continuation  and an  extension of  the
PrimeNet network which  was active 2009-2011 and the  summaries of the
similar workshops are available \cite{Hoistadt:2011iv,Adlarson:2012bi}.

The detailed program, which consisted of 47 talks and 15 posters, 
was arranged by a program committee having the members: 
Reinhard Beck, Johan Bijnens, Simon Eydelman, Ingo Froehlich, Simona Giovannella, Dieter Grzonka, Christoph Hanhart, Volker Hejny, Bo H\"oistad, Tord Johansson, Karol Kampf, Bernd Krusche, Bastian Kubis, Andrzej Kupsc, Stefan Leupold, Pawel Moskal, Michael Ostrick, Teresa Pe\~na, Piotr Salabura, Susan Schadmand.

The workshop was held in June 17-19, 2013, at the campus of Faculty of Mathematics and Physics 
of the Charles University at Malostransk\'e n\'am\v{e}st\'{\i} 2 in Prague, enjoying
hospitality from Institute of Particle and Nuclear Physics, Prague, Czech Republic.

Webpage of the conference is 
\begin{center}
\url{http://ipnp.mff.cuni.cz/mesonnet13} 
\end{center}
\noindent
where detailed program and talks can be found.

Financial support is gratefully acknowledged from the European HadronPhysics3  project. 

\begin{figure}[h!]
\begin{center}
\begin{subfigure}[b]{0.4\textwidth}
\centering
\includegraphics[width=2cm]{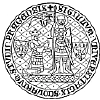}\\
\includegraphics[width=5cm]{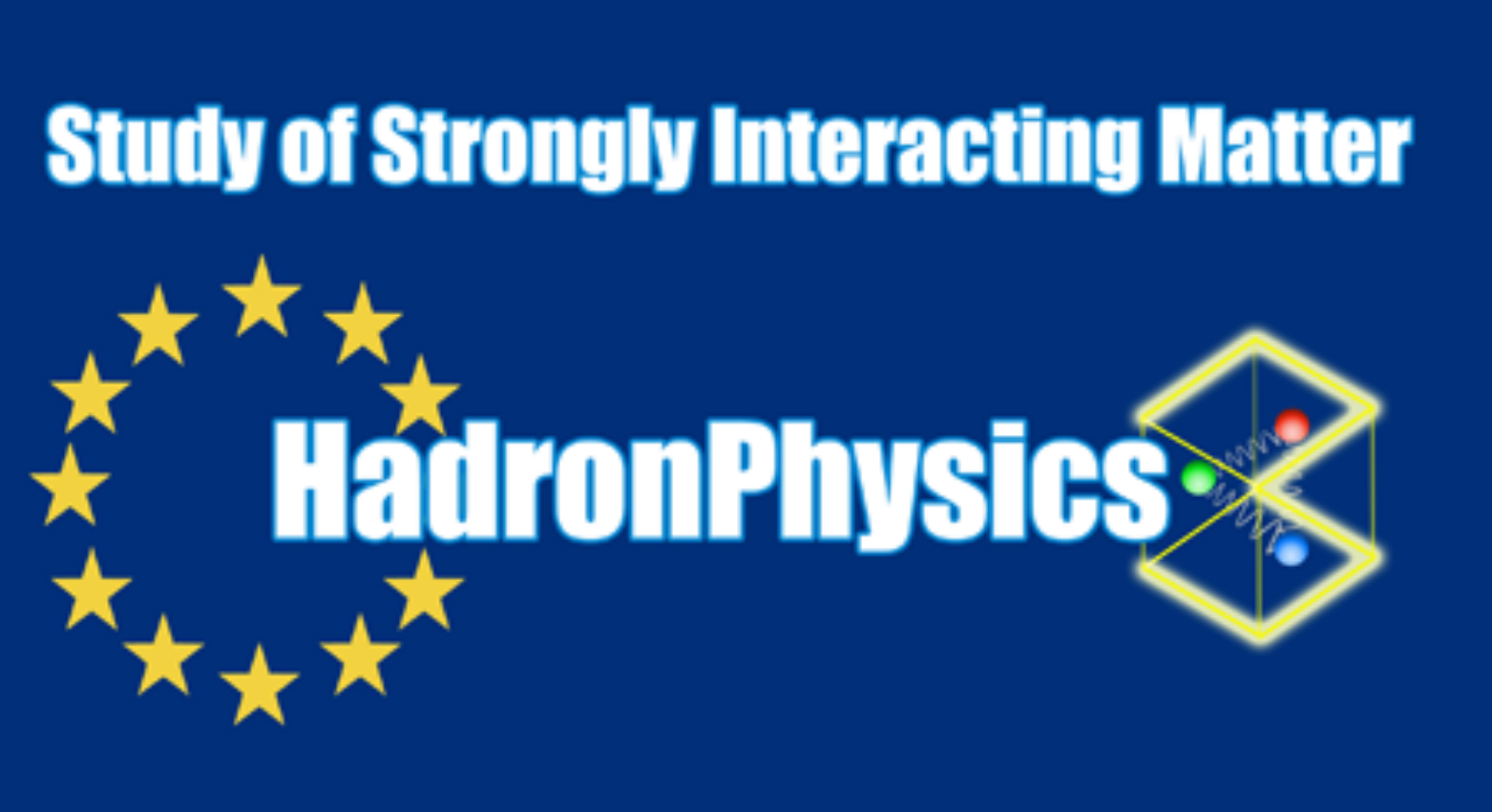}
\end{subfigure}
\begin{subfigure}[b]{0.5\textwidth}
\includegraphics{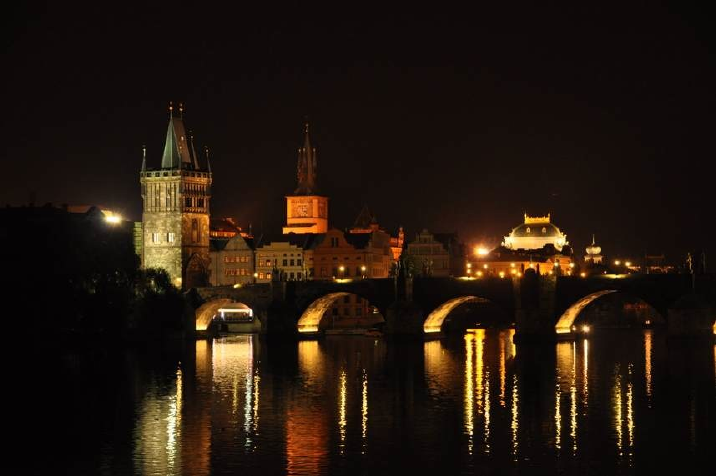}
\end{subfigure}
\label{default}
\end{center}
\end{figure}

\newpage

\section{Summaries of the talks}

\subsection{Light meson decays in CLAS. Current status and future plans}
\addtocontents{toc}{\hspace{2cm}{\sl M.J.~Amaryan}\par}

\vspace{5mm}

M.J.~Amaryan

\vspace{5mm}

\noindent
Old Dominion University\\Norfolk, Virginia, USA

\vspace{5mm}

We present preliminary results on a decay of light mesons in CLAS setup at Thomas Jefferson National Accelerator Facility.
Experimental data are collected in photoproduction reactions on hydrogen target. The main particles we have studied
are pseudoscalar - $\pi^0, \eta$, $\eta^{\prime}$ and $\rho, \omega$ and $\phi$ vector mesons. Decay channels can be classified
into three groups: A) Dalitz decays with $e^+e^-\gamma$ for $\pi^0, \eta$ and for the first time for $\eta^{\prime}$, $e^+e^-\pi^+\pi^-$ for $\eta$ 
and $\eta^{\prime}$, $e^+e^-\pi^0$ for $\omega$ meson; B) radiative decays of $\eta$, $\eta^{\prime}$ and $\rho/\omega$ to $\pi^+\pi^-\gamma$; 
C) Hadronic decays $\eta(\eta^{\prime} \to \pi^+\pi^-\pi^0$, $\eta^{\prime}\to \pi^+\pi^-\eta$. The physics topics include precise measurement of transition form factors, Dalitz plot analysis, as well as C-parity and CP violation processes. The data already on tape exceed the world data in some of these channels 
by an order of magnitude. Besides the traditional topics, existing data may improve on the upper limit of  the $A^{\prime}-\gamma$ mixing strength parameter of hypothetical new gauge  field boson with electromagnetic field $\gamma$.  The latter may be accomplished via study of the invariant mass of $e^+e^-$ pairs from Dalitz decays of pseudoscalar mesons up to the mass of $\eta$. Finally, it should be mentioned that huge statistics collected at CLAS may provide an excellent opportunity to  improve on the upper limit of dark matter by searching for invisible decays of pseudoscalar mesons reconstructed using missing mass technique.

\newpage

\subsection{New results on the ABC Resonance Structure }
\addtocontents{toc}{\hspace{2cm}{\sl M.~Bashkanov}\par}

\vspace{5mm}

M.~Bashkanov and H.~Clement

\vspace{5mm}

\noindent
Physikalisches Institut \\ Kepler Center for Astro and Particle Physics \\
Universt\"at T\"ubingen, Germany\\ 

\vspace{5mm}

The question whether the two-baryon system possesses more 
eigenstates than just the deuteron groundstate awaits an answer since
decades. Based on quantum chromodynamics (QCD) the possible existence of
so-called dibaryons was first envisaged by Dyson~\cite{Dyson1} in 1964 and then
by Jaffe~\cite{Jaffe1} in 1977. The latter publication initiated worldwide
dibaryon searches, which, however, have been finally unsuccessful. 

Now, in high-precision measurements of the WASA-at-COSY collaboration in
Juelich a narrow isoscalar resonance-like structure has been observed recently
in neutron-proton collisions in the case that a deuteron and a neutral pion
pair are formed~\cite{PRL2011}. The differential distributions are consistent
with spin-parity 3+ and with a two-Delta decay mode peaking significantly
below the nominal two-Delta mass. In subsequent experimental searches by the
WASA-at-COSY collaboration this resonance structure has been also found in the
double-pionic fusion reactions $pn \to d\pi^+\pi^-$~\cite{isoabc1}, $pd \to
^3$He$\pi^0\pi^0$~\cite{EP} and $dd \to ^4$He$\pi^0\pi^0$ \cite{AP} (in the
latter two as a dibaryon excitation within the 
nuclear medium) as well as in the non-fusion reaction $pn \to pp\pi^0\pi^-$
\cite{pppi0pi-1}. The analysis of data for the $pn \to pn \pi^0\pi^0$ reaction
and for the isoscalar part of single pion production in $NN$ collisions is in
progress. In the first reaction the resonance structure should also show up,
whereas in the latter reaction the appearance of this resonance would be a
surprise. Note that since no evidence for this structure has been found in
$pp$ initiated two-pion production
\cite{hcl,JJ,WB,JP,TS,iso,FK,deldel,nnpipi,JK,evd,tt}, it must be of purely
isoscalar nature. 

The resonance structure with M = 2370 MeV,  
$\Gamma$ = 70 MeV and $I(J^P) = 0(3^+)$ appears to be tightly correlated with
the so-called ABC effect~\cite{ABC}, an intriguing low-mass enhancement in
isoscalar $\pi\pi$ invariant mass distributions. This effect points to a
charcteristic feature of the decay of this resonance. The Dalitz plots of the
double-pionic fusion reactions exhibit the dominant decay to proceed via an
intermediate $\Delta\Delta$ configuration. 

The {\it experimentum crucis} for solving the question, whether this observed
resonance structure denotes truly a $s$-channel resonance in the $NN$ sytem,
is high-precision $np$ scattering, where in the energy region of interest
the partial waves $^3D_3$ and / or $^3G_3$ should exhibit a resonance-like
behavior. A theoretical investigation of $pn$ observables shows that the total
cross section and the analyzing power give the highest sensitivity to these
resonating partial waves~\cite{proposal1,PBC1}. Hence measurements of polarized
quasielastic neutron scattering on hydrogen have been carried out with the
WASA setup at COSY. Preliminary results are very encouraging and are reported
at the workshop.

Implications of the observation of this resonance structure on further
dibaryon searches are discussed in a subsequent contribution to this workshop
\cite{mesonnetHCl1}.

This work has been supported by BMBF and COSY-FFE (FZ J\"ulich).

\newpage

\subsection{Estimates for the muon $g-2$ using  VMD Constraints And Global Fit Methods}
\addtocontents{toc}{\hspace{2cm}{\sl M.~Benayoun}\par}

\vspace{5mm}

M.~Benayoun

\vspace{5mm}

\noindent
LPNHE des Universit\'es Paris 6 et 7, Paris, France\\
\vspace{5mm}

The anomalous magnetic moment of the muon $a_\mu=(g-2)/2$ is one of the most precisely measured
particle physics quantity as its numerical value ($[11\,659\,208.9 \pm 6.9]\times 10^{-10}$)
 is known with an accuracy of 0.5 ppm~\cite{BNL}.  Most of the contributions
to its theoretical value can be computed with an accuracy at the level of a few $10^{-11}$
or better (see~\cite{Jegerlehner}, for instance); this covers the QED and electroweak contributions
but also the hadronic contribution from the onset of the perturbative regime of QCD on. However, at lower energies,
one is in the realm of the non--perturbative regime of QCD and, then, the corresponding contributions 
to $a_\mu$ have to be evaluated using other methods. Among these, 
the most important piece is  by far the 
so--called hadronic vacuum polarization (HVP) which is estimated by integrating the 
annihilation cross--sections to definite hadronic final states $\{H_i, i=1 \cdots n\}$,
$e^+e^- \to H_i$ weighted by some kernel enhancing the very low energy 
region. 

Up to very recently, the single operating method to evaluate these HVP pieces  was to 
numerically integrate the  $experimental $ spectra, relying on more or 
less sophisticated interpolation  and extrapolation methods. 

However, another method has been proposed and successfully applied in our~\cite{ExtModxx}. 
Basically, the paradigm is to rely on the well--defined Hidden Local Symmetry (HLS) Model~\cite{HLS}
and construct an effective Lagrangian  which preserves the symmetries of QCD. The HLS  Model corresponds to 
an extension of the Chiral Perturbation Theory which includes  vector mesons and thus allows to go deeper
inside the energy region dominated by resonances. Nevertheless, in order to be really operating with
experimental data, the original HLS  
Model should be complemented with symmetry breaking mechanisms.
This  is performed is two steps~: One first implements 
a variant of the original (BKY) scheme as defined in~\cite{BKY}; a second mechanism happens to
be crucial, the vector meson mixing~\cite{ExtModxx}
generated by the breaking of isospin symmetry in the pseudoscalar meson sector.

Equipped with these breaking schemes, the HLS Lagrangian~\cite{ExtModxx}
encompasses most of the physics involving
the basic nonet of vector mesons and is expected to improve the account of all annihilation
channels up to -- and including -- the $\phi$ meson region, {\it i.e.} about 85 \% of the  HVP.  
In particular, the realm covered by the broken HLS model (BHLS) incorporates six annihilation channels~:
$e^+ e^- \to \pi^+ \pi^-/K^+ K^-/K^0 \overline{K^0}/\pi^+ \pi^-\pi^0/\pi^0 \gamma/\eta\gamma$ and
the $\tau^\pm \to \pi^\pm \pi^0 \nu$ decay spectrum up to $\simeq 1.05 $ GeV. 

The BHLS model substantiates several important properties~: {\bf 1/} 
Because of the  $physics$ correlations implied by the unified BHLS framework,
the information provided by data sets covering any channel ($e^+ e^- \to \pi^0 \gamma$
or $\tau^\pm \to \pi^\pm \pi^0 \nu$) are equivalent to having larger statistics covering
$any$ other channel ($e^+ e^- \to \pi^+ \pi^-$ or $e^+ e^- \to \pi^+ \pi^- \pi^0$, for instance),
{\bf 2/} The statistical consistency issue of the various data samples can be addressed and
is reflected by the global fit probability, {\bf 3/} If several data samples covering the same 
physics channel are conflicting, the probability of the global fits including in turn each of these 
in isolation gives a hint of its consistency with the rest of the physics covered by the global model,
{\bf 4/} Finally, the fit results (parameter values and error covariance matrix) and the BHLS
theoretical cross--sections provide a more optimal and refined method
to compute the contribution of any given channel to $a_\mu$, instead of evaluating the numerical 
integration of the  experimental spectra. Above the  $\phi$ mass region, however, one cannot
presently avoid the use of this usual method. 

It has been shown previously~\cite{ExtModxx} that almost all existing data samples covering the collection 
of channels quoted above are statistically consistent with each other. Focussing on the most important channel,
$e^+ e^- \to \pi^+ \pi^-$, it was thus shown
that all data collected -- in scan mode --  by CMD--2 and SND \cite{NSK} on $e^+ e^- \to \pi^+ \pi^-$
lead already to a global fit probability
larger than 90\%. and provide a significance for $a_\mu^{exp}-a_\mu^{th}$ of 4.5 $\sigma$,
larger by 1. $\sigma$ than usual estimates.

The consistency issue has been examined for the 4 data samples covering the $e^+ e^- \to \pi^+ \pi^-$
channel 
collected by the KLOE and BaBar detectors using the Initial State Radiation (ISR) method. It was thus 
shown that two among these, namely the samples~ \cite{KLOE++} behave in perfect accord with all
the other --more than 40 -- data sets, in particular with the scan data  from~ \cite{NSK}.
However, the samples in~\cite{Discard} reveal clear  inconsistencies which forces us to discard them
from the global fit treatment, as explained in~\cite{ExtModxx}.

Using together the ISR~\cite{KLOE++} and scan~\cite{NSK} data for $e^+ e^- \to \pi^+ \pi^-$ within the
BHLS framework leads to a global fit probability of 93\% and enlarges the significance
of $a_\mu^{exp}-a_\mu^{th}$ to 5 $\sigma$. This provides the first serious hint that new
physics is at work in the muon $g-2$.

\newpage

\subsection{Investigation of Total and Differential Cross Sections of the Reaction $pd\rightarrow\, ^3He\,+\,\eta$ at 49 and 60 MeV Excess Energy}
\addtocontents{toc}{\hspace{2cm}{\sl F.~Bergmann}\par}

\vspace{5mm}

Florian Bergmann*, Kay Demmich, Paul Goslawski, Nils H\"{u}sken, Alfons Khoukaz, Annika Passfeld, Alexander T\"{a}schner\\
for the WASA-at-COSY Collaboration

\vspace{5mm}
\noindent
Institut f\"{u}r Kernphysik, Westf\"{a}lische Wilhelms-Universit\"{a}t M\"{u}nster\\
\vspace{5mm}

The $\eta$-meson production reaction $\mathrm{p}+\mathrm{d}\rightarrow{^3{\mathrm{He}}}+\eta$ has been used recently for various precision experiments at COSY-J\"{u}lich, {\it e.g.}\ for the investigation of the $\eta$-nucleus final state interaction, the search for possible $\eta$-mesic nuclei as well as for the $\eta$-mass determination with highest accuracy \cite{Mers2007,Raus2009,Gosl20121}. A remarkable feature of this reaction is the unexpected shape of the excitation function which is strongly influenced by the $\eta$-${^3{\mathrm{He}}}$ final state interaction. Different to the close-to-threshold region, where a rich data sample exists~\cite{Mers2007,Smyrski2007,Berger1988,Mayer1996,Adam2007}, only limited information are available at higher excess energies~\cite{Raus2009,Bilger2002,Betigeri2000}. Hence, at the WASA-at-COSY experiment new measurements with high statistics have been performed at $Q =$ 49 and $60\,\mathrm{MeV}$~\cite{Pass2010,Jany2010}. The large angular acceptance of the detector allows for detailed investigations on angular distributions as well as on their energy dependence~\cite{Adam2004}. Final results on total and differential cross sections have been presented and discussed.

*Supported by COSY-FFE grants

\newpage

\subsection{Status of the muon $g-2$ hadronic light-by-light contribution}
\addtocontents{toc}{\hspace{2cm}{\sl J.~Bijnens}\par}

\vspace{5mm}

Johan Bijnens

\vspace{5mm}

\noindent
Department of Astronomy and Theoretical Physics, Lund University, Sweden\\

\vspace{5mm}

The measured muon anomalous magnetic moment \cite{gm2exp3}
shows a discrepancy with the standard model prediction
of 3--4$\sigma$, see e.g. the PDG report \cite{PDGBijnens}.
The new experiments at Fermilab and J-PARC plan to decrease the uncertainty
further by a factor of four. The theoretical uncertainty is at present
dominated by the lowest order hadronic vacuum polarization but this error
can be systematically reduced by experiment. A recent treatment can be found
in the next talk \cite{Benayoun}. The main uncertainty in the future
will be the hadronic light-by-light contribution which is
roughly summarized by $a_\mu^{LbL} = (10.5\pm2.6)\,10^{-10}$ \cite{PdRV}.
Most summaries quote values between 10--14 and an error of
2.6--4 in units of $10^{-10}$.

The underlying hadronic object is the Green function of four
electromagnetic currents integrated over two of the photon momenta and the
third photon momentum set to zero after taking the derivative w.r.t. it.
This object has 138 Lorentz-structures and each of them is after the above
reduction still a function of three independent variables, the off-shellness
or mass of the three photons connecting to the muon line.
This is what makes the hadronic light-by-light so hard to calculate and the
mixing of long and short distances makes possible double counting of quark-gluon
and hadron contributions a very difficult problem.
Using de Rafael's \cite{deRafael} suggestion of using $N_c$ and chiral counting
as a guide two groups did a full estimate with similar final numbers
\cite{HKS,BPP}. A sign mistake was found in both by
\cite{KN} and the main contribution,
$\pi^0,\eta,\eta^\prime$-exchange, has been recalculated many times with all
results fitting in the range (8--10)$\,10^{-10}$, for references see
\cite{BP,JNreview}. A short distance constraint was found in \cite{MV}
which increased the result. This is discussed more in \cite{PdRV,BP}.

The main new part I discussed is the pion loop contribution.
The models originally used were the hidden local symmetry (HLS) model
and the ENJL model were all photon propagators are modulated with a factor
resembling $m_V^2/(m_V^2+Q^2)$. The full VMD model uses exactly that factor.
These gave $-0.45\,10^{-10}$,$-1.9\,10^{-10}$, $-1.6\,10^{-10}$ respectively.
The large difference between the first and latter two is disturbing.
In \cite{talk,Mehranthesis,BR} this was studied further where it was shown
that the HLS model has contributions of the opposite sign at higher
photon masses. \cite{Ramsey-Musolf} suggested that pion polarizability
effects might be important. Pure ChPT can only be used here up to
500 MeV cut-off or so, so to fully study the effects models with an $a_1$
have to be introduced. However, even with including many more couplings,
no satisfying model with the $a_1$ that actually gives a finite result
for the muon $g-2$ was found \cite{BR}. However all models that gave a
reasonable low-energy behaviour when integrated up to about 1~GeV gave similar
answer with a result $a_\mu^{LbL\,\pi\mathrm{-loop}} = (-2.0\pm0.5)\,10^{-10}$
which is the new preliminary result for this contribution.

A recent evaluation of the light-by-light contribution using the
Dyson-Schwinger equations as an overall model \cite{SDE}
has a much larger contribution from the quark-loop then found earlier.
Other estimates using a smaller quark mass for at least part
of the low-energy domain are \cite{BM,GR,Masjuan}. However, the question
whether double-counting is involved when using such low masses is not
fully solved. 
A final remark is that models with a finite number of resonances
will have to make compromises between various QCD constraints \cite{BGLP}.

\newpage

\subsection{Experimental Dalitz Plot analysis for $\eta \rightarrow \pi^+ \pi^- \pi^0$ }
\addtocontents{toc}{\hspace{2cm}{\sl L.~Caldeira Balkest\aa{}hl}\par}

\vspace{5mm}

 L.~Caldeira Balkest\aa{}hl$^1$ (on behalf of the KLOE-2 collaboration)\\
 P.~Adlarson$^2$ (on behalf of the WASA-at-COSY collaboration)

\vspace{5mm}

\noindent
$^1$Department of Physics and Astronomy, Uppsala University\\
$^2$Department of Physics and Astronomy, Uppsala University\\
\vspace{5mm}

The isospin-violating process $\eta\rightarrow \pi^+\pi^-\pi^0$ is sensitive to the light quark mass ratio:
\[
Q^2= \frac{m_s^2-m^2}{m_d^2-m_u^2}, \qquad  m=\frac{m_d+m_u}{2}
\]
in that the decay rate is proportional to $Q^{-4}$. The values for this decay rate, calculated at leading order ($\Gamma_{LO} \sim 70$ eV) and next-to-leading order ($\Gamma_{NLO}=160 \pm 50$ eV) in chiPT are significantly lower than the experimental value $\Gamma_{exp}=296\pm 16$ eV, obtained from a fit to all the available data \cite{pdg}. This points to a slow convergence of the chiPT series, due to strong pion rescattering effects in the final state, which  can be  treated by means of dispersion relations \cite{colangelo}.

In 2008, the KLOE collaboration published the Dalitz plot analysis of the $\eta\rightarrow \pi^+\pi^-\pi^0$   with the most statistics to date \cite{kloep}. The results have been used to fix the parameters of the dispersion relations \cite{lanz} and  in analytic dispersive analysis~\cite{zdrahal}. More data on $\eta\rightarrow \pi^+\pi^-\pi^0$ Dalitz plot are needed to understand the origin of the residual tension between data and theoretical calculations.

A new analysis by KLOE is in progress to improve on the statistical sample and to overcome some limitations of the previous analysis. For this, a new selection scheme is used. The selection efficiency is planned to be  measured directly from minimum bias events, to reduce systematic errors.  The analysis is performed on an independent, larger data sample. The preliminary results are presented in Table~\ref{tab:results}.

An analysis by the WASA-at-COSY collaboration is also in progress. The WASA preliminary results from the analysis with 4 weeks of data taking in  $pd \rightarrow \eta ^3He$ are presented in table  \ref{tab:results}. The WASA analysis is described in more detail in \cite{patrik}.

The Dalitz plot variables $X$ and $Y$ are defined in the $\eta$ rest frame as
\[
X= \sqrt{3} \frac{T_+ - T_-}{Q_\eta} = \frac{\sqrt{3}}{2 m_\eta Q_\eta}(u-t)
\]

\[
Y= \frac{3T_0}{Q_\eta} -1 =  \frac{\sqrt{3}}{2 m_\eta Q_\eta} \left[ \left(  m_\eta - m_{\pi^0} \right)^2 - s \right] -1
\]
where $Q_\eta =T_+ +  T_- + T_0 = m_\eta - 2m_{\pi^+} -m_{\pi^0}$;  $T_+$,$T_-$, $T_0$ are  kinetic energies of the $\pi^+$, $\pi^-$, $\pi^0$ and  $s$, $u$, $t$ are the Mandelstam variables.

The Dalits plot amplitude is parametrized as $|A(X,Y)|^2=N(1+aY+bY^2+cX+dX^2+eXY + fY^3 + gX^2Y) $. The results for both analysis, and well as the previous KLOE measurement \cite{kloep} are shown in Table~\ref{tab:results}. Both analysis find $c$ and $e$ consistent with zero in accordance with C-invariance.

\begin{table}[!h]
\begin{center}
\begin{tabular}{|@{}l@{} |l l l l|}
\hline
 \textbf{\,Experiment\,} & $-a$ &$ b$ & $d$ & $f$ \\
\hline \hline
\,KLOE 08 & $1.090(5)(^{+8}_{-19})$ & $0.124(6)(10)$ & $0.057(6)(^{+7}_{-16})$ &$ 0.14(1)(2)$\\
\,WASA prel. & $1.074(23)(3)$ & $0.179(27)(8)$ & $0.059(25)(10)$ &$ 0.089(58)(110)$\\
\,KLOE prel. & $1.104(3)$ & $0.144(3)$ & $0.073(3)$ & $0.155(6)$ \\
\hline
\end{tabular}
\caption{Results for the fit to the Dalitz plot.\label{tab:results}}
\end{center}
\end{table}

\newpage

\subsection{Resonance Multiplets in the Two-Baryon System --- Dibaryons
 revisited}
\addtocontents{toc}{\hspace{2cm}{\sl  H.~Clement}\par}

\vspace{5mm}

 H.~Clement and M.~Bashkanov

\vspace{5mm}

\noindent
Physikalisches Institut \\ Kepler Center for Astro and Particle Physics \\
Universt\"at T\"ubingen, Germany\\ 

\vspace{5mm}

Only few weeks after Gell-Mann's famous paper on the quark model in 1964
\cite{Gell-Mann} Dyson published a prediction for six non-strange states 
in the two-baryon system based on symmetry breaking in SU(6)
\cite{Dyson}. Denoting these states by $D_{IJ}$ with (isospin I, spin J) = (0,1), (1,0),
(1,2), (2,1), (0,3) and (3,0) he associated the two lowest-lying states with
the deuteron ground state ($D_{10}$) and the virtual $^1S_0$ state ($D_{01}$),
respectively. The remaining four states were
predicted to be excited states with masses up to 2350 MeV. 
Whereas this prediction did not receive much attention, it was in 1977  
Jaffe's note on the possible existence of a bound six-quark system,
the H-dibaryon (denoting asymptotically a bound $\Lambda\Lambda$
system) \cite{Jaffe}, which initiated numerous theoretical investigations
\cite{goldman,huang,mulders,maltman,barnes,kamae,oka,sato,li,kukulin,mota}
predicting a vast number of states in the system of two baryons. In the 
subsequent experimental hunt for dibaryons many claims have been announced,
however, none survived careful experimental investigations. 
 
The interest in dibaryons revived recently, when two groups announced that
lattice QCD calculations provide evidence for a bound H-dibaryon. Also 
recently it has been noted that the double-pionic fusion reactions $pn \to
d\pi^0\pi^0$  and $pn \to d\pi^+\pi^-$ proceed dominantly via a resonance
structure observed in the total cross section at $\sqrt s$ = 2.37 GeV with
$\Gamma \approx$ 70 MeV and $I(J^P) = 0(3^+)$ \cite{prl2011,isoabc}, {\it
  i.e.} corresponding to $D_{03}$ in Dyson's notation. Further
evidence for this resonance has meanwhile been found in the $pn \to
pp\pi^0\pi^-$ reaction \cite{pppi0pi-} and -- most importantly -- in $np$
scattering \cite{proposal,PBC,mesonnetMB}, which is so to speak the {\it experimentum
  crucis} for the confirmation of a $s$-channel resonance in the $np$ system. 
Since its decay proceeds dominantly via an intermediate $\Delta\Delta$ system,
this resonance constitutes asymptotically a $\Delta\Delta$ system bound by
nearly 100 MeV. 

In recent years also another resonance got established by SAID phase shift
analyses, which resonates in the $^1$D$_2$ $pp$ partial wave at $\sqrt s$ =
2.144 GeV with $\Gamma \approx$ 110 MeV
\cite{arndt,hoshizaki1,hoshizaki2}, {\it i.e.} $D_{12}$ in Dyson's
notation. Since it resides just near the $N \Delta$ 
threshold with a width compatible to that of the $\Delta$, it is assumed to be
a loosely bound (molecular) $\Delta N$ configuration. 

None of the many dibaryon predictions is able to predict both resonances at the
proper energies -- with the remarkable exception of Dyson's multiplet
prediction. If this puts confidence into the predictive power, then there are
two more states awaiting its discovery, one with isospin I = 2 ($D_{21}$) and
one with isospin I = 3 ($D_{30}$). Since both are decoupled from $NN$, this
means search in dedicated two-pion and four-pion production channels. The
available WASA-at-COSY data base provides the posibilty to search for these
states of putative exotic nature.

This work has been supported by BMBF and COSY-FFE (FZ J\"ulich)

\newpage

\subsection{Dispersive Analysis of Scalar Pion and Kaon Form Factors}
\addtocontents{toc}{\hspace{2cm}{\sl  J.~Daub}\par}

\vspace{5mm}

 J.~Daub$^{a,c}$, C.~Hanhart$^b$, B.~Kubis$^{a,c}$

\vspace{5mm}

\noindent
${}^a$ Helmholtz-Institut f\"ur Strahlen- und Kernphysik, Universt\"at Bonn, Germany\\
${}^b$ Forschungszentrum J\"ulich GmbH, Germany\\
${}^c$ Bethe Center for Theoretical Physics, Universit\"at Bonn, Germany
\vspace{5mm}

The scalar pion and kaon form factors are investigated in a dispersive formalism. Due to strong inelastic effects in the scalar sector---namely the coupling of two S-wave pions to $\bar{K}{K}$ near the $f_0(980)$ resonance---the treatment by means of a coupled-channel approach is mandatory. Therefore the Omn\`{e}s problem is generalized to a two-channel Muskhelishvili--Omn\`{e}s problem~\cite{MuskhOm}, which is solved numerically~\cite{DGL90+Moussallam2000}.
Three input functions are required: the $\pi\pi$ S-wave phase shift, known from a Roy equation analysis~\cite{CCL2012}, and the $\pi\pi \to \bar{K}K$ amplitude, whose modulus is known from a Roy--Steiner analysis of $\pi K$ scattering~\cite{BDM04}, and the phase from a partial-wave analysis~\cite{CohenEtkin}. The normalizations of strange and non-strange scalar form factors are related to the pion and kaon masses according to the Feynman--Hellmann theorem~\cite{Hellmann},
employing low-energy constants from lattice simulations~\cite{FLAG}.

We present two applications: a study of the decays $J/\psi \to \phi \pi^+ \pi^-$, $\phi K^+ K^-$,  $\omega \pi^+ \pi^-$, $\omega K^+ K^-$, based on an SU(3) invariant chiral Lagrangian, and an analysis of the decay $\tau \to \mu \pi^+ \pi^-$  to improve bounds on R-parity violation. 

The adoption of the scalar-form-factor analysis to the decay $J/\psi$ into a vector and two pseudoscalar mesons
is suggestive due to identical final-state interactions when neglecting left-hand-cut structures~\cite{MeiOll}. 
However, the model thus strongly constrained by dispersion relations seems to systematically underpredict the 
$J/\psi \to \phi K^+ K^-$ decay distribution,  when compared to data from the BES collaboration~\cite{Ablikim}.
Since it is not obviously justified to entirely ignore crossed-channel effects, left-hand-cut structures from sequential axial-vector- and vector-meson exchanges are also investigated~\cite{Roca}.

We study the hadron physics in the lepton-flavor-violating decay $\tau \to \mu\pi^+\pi^-$,
based on the example of an R-parity-violating superpotential~\cite{DDHKM}. 
The hadronization of the effective quark operators into pions proceeds precisely through pion scalar and vector form factors,
thus describing the strong final-state interaction of the pion pair in a model-independent way,
eschewing any assumptions on the nature of (in particular) the scalar resonances.
Experimental upper limits on these decays (from the Belle collaboration~\cite{Belle}) can therefore be translated into upper limits on 
certain underlying (in this case: supersymmetric) operators with very little hadronic uncertainty.

\newpage

\subsection{Meson Studies at Belle}
\addtocontents{toc}{\hspace{2cm}{\sl S.~Eidelman}\par}

\vspace{5mm}

S.~Eidelman

\vspace{5mm}

\noindent
Budker Institute of Nuclear Physics SB RAS \\
and Novosibirsk State University, \\ 
 Novosibirsk, Russia

\vspace{5mm}

In addition to the main task of studying CP violation in the $B$ meson system,
Belle uses its large integrated luminosity of about 1 ab$^{-1}$ to solve 
various other problems. In particular, the recorded data samples can be
used to study mesons made of light quarks. 

Recently  the process 
$\gamma\gamma \to \omega\omega,~\omega\phi,~\phi\phi$ 
was investigated  with 870 fb$^{-1}$~\cite{Liu:2012eb}.
The authors observe signals of the $\eta_c(1S),~\chi_{c0}(1P),~\chi_{c2}(1P)$
in the $\omega\omega,~\phi\phi$ modes and also find evidence for 
additional structures. The $2D$ angular analysis gives an
$\omega\omega$ state at 1.91 GeV, which is a mixture of $0^+$- and $2^+$-waves, 
an $\omega\phi$ state at 2.2 GeV ($0^+$- and $2^+$-waves) and
a $\phi\phi$ state at 2.35 GeV ($0^+$- and $2^-$-waves). 

With a data sample of 673 fb$^{-1}$ they study the process
$\gamma\gamma \to \eta^\prime\pi^+\pi^-$ in a search for the
state $X(1835)$ observed in various final states by BES~\cite{Zhang:2012t}. 
They find first evidence for the decay mode 
$\eta(1760) \to \eta^\prime \pi^+\pi^-$ and 
determine its mass and width to be $1768^{+24}_{-25}\pm10$ MeV and
$224^{+62}_{-56}\pm 25$ MeV, respectively. Also measured 
for the $\eta(1760)$  is the product of its 
two-photon width and branching fraction to
$\eta^\prime \pi^+\pi^-$. They do not find 
evidence for the $X(1835)$ and place corresponding upper limits for its
two-photon width times the branching fraction to
$\eta^\prime \pi^+\pi^-$ for  constructive and  destructive interference.

The Belle Collaboration also investigated with initial-state radiation
(ISR) the process $e^+e^- \to K^+K^-\pi^+\pi^-$ based on 
673 fb$^{-1}$~\cite{Shen:2009zze}. They observed for the first time
a clear signal of the $\phi(1680) \to \phi\pi^+\pi^-$ decay mode,
showed that in the vicinity of the $\phi(2170)$
this final state is dominated by the $\phi f_0(980)$ and determined
mass, total width and the leptonic width times the branching fraction
to the final state for both  $\phi(1680)$ and   $\phi(2170)$. 

Various two-meson final states were studied with Belle in 
$\gamma\gamma$ collisions in a zero-tag mode:
$\pi^+\pi^-$~\cite{Nakazawa:2004gu,Mori:2006jj,Mori:2007bu}, 
$\pi^0\pi^0$~\cite{Uehara:2008ep,Uehara:2009cka}, 
$K^+K^-$~\cite{Abe:2003vn,Nakazawa:2004gu}, 
$K^0_SK^0_S$~\cite{Chen:2006gy}, $\eta\pi^0$~\cite{Uehara:2009cf}, 
$\eta\eta$~\cite{Uehara:2010mq} and $p\bar{p}$~\cite{Kuo:2005nr}.
PWA of different waves in these processes revealed various light mesons: 
$f_0(980),~a_0(980),~f_2(1270)$, \\ $a_2(1320),~f_0(1370/1500),~f^\prime(1525)$ 
and measured their $\Gamma_{\gamma\gamma}{\cal B}_f$. This quantity has
also been determined for the charmonium states 
$\eta_c(1S),~\chi_{c0}(1P),~\chi_{c2}(1P)$ observed at high 
$\gamma\gamma$ masses.

Belle has also showed that important information on various $K^*$
states can be obtained from $\tau$ lepton decays. For example,
a study of the $\tau^- \to \bar{K}^0\pi^-\nu_{\tau}$ decay mode 
with 351 fb$^{-1}$ allowed a number of various measurements to be performed:
determination of the mass and width of the $K^*(892)$,    
observation of higher mass $K^*$'s near 1400 MeV, evidence for
existence of the $K^*_0(800)$~\cite{Epifanov:2007rf}.

One can conclude that $B$ factories have high potential for observing 
light mesons with different quantum numbers via various mechanisms --
in $B$ meson and $\tau$ lepton decays, in continuum and ISR as well as in 
$\gamma\gamma$ collisions. These mechanisms were also successfully used
in the Belle experiment to discover and study a number of charmonium-
and bottomonium-like states~\cite{Beringer:2012zz}.

\newpage

\subsection{$\eta$ and $\eta^\prime$ physics at BESIII}
\addtocontents{toc}{\hspace{2cm}{\sl S.~Fang}\par}

\vspace{5mm}

S.~Fang (on behalf of the BESIII collaboration)

\vspace{5mm}

\noindent
Institute of High Energy Physics, Beijing, China\\
\vspace{5mm}

Both $\eta$ and $\eta^\prime$,
discovered about half of a century ago, are two important states in the lightest pseduoscalar nonet,  which attracts considerable interest in the decays both theoretically and experimentally because of their special roles in low energy scale quantum chromodynamics 
theory.  Their dominant radiative and hadronic decays were observed and well measured, but the study of their anomalous  decays is still an open field.  A sample of 225.3 million $J/\psi$ events taken at the BESIII detector in 2009  offers a unique opportunity to study $\eta$ and $\eta^\prime$ decays via $J/\psi \rightarrow \gamma \eta(\eta^\prime)$ or $J/\psi\rightarrow\phi\eta(\eta^\prime)$.

 With a new level of precision,
the Dalitz plot parameters for $\eta^\prime\rightarrow \pi^+\pi^-\eta$ are determined in a generalized and a linear representation~\cite{bes3_dalitz}. In general the results are in reasonable agreement with the previous measurements and the C-parity violation is not evident.  The statistical error of the parameters are still quite large,
much more data strongly needed to provide a more stringer test of the chiral theory. The decays of $\eta^\prime\rightarrow \pi^+\pi^-e^+e^-$ and $\eta^\prime\rightarrow \pi^+\pi^-\mu^+\mu^-$ were also studied via $J/\psi\rightarrow \gamma\eta^\prime$~\cite{bes3_ppmm}. A clear $\eta^\prime$ peak is observed in 
the $M_{\pi^+\pi^-e^+e^-}$mass spectrum, and the branching fraction is measured to be $B(\eta^\prime\rightarrow\pi^+\pi^-e^+e^-)=(2.11 \pm 0.12\pm 0.14)
\times 10^{-3}$, which is in good agreement with theoretical predictions~\cite{theo} and the previous measurement~\cite{cleo}, but is determined with much higher precision. The mass spectra of $M_{\pi^+\pi^-}$ and $M_{e^+e^-}$ are also consistent with the theoretical predictions~\cite{theo} that $M_{\pi^+\pi^-}$ is dominated by $\rho^0$ , and $M_{e^+e^-}$ has a peak just above 2$m_e$ .   No $\eta^\prime$ signal is found in the $M_{ \pi^+\pi^-\mu^+\mu^-}$ mass spectrum, and the upper limit is determined to be 
$B(\eta^\prime\rightarrow  \pi^+\pi^-\mu^+\mu^-) < 2.9\times10^{-5} $ at the 90\% confidence level.  To test the fundamental symmetries, a search for P and CP violation decays of
$\eta/\eta^\prime\rightarrow \pi^+\pi^-,\pi^0\pi^0$ was performed~\cite{bes3_cp}.  No evident signals were observed, and then the branching fraction upper limits,
$B(\eta\rightarrow\pi^+\pi^-)<3.9\times 10^{-4}$,$B(\eta\rightarrow\pi^0\pi^0)<6.9\times 10^{-4}$,
$B(\eta^\prime\rightarrow\pi^+\pi^-)<5.5\times 10^{-5}$ and $B(\eta^\prime\rightarrow\pi^0\pi^0)<4.5\times 10^{-4}$,
 are presented at the 90\% confidence level.
 
  In addition we  made an attempt to search for their invisible and weak decays via $J/\psi\rightarrow\phi\eta$ and $J/\psi\rightarrow\phi\eta$~\cite{bes3_inv,bes3_weak}.
 These two-body decays provide a very simple event topology, in which the $\phi$ meson can be reconstructed easily and cleanly with its dominant decays
 of $\phi\rightarrow K^+K^-$ .  Since the $\phi$ and $\eta (\eta^\prime)$ are given strong boosts in the $J/\psi$ decay,  
 the invisible decays of the $\eta$  and $\eta^\prime$ were investigated with the mass spectra recoiling against $\phi$.  We find no signal above background for the invisible decays of $\eta$ and $\eta^\prime$. To reduce the systematic uncertainty, the upper limits of the ratios, 
 $\frac{B(\eta\rightarrow invisible)}{B(\eta\rightarrow\gamma\gamma)}<2.6\times 10^{-4}$ and 
  $\frac{B(\eta^\prime\rightarrow invisible)}{B(\eta^\prime\rightarrow\gamma\gamma)}<2.4\times 10^{-2}$, were obtained first at the 90\% confidence level. Then, using the branching fractions of $\eta(\eta^\prime)\rightarrow\gamma\gamma$, the branching fraction upper limits at the 90\% confidence level were determined to be $B(\eta\rightarrow invisible)<1.0\times 10^{-4}$ and $B(\eta^\prime\rightarrow invisible)<5.3\times 10^{-4}$. For the first time  a search for
 the semileptonic weak decay modes $\eta(\eta^\prime)\rightarrow\pi^+e^-\bar{\nu_e}$ was performed and no signal was observed.  At the 90\% confidence level, the semileptonic weak rates were given to be $B(\eta\rightarrow\pi^+e^-\bar{\nu_e}+c.c.)<1.7\times 10^{-4}$ and $B(\eta^\prime\rightarrow\pi^+e^-\bar{\nu_e}+c.c.)<2.2\times 10^{-4}$.

Based on the 225.3 million $J/\psi$ events,  we present the recent results on $\eta$ and $\eta^\prime$ decays in this talk.  To precisely test the fundamental symmetries and theoretical predictions,    the larger statistics of $\eta(\eta^\prime)$ decays are strongly needed. In 2012 the BESIII  detector collected about 1 billion $J/\psi$ events,  four times larger than the sample taken in 2009, which allows us to update the study of $\eta^\prime$, including the Dalitz plot analysis, the search for new decays, as well as the test to the fundamental symmetries. We believe that more interesting results will be coming soon in the near future.

\newpage

\subsection{A novel neutral vertex reconstruction algorithm for regeneration events in the $\mathrm{K_SK_L} \to \pi^+\pi^-\pi^0\pi^0$ channel}
\addtocontents{toc}{\hspace{2cm}{\sl A.~Gajos}\par}

\vspace{5mm}

A.~Gajos for the KLOE/KLOE-2 Collaboration\footnote{D.~Babusci, D.~Badoni, I.~Balwierz-Pytko, G.~Bencivenni, C.~Bini, C.~Bloise, F.~Bossi, P.~Branchini, A.~Budano, L.~Caldeira~Balkest\aa hl, G.~Capon, F.~Ceradini, P.~Ciambrone, F.~Curciarello, E.~Czerwi\'nski, E.~Dan\`e, V.~De~Leo, E.~De~Lucia, G.~De~Robertis, A.~De~Santis, A.~Di~Domenico, C.~Di~Donato, R.~Di~Salvo, D.~Domenici, O.~Erriquez, G.~Fanizzi, A.~Fantini, G.~Felici, S.~Fiore, P.~Franzini, A.~Gajos, P.~Gauzzi, G.~Giardina, S.~Giovannella, E.~Graziani, F.~Happacher, L.~Heijkenskj\"old, B.~H\"oistad, L.~Iafolla, M.~Jacewicz, T.~Johansson, K.~Kacprzak, A.~Kupsc, J.~Lee-Franzini, B.~Leverington, F.~Loddo, S.~Loffredo, G.~Mandaglio, M.~Martemianov, M.~Martini, M.~Mascolo, R.~Messi, S.~Miscetti, G.~Morello, D.~Moricciani, P.~Moskal, F.~Nguyen, A.~Palladino, A.~Passeri, V.~Patera, I.~Prado~Longhi, A.~Ranieri, C.~F.~Redmer, P.~Santangelo, I.~Sarra, M.~Schioppa, B.~Sciascia, M.~Silarski, C.~Taccini, L.~Tortora, G.~Venanzoni, W.~Wi\'slicki, M.~Wolke, J.~Zdebik}

\vspace{5mm}

\noindent
Jagiellonian University in Cracow, Poland

\vspace{5mm}

The KLOE experiment located at the DA$\Phi$NE $\phi$-factory provides pairs of neutral kaons produced in $\phi$ meson decays which allow for numerous investigations of $CP$ symmetry violation \cite{DiDomenico:2012pxa} such as in the $\Ks \to 3\pi^0$ decay \cite{Babusci:2013tr} and through a study of quantum interference of entangled kaon states \cite{Czerwinski:2013kra}. An interferometric analysis of the $\phi \to \Ks\Kl \to \pi^+\pi^-\pi^0\pi^0$ decay chain may be used to measure the $\frac{\epsilon'}{\epsilon}$ parameter whose non-zero value would be an indication of CP violation \cite{D'Ambrosio:1994wx,Ciuchini:1995qx}. The value of $\frac{\epsilon'}{\epsilon}$ parameter can be determined through the distribution of time differences $\Delta t$ between decay times of two kaons. If the CP symmetry is violated, this distribution will exhibit an asymmetry around $\Delta t = 0$ which is sensitive to $\Im \left(\frac{\epsilon'}{\epsilon}\right)$ in the region of $|\Delta t|<5\tau_s$ and to $\Re \left(\frac{\epsilon'}{\epsilon}\right)$ for $\Delta t \gg \tau_s$ \cite{DiDomenico:2007zza} where $\tau_s$ is the life time of the short-lived neutral kaon.

Neutral K mesons can, however, undergo the regeneration process as a result of interaction with nucleons of medium. Several elements of the KLOE detector cause a significant part of produced long-lived kaons to be incoherently regenerated to $\Ks$ \cite{Balwierz:2012np1} which leads to appearance of excesses of events in the decay time difference distribution with $\Delta t$ corresponding to time needed by $\Kl$ to reach these detector parts. Such events constitute background for the interferometric study of the $\phi \to \Ks\Kl \to \pi^+\pi^-\pi^0\pi^0$ process since quantum entanglement between $\Ks$ and $\Kl$ is lost at the moment of kaon regeneration.

Even though incoherent kaon regeneration is predominantly associated with a change in the kaon momentum direction, it is especially difficult to detect in case of the $\mathrm{K} \to \pi^0\pi^0 \to 4\gamma$ decay which only involves neutral particles and thus where momentum of the decaying kaon cannot be precisely determined. Standard reconstruction method used for this decay vertex at KLOE assumes the vertex to lie along original direction of the kaon flight obtained using pion DC tracks from the $\mathrm{K} \to \pi^+\pi^-$ decay  of the second kaon. This assumption, however, renders this method insensitive to regeneration which raises a need for an auxiliary reconstruction algorithm which would properly reconstruct kaon neutral vertices both in case of signal and regeneration and provide means to reject the latter.

A novel algorithm for reconstruction of the $\mathrm{K} \to \pi^0\pi^0 \to 4\gamma$ decay vertex is devised which utilizes only information on clusters in the electromagnetic calorimeter created by photons originating from $\pi^0$ decays. The idea of reconstruction is based upon observation that a set of possible origin points of a photon hitting the calorimeter constitutes a sphere centered in the cluster position with radius dependent on the difference of cluster time and life time of kaon before its decay. The kaon neutral decay vertex may then be found as an intersection of such spheres for each of four photons involved in the process.

In addition to rejection of regeneration background, the new reconstruction method may find an application in case of other neutral decay vertices such as $\mathrm{K_L} \to 3\pi^0$ in the $\Ks\Kl \to 3\pi^0\pi l\nu$ process which can be studied at the KLOE-2 experiment in a test of T and CPT symmetries \cite{Bernabeu:2012nu}.

This work is supported by the European Community-Research Infrastructure Integrating Activity "Study of Strongly Interacting Matter" (acronym HadronPhysics3, Grant Agreement n. 283286) under the Seventh Framework Programme of EU; by the Polish National Science Centre through the Grants No. 0469/B/H03/2009/37, 0309/B/H03/2011/40, DEC-2011/03/N/ST2/02641, 2011/01/D/ST2/00748 and by the Foundation for Polish Science through the MPD programme and the project HOMING PLUS BIS/2011-4/3.

\newpage

\subsection{Status of LNF}
\addtocontents{toc}{\hspace{2cm}{\sl S.~Giovannella}\par}

\vspace{5mm}

S.~Giovannella

\vspace{5mm}

\noindent
Laboratori Nazionali di Frascati dell'INFN, Italy\\

\vspace{5mm}

The Frascati National Laboratory of INFN (Istituto Nazionale di 
Fisica Nucleare) is located 20 km south-east of Rome. There are
several ongoing research facilities:
SPARC, a test-facility aimed to produce high brightness electron 
beams \cite{SPARC}, DA${\Phi}$NE, an $e^+e^-$ collider running 
at the $\phi$ resonance \cite{DAFNE} and the enclosed sincroton 
radiation \cite{Luce} and test beam \cite{BTF} facilities.

DA${\Phi}$NE ran in the period 1999-2007, providing a peak 
luminosity of $1.4\times 10^{32}$ cm$^{-2}$ s$^{-1}$, corresponding 
to an integrated luminosity of 8.5 pb$^{-1}$/day. 
In 2008, a new crossing scheme, allowing a reduced beam size 
and increased luminosity \cite{CrabWaist} was installed.
%
The KLOE experiment operated at DA${\Phi}$NE from 2000 to 2006, 
collecting 2.5 fb$^{-1}$ at the $\phi$ meson peak and about 240 
pb$^{-1}$ below the $\phi$ resonance ($\sqrt{s}=1$ GeV).
The $\phi$ meson predominantly decays into charged and neutral 
kaons, thus allowing KLOE to make precision studies in the fields 
of flavor physics, low energy QCD and test of discrete symmetries 
\cite{Cimento}. 

The increase in luminosity motivated an upgrade of the KLOE detector
with small angle tagging devices to detect both high and low energy 
electrons or positrons in $e^+e^-\to e^+e^-X$ events \cite{LET,HET} and
an inner tracker \cite{IT} and small angle calorimeters \cite{CCALT,QCALT}
to provide larger acceptance both for charged particles and photons.
A detailed description of the KLOE-2 physics program can be found in 
Ref.~\cite{KLOE2}.


\newpage

\subsection{Neutral pion decay physics in kaon experiments at CERN}
\addtocontents{toc}{\hspace{2cm}{\sl E.~Goudzovski}\par}

\vspace{5mm}

E.~Goudzovski\footnote{On behalf of the NA62 collaboration: Birmingham, Bratislava, Bristol, CERN, Dubna, Fairfax, Ferrara, Florence, Frascati, Glasgow, Liverpool, Louvain-La-Neuve, Mainz, Merced, Moscow, Naples, Padua, Perugia, Pisa, Prague, Protvino, Rome La Sapienza, Rome Tor Vergata, San Luis Potos\'i, Sofia, Stanford, Turin}

\vspace{5mm}

\noindent
School of Physics and Astronomy, University of Birmingham, United Kingdom\\

\vspace{5mm}

The charged kaon decay sample collected by the NA48/2 experiment at CERN in 2003--04 corresponds to $\sim2\times 10^{11}$ $K^\pm$ decays in flight~\cite{ba07}. Its successor, the NA62 experiment, aims to collect a $\sim50$ times larger sample of $K^+$ decays in 2014--17 with improved particle identification and resolution~\cite{ru13}. Kaons represent sources of tagged neutral pions via their hadronic decays such as $K^\pm\to\pi^\pm\pi^0$, $K_L\to3\pi^0$. Therefore these and other kaon experiments are capable of studying rare and forbidden $\pi^0$ decays at the highest precision.

The NA48/2 data set contains a sample of about $2\times 10^7$  $\pi^0_D\to e^+e^-\gamma$ decays with a negligible background. A search for the dark photon ($U$), a hypothetical light vector particle~\cite{po09}, is on-going with this data set via the decay chain $\pi^0\to U\gamma$, $U\to e^+e^-$: for $M_U<2M_\mu$, the dark photon is expected to decay exclusively into $e^+e^-$ pairs~\cite{batell09}. Due to the short dark photon mean free path in the NA48/2 conditions ($\sim 1$~mm), the $\pi^0_D$ decay represents an irreducible background and limits the sensitivity. This is mitigated by a good electron pair mass resolution ($\delta M_{ee}/M_{ee}\approx 0.012$). Preliminary studies indicate that the expected upper limit on the mixing parameter $\varepsilon^2$ is below $10^{-6}$ in the $U$ mass range from 10 MeV/$c^2$ to 100 MeV/$c^2$. That would be an improvement over the recent KLOE-2~\cite{kloe2} and WASA~\cite{wasa1} limits in this mass range, and could shed light on the muon $g-2$ anomaly~\cite{en12}.

The NA62 experiment aiming at $K^+\to\pi^+\nu\bar\nu$ measurement, with $\sim10^{-11}$ single event sensitivities for $\pi^0$ decays, should be capable of improving the $\varepsilon^2$ limit by an order of magnitude with respect to NA48/2. Moreover, it can improve the upper limits for rare and forbidden decays such as $\pi^0\to3\gamma$, $\pi^0\to4\gamma$, $\pi^0\to\mu^\pm e^\pm$ and $\pi^0\to\nu\bar\nu$. The prospects for these modes (except $\pi^0\to\nu\bar\nu$) depend on the implementation of dedicated trigger chains.

\newpage

\subsection{Status from COSY}
\addtocontents{toc}{\hspace{2cm}{\sl D.~Grzonka}\par}

\vspace{5mm}

D.~Grzonka

\vspace{5mm}

\noindent
Forschungszentrum J\"ulich, Institut f\"ur Kernphysik, J\"ulich, Germany\\

\vspace{5mm}

The cooler synchrotron COSY at the research center J\"ulich delivers unpolarized and polarized proton and deuteron beams 
with momenta up to 3.7 GeV/c \cite{cosy}. Electron and stochastic cooling systems are installed for a phase space cooled beam which can be used at internal and external experiments.
The main hadron physics experiments are the internal facilities ANKE \cite{anke} and WASA \cite{wasa} and the 
external TOF experiment.

ANKE is a flexible magnetic spectrometer consisting of three dipoles where the central dipole acts as spectrometer for charged ejectiles. Various detector components like wire chambers and scintillators are located at the exit windows and in addition close to the target silicon detectors are installed. A cluster target as well as a polarized atomic beam target can be used.
The actual physics program is mainly devoted to nucleon-nucleon scattering and pion production studies with polarization \cite{anke2}.

WASA is a close to 4$\pi$ detector for charged and neutral particles. It consists of a central detector with a superconducting coil for magnetic fields up to 1 T including straw tubes and scintillators, which is surrounded by an electromagnetic calorimeter and a forward detector part with several layers of scintillators and wire chambers to measure fast forward emitting ejectiles. The physics program at WASA focusses on the investigation of symmetries by decay studies of light mesons, ($\pi ^0$, $\eta$, $\omega$)
where rare decays are of special interest.  
The WASA activities are described in various contributions to these proceedings \cite{wasa2}.

TOF is an external large acceptance non magnetic spectrometer consisting of a 3 m long vacuum tank with a diameter of about 3 m equipped with tracking detectors and scintillators.  The latest measurements deal with $\Lambda$ and $\Sigma$ hyperon production studied in pp and pn induced reactions. A straw tube detector allows for a precise track and vertex reconstruction resulting in a full event reconstruction including the decay particles. With the use of polarized beam and the self analyzing $\Lambda$ decay also polarization observables are accessible. For details on the detector system and first result of the hyperon production studies see ~\cite{tof}, \cite{tof2} and references therein. The TOF experiment will be removed end of 2013 in order to create space for the planned partly pre-assembly of the PANDA experiment for FAIR.

Other activities at COSY are PAX and EDM/JEDI. 
The PAX experiment~\cite{pax} studies the polarization of a stored beam by the spin filtering technique in view of a polarized antiproton beam for the future FAIR facility. The feasibility of the spin filtering method has been demonstrated at TSR and at COSY \cite{pax2} for the transverse polarization and will be continued with measurements of longitudinal polarization. As next step studies of  the spin dependent scattering cross sections with antiprotons were proposed \cite{pax3}.
Another experiment to be done at the PAX facility will be the TRIC experiment \cite{tric} to look for time invariance violations.

The EDM activity has the goal to search for electric dipole moments of charged particles in a storage ring
\cite{edm} .
This activity is at the very beginning and is doing basic studies to show the feasibility of such experiments where one of the first steps is the extension of the spin coherence time, a prerequisite for measuring possible tiny spin rotations by an edm effect.
Its a long standing program which has to be performed in several steps before a reasonable precision can be achieved.
The JEDI collaboration has been formed to work into this direction \cite{jedi}.

\newpage

\subsection{Search for a new gauge boson in $\pi^{0}$ decays with WASA-at-COSY }
\addtocontents{toc}{\hspace{2cm}{\sl Carl-Oscar Gullstr\"om}\par}

\vspace{5mm}

Carl-Oscar Gullstr\"om

\vspace{5mm}

\noindent
Institution of Physics and Astronomy, Uppsala University, Sweden\\

\vspace{5mm}

In some recent astrophysical observations there is an unexplained excess of positrons \cite{Jean:2003ci,Adriani:2008zr}. A possible 
explanation is that there exist a 
new light boson U with weak couplings and the source of the boson is dark matter annihilation \cite{Boehm:2003hm}. If the U boson is vector like it could couple to the normal photon with strength $\epsilon$.
This new boson could also account for the muon g-2 anomaly \cite{Pospelov:2008zw}. The virtual photon in the decay $\pi^{0}\rightarrow \gamma e^{+}e^{-}$ is a suitable place to look for this new hypothetical "dark photon".
An analysis was carried out on a 500.000 $\pi^{0}\rightarrow \gamma e^{+}e^{-}$ data sample recorded in 2010 with the WASA-at-COSY detector, for the future there is a
recent approx. ten times larger data sample to be analyzed. No indication of a new boson was found and a new upper limit was set for the
decay $\pi^{0}\rightarrow \gamma U$. Then it was possible to set an improved upper limit in the range 30-100 MeV for the coupling $\epsilon$ \cite{Adlarson:2013eza}.

\newpage

\subsection{Two-pion production in pp and np reactions with HADES}
\addtocontents{toc}{\hspace{2cm}{\sl M.~Gumberidze}\par}

\vspace{5mm}

M.~Gumberidze for HADES collaboration

\vspace{5mm}

\noindent
TU Darmstadt, Germany\\

\vspace{5mm}

The High Acceptance Di-Electron Spectrometer (HADES) \cite{Hades} installed at GSI \linebreak Helmholtzzentrum fuer Schwerionenforschung 
in Darmstadt was designed to investigate dielectron production in heavy-ion collisions in the range of kinetic beam energies 1-2 A GeV. 
The main goal of the HADES experiment is to study properties of hadrons inside the hot and dense nuclear medium via their dielectron decays.  

One specific issue of heavy-ion reactions in the 1-2 AGeV regime is the important role played by the baryonic resonances, 
which propagate and regenerate, due to the long life-time of the dense hadronic matter phase. The $\Delta$(1232)  
is the most copiously produced resonance but with increasing incident energy, higher lying resonances also contribute to pion production. 
A detailed description of the resonance excitation and coupling to the pseudoscalar and vector mesons is important for the interpretation 
of the dielectron spectra measured by HADES. Baryonic resonances are important sources of dileptons through two mechanisms: 
(i) Dalitz decays (e.g. $\Delta /\ N \rightarrow N e^{+}e^{-}$), and (ii) mesonic decays with subsequent dielectron production. 

Pion production in NN collisions is one of the sources of information on the NN interaction and on nucleon resonance properties. 
Two-pion production, in particular, is an outstanding subject, since it connects $\pi \pi$ dynamics with baryon 
and baryon-baryon degrees of freedom.  In our energy regime, $\Delta \Delta$ excitation becomes the leading process. 
The one- and two-pion production in pp and np reactions has been investigated with HADES in exclusive measurements 
for beam kinetic energies of 1.25 and 2.2 GeV. 

Several theoretical models for double pion production have been suggested in the energy range from the production threshold up to several GeV.
A full reaction model describing the double pion production in NN collisions has been developed by Valencia group \cite{Valencia}. 
More advanced calculations by Cao, Zou and Xu including relativistic corrections have been published in \cite{XuCao1}. These models include 
and study both the resonant and non-resonant terms of $\pi\pi$ production. The models predict that at our energy regime the dominant
contribution for $pp$ and $np$ reaction is $N^{*}(1440) \rightarrow N \sigma$, $N^{*}(1440) \rightarrow \Delta \pi$ and double-$\Delta$.

Thanks to the large cross sections and t the large acceptance of HADES detector, the differential spectra can be measured with high statistics hence providing
strong constraints on the production mechanisms and on the various resonance contributions ($\Delta(1232)$, N*(1440)).  In this talk, a comparison of HADES 
results on two-pion production in np and pp reaction at 1.25 GeV kinetic energies with several model predictions  (Valencia \cite{Valencia}, 
OPER \cite{Oper1, Oper2, Oper3} and Xu Cao et al. \cite{XuCao}) has been presented.

\newpage

\subsection{Dalitz plot analysis of $\omega \to \pi^+\pi^-\pi^0$ with the KLOE detector}
\addtocontents{toc}{\hspace{2cm}{\sl  L.~Heijkenskj\"old}\par}

\vspace{5mm}

 L. Heijkenskj\"old for the KLOE-2 collaboration\renewcommand{\thefootnote}{\fnsymbol{footnote}}\footnote{D. Babusci, D. Badoni, I. Balwierz-Pytko, G. Bencivenni, C. Bini, C. Bloise, F. Bossi, P. Branchini, A. Budano, L. Caldeira Balkest\aa hl, G. Capon, F. Ceradini, P. Ciambrone, F. Curciarello, E. Czerwi\'nski, E. Dan\`{e}, V. De Leo, E. De Lucia, G. De Robertis, A. De Santis, A. Di Domenico, C. Di Donato, R. Di Salvo, D. Domenici, O. Erriquez, G. Fanizzi, A. Fantini, G. Felici, S. Fiore, P. Franzini, P. Gauzzi, G. Giardina, S. Giovannella, F. Gonnella, E. Graziani, F. Happacher, B. H\"oistad, L. Iafolla, M. Jacewicz, T. Johansson, K. Kacprzak, A. Kupsc, J. Lee-Franzini, B. Leverington, F. Loddo, S. Loffredo, G. Mandaglio, M. Martemianov, M. Martini, M. Mascolo, R. Messi, S. Miscetti, G. Morello, D. Moricciani, P. Moskal, F. Nguyen, A. Passeri, V. Patera, I. Prado Longhi, A. Ranieri, C.F. Redmer, P. Santangelo, I. Sarra, M. Schioppa, B. Sciascia, M. Silarski, C. Taccini, L. Tortora, G. Venanzoni, W. Wi\'slicki, M. Wolke, J. Zdebik}

\vspace{5mm}

\noindent
Department of Physics and Astronomy, Uppsala University\\

\vspace{5mm}

The decay mechanism of a three particle final state can be studied using the density distribution in a Dalitz plot. For the $\omega\to\pi^+\pi^-\pi^0$ reaction the symmetry of the distribution in the $\Phi$ Dalitz variable, indicating a P-wave of the final state particles,  have already been experimentally verified \cite{Maglic:1961nz}. Vector meson dominance (VMD) model predicts an intermediate $\rho$ meson, which would show as an onset towards the edges of the phase space. There are also two more recent theoretical calculations of the decay mechanism and final state interactions with testable predictions of this Dalitz plot \cite{Terschlusen:2013iqa} \cite{Niecknig:2012sj}. Both theoretically predicted distributions have been parameterised using a polynomial in the Z and $\Phi$ Dalitz variables.

We will provide Dalitz plot parameters from an experimental distribution of high statistics to use as convincing tests of the theoretical predictions of this decay process.\\

The KLOE collaboration has performed a study of the $e^+e^-\to\pi^0\omega$ reaction in the $\Phi$-meson mass region where $1.3\times10^6$ $\omega\to3\pi$ events were recorded \cite{Ambrosino:2008gb}. We intend to use this data to perform a Dalitz plot analysis.

To fully describe the reaction which we measure, $e^+e^-\to\pi^0\omega\to\pi^0\pi^+\pi^-\pi^0$, we use a complete VMD matrix element obtained by summing over all possible values of the $\rho$ meson charge and permutations of $\pi^0$ mesons \cite{Akhmetshin:1998df}. This matrix element includes a term corresponding to the interference between the two neutral pions in the final state. We have studied the impact this interference might have on the kinematics of the final state pions, and hence the density distribution in the resulting Dalitz plot. This was done by comparing the pion-kinematics between simulated data weighted with two different matrix elements, one including the fully permuted VMD diagrams and one where the $\pi^0\pi^0$ interference term was left out. The result was a $\sim10$\% difference in the density distribution of the Dalitz plot. The discrepancy between the two different distributions was mainly located in the region where the $\Phi$-variable is negative.\\

\newpage

\subsection{Charge Symmetry Breaking in $dd$ collissions with WASA-at-COSY}
\addtocontents{toc}{\hspace{2cm}{\sl V.~Hejny}\par}

\vspace{5mm}

V.~Hejny for the WASA-at-COSY Collaboration

\vspace{5mm}

\noindent
Institut f\"ur Kernphysik and J\"ulich Center for Hadron Physics,\\ 
Forschungszentrum J\"ulich GmbH, J\"ulich, Germany\\

\vspace{5mm}

Charge symmetry is a special case of isospin symmetry, and, thus, a
fundamental symmetry of QCD~\cite{Miller:1990iz}.
Isospin symmetry is broken by the different masses of the up and
down quarks. In addition, quarks are distinguished by the electromagnetic interaction.
Nowadays, these contributions can be treated theoretically within the
framework of chiral perturbation theory. Thus, a detailed study of isospin
violation in low energy hadron physics is a unique window to the quark
masses and to fundamental parameters of the standard model. To get
access to the quark mass difference it is advised to look at charge
symmetry breaking observables. Here, the relative $\pi$-mass
difference, which is of electromagnetic origin, does not contribute.
As the reaction $dd\,\to\,\mathrm{^4He\pi^0}$ violates charge
symmetry, the cross section is directly proportional to the square of
the charge symmetry breaking amplitude.
Triggered by first high-precision experiments performed at
TRIUMF~\cite{Opper:2003sb} and IUCF~\cite{Stephenson:2003dv} an international
collaboration has been formed aiming at a consistent description within chiral perturbation theory. In the
course of the ongoing analysis~\cite{miller06,filin,Hanhart04,Gardestig04,
Nogga06,timo,jerrynew}, a set of
essential observables has been identified, which are now being
addressed by the experimental program of WASA-at-COSY.

In a first step the focus was on an exclusive measurement of the reaction
$dd\,\to\,\mathrm{^3He}$$n\pi^0$ at $p_d = 1.2\,\mathrm{GeV/c}$. In order to
provide quantitative results the data have been compared to a
quasi-free reaction model based on existing data for the
two-body reaction $dp\,\to\,\mathrm{^3He}\pi^0$ and a partial-wave expansion
for the three-body reaction limited to at most one $p$-wave in the system,
both added incoherently.
The $\mathrm{^3He}$$n\pi^0$ final state is described by the two Jacobi momenta
$\vec{q}$ and $\vec{p}$, where $\vec{q}$ is the $\pi^0$ momentum in the overall
c.m. frame and $\vec{p}$ the relative momentum in the $\mathrm{^3He}$$n$ subsystem
\cite{prc-sub}.

In a subsequent run first data for the charge-symmetry breaking reaction
$dd\,\to\,\mathrm{^4He}\pi^0$ at $p_d = 1.2\,\mathrm{GeV/c}$ have been taken.
In total $120\pm20$ signal events were identified and a total cross section as
well as a first angular distribution was extracted. Data are currently being 
finalized \cite{annrep12}.

Based on these results a new measurement using a modified detector
setup will be carried out beginning of 2014: removing the Forward Range Hodoscope
and other detector components will introduce a 1.5~m time-of-flight
path in order to improve $^3$He-$^4$He separation and kinetic energy
reconstruction. The goals of the measurement are to increase statistics
and to minimize systematic errors for data taken at
$p_d = 1.2\,\mathrm{GeV/c}$ in order to provide a decisive
angular distribution.

\newpage

\subsection{Dispersion theory methods for transition form factors}
\addtocontents{toc}{\hspace{2cm}{\sl M.~Hoferichter, F.~Niecknig, S.~P.~Schneider}\par}

\vspace{5mm}

M.~Hoferichter,$^{1}$ F.~Niecknig,$^{2}$ S.~P.~Schneider$^{2}$

\vspace{5mm}

\noindent
$^{1}$ Albert Einstein Center for Fundamental Physics and Institute for Theoretical Physics,
	    Universit\"at Bern, Switzerland\\
$^{2}$ Helmholtz-Institut f\"ur Strahlen- und Kernphysik (Theorie) and
             Bethe Center for Theoretical Physics,
             Universit\"at Bonn, Germany\\

\vspace{5mm}

Dispersion theoretical analyses of transition form factors are stepping
stones to a model-independent determination of the light-by-light
scattering contribution to the anomalous magnet moment of the muon $(g-2)_\mu$.
The strength of the numerically dominant class of diagrams, the pion-pole term,
is governed by the doubly-virtual $\pi^0$ transition form factor $F_{\pi^0\gamma^*\gamma^*}(q_1^2,q_2^2)$~\cite{JF,Czerwinski}.
Its full dispersive construction
relies on many ingredients,
with the anomalous process $\gamma\pi\to\pi\pi$, the vector meson decays $\omega/\phi\to3\pi$, and the precisely constrained pion
vector form factor $F_\pi^V(s)$ as the most prominent amongst them.
In addition, these amplitudes and form factors play a crucial role also in the non-pole contributions to light-by-light scattering,
as they are needed for the incorporation of the photon virtualities in $\gamma^*\gamma^*\to\pi\pi$~\cite{GM, HPS,Moussallam}.

An accurate representation of  $\gamma\pi\to\pi\pi$ requires knowledge
of the normalization of the pertinent amplitude, which is intimately related to the $\gamma3\pi$ chiral anomaly.
A dispersive representation of the process in terms of just two parameters
can be used to extract the chiral anomaly from the recent Primakoff measurements at
COMPASS~\cite{HKS}. With the $\rho$ meson included model-independently in terms of
the $\pi\pi$ $P$-wave phase shift, all data up to $1\,$GeV may be
exploited in the analysis, dramatically increasing the statistics of the
anomaly extraction.
Once the subtraction constants are fixed by fitting to the measured cross section, this will provide
the desired precise determination of the $\gamma\pi\to\pi\pi$ partial-wave amplitude $f_1^{\gamma\pi\to\pi\pi}(s)$.

The vector meson decays $\omega/\phi\to3\pi$ represent the next step in complexity. 
While the quantum numbers involved are identical to those of $\gamma\pi\to\pi\pi$,
the calculation of these processes
is complicated by the decay kinematics~\cite{Khuri,Aitchison}. 
In fact, a rigorous theoretical description of vector-meson interactions 
remains a challenge, while the ongoing experimental interest in
$\omega/\phi\to 3\pi$ decays emphasizes the need for equally precise theoretical
predictions. However, the prevailing treatments all lack a thorough
inclusion of final-state interactions and are often at odds with unitarity.
A dispersive approach to these decays offers a framework that is consistent with the requirements of unitarity and analyticity and facilitates 
the study of crossed-channel rescattering effects on the shape of the Dalitz plots~\cite{Niecknig}. Using a twice-subtracted dispersion relation
and fitting the only non-trivial subtraction constant (besides the overall normalization) we obtain a very precise representation of the KLOE data set for $\phi\to3\pi$~\cite{KLOE1}, which currently offers the best statistics. 

The partial-wave amplitudes obtained from the $\omega/\phi\to3\pi$ analysis in combination with the pion vector form factor can be used to determine the pertinent vector-meson transition form factors~\cite{SKN}, as measured in the conversion decays $\omega/\phi\to\pi^0\ell^+\ell^-$. While a successful description of the $\omega\to\pi^0\mu^+\mu^-$ transition-form-factor data towards the end of the physical decay region remains elusive, a measurement of $\phi\to\pi^0\ell^+\ell^-$ could provide insights into the nature of the steep rise in the $\omega\to\pi^0\mu^+\mu^-$ spectrum.

Finally, we give an outline on how to extend the framework to obtain a description of the singly-virtual $\pi^0$ transition
form factor $F_{\pi^0\gamma^*\gamma}(q^2,0)$~\cite{inprogress}, as measured in $e^+e^-\to\pi^0\gamma$. Fixing the normalization by
fits to $e^+e^-\to3\pi$ data gives access to the partial-wave amplitude for $\gamma^*\to3\pi$, $f^{\gamma^*\to3\pi}_1(s,q^2)$. 
At $q^2=0$, this is related to the $\gamma\pi\to\pi\pi$ partial wave, 
$f^{\gamma^*\to3\pi}_1(s,q^2=0)=f^{\gamma\pi\to\pi\pi}_1(s)$.
Adding isoscalar and isovector contributions, we derive the representation
\[
 F_{\pi^0\gamma^*\gamma}(q^2,0)=F_{\pi\gamma\gamma}
+\int\limits_{4\mpi^2}^\infty ds'\frac{q_\pi^3(s')F_\pi^{V*}(s') }{12\pi^2s'^{3/2}}
\bigg\{f^{\gamma^*\to3\pi}_1\big(s',q^2\big)-f^{\gamma\pi\to\pi\pi}_1(s')
 +\frac{q^2}{s'-q^2}f^{\gamma\pi\to\pi\pi}_1(s')\bigg\}\,,
\]
where
\[
 F_{\pi\gamma\gamma}=\frac{e^2}{4\pi^2F_\pi}\,,\qquad q_\pi(s)=\sqrt{\frac{s}{4}-\mpi^2}\,.
\]

\newpage

\subsection{Bremsstrahlung in $\piee$ process beyond soft-photon approximation }
\addtocontents{toc}{\hspace{2cm}{\sl T.~Husek}\par}

\vspace{5mm}

T.~Husek, K.~Kampf, J.~Novotn\'y

\vspace{5mm}

\noindent
Institute of Particle and Nuclear Physics, Faculty of Mathematics and Physics\\Charles University in Prague, Czech Republic\\

\vspace{5mm}

Recently, the rare decay $\piee$ engrossed the attention of the theorists in connection with a new precise branching ratio measurement. The KTeV-E799-II experiment at Fermilab~\cite{Abouzaid} has observed $\piee$ events (altogether 794 candidates), where $K_L\rightarrow3\pi^0$ decay was used as a source of neutral pions. The branching ratio of the neutral pion decay into an electron-positron pair was determined to be
\begin{equation}\nonumber
B(\piee(\gamma),\,x_\text{D}>0.95)=(6.44\pm0.25\pm0.22)\times10^{-8}\,,
\label{eq:B}
\end{equation}
where the first error is from data statistics alone and the second is the total systematic error.
Using a model for the radiative corrections based on the calculation by Bergstr\"{o}m \cite{Berg}, this result has been extrapolated to the full radiative tail beyond $x_\text{D}>0.95$ and scaled back up by the overall radiative corrections of 3.4\% to get the lowest order rate (with the final state radiation removed) for $\piee$ process. The finale result is
\begin{equation}\nonumber
B^{\text{no-rad}}_{\text{KTeV}}(\piee)=(7.48\pm0.29\pm0.25)\times10^{-8}\,.
\end{equation}
Subsequent comparison with theoretical predictions of the SM were made, e.g. in~\cite{Dorokhov}, using pion transition form factor data from CELLO and CLEO experiments. Finally, it has been found, that according to SM the result should be
\begin{equation}\nonumber
B^{\text{no-rad}}_{\text{SM}}(\piee)=(6.23\pm0.09)\times10^{-8}\,.
\end{equation}
This can be interpreted as a 3.3\,$\sigma$ discrepancy between the theory and the experiment. Aside from the attempts to find the corresponding mechanism within the physics beyond the SM, also the possible revision of the SM predictions has been taken into account.

The full two-loop virtual radiative corrections (pure QED) and soft-photon bremsstrahlung were determined recently in~\cite{Novotny} with the result $\delta^{\text{virt.+soft}\gamma}_\text{\cite{Novotny}}(0.95)=(-5.8\pm0.2)\,\%$, which differs significantly from the previous approximative calculations
$\delta^{\text{virt.+soft}\gamma}_\text{\cite{Berg}}(0.95)=-13.8\,\%$ and $\delta^{\text{virt.+soft}\gamma}_\text{\cite{Dorokhov4}}(0.95)=-13.3\,\%$.

Here, we calculate analytically the real radiative corrections beyond the soft-photon region in order to test the reliability of the soft-photon approximation within the regime used by the KTeV collaboration. We have found out, that the relative contribution to the leading order in the region of KTeV experiment stands
\begin{equation}\nonumber
\delta^\text{BS}(0.95)=\frac{\Gamma^\text{BS}|_{x>0.95}}{\Gamma^\text{LO}\left(\piee\right)}=(0.30\pm0.01)\,\%\,.
\label{eq:BS95}
\end{equation}
In other words, using this cut of Dalitz variable in KTeV experiment, the soft-photon approximation is relevant.

We also estimate the two-loop contribution in QCD sector. Using Weinberg consistency relation, we can calculate the term proportional to the double logarithm ($\ln^2\mu^2$), when summing contributions up to the 2-loop order of the $\piee$ process in QCD sector.
The only contribution, which is not canceled out, is coming from the anomalous $\mathcal{O}\left(p^6\right)$ Lagrangian (monomial $o_{13}$) and preliminary calculations indicate that this contribution is negligible

For the contact interaction counter-term used in the leading order calculation, coupling $\chi$ is essential, which effectively substitutes for the transition formfactor in the region of  high loop momenta. Usually, it is determined from the lowest meson dominance (LMD) approximation to the large-$N_C$ spectrum of vector meson resonances to be $\chi^\text{(r)}$=2.2$\pm$0.9~\cite{Knecht}.
Since it was not possible to solve the discrepancy by including previously discussed contributions, we can fit the value of the coupling $\chi^\text{(r)}$ to meet the experiment with the result
\begin{equation}\nonumber
\chi^\text{(r)}(M_\rho)=4.4\pm1.0\,.
\end{equation}
Previous results may be used in further investigations, such as muon $g$-2 experiments. Numerical outputs are under development.

To sum it briefly up, we have calculated the bremsstrahlung contribution to the $\piee$ process. It has been shown, that soft-photon approximation is adequate approach in the region of KTeV experiment. We have also estimated the contribution of the two-loop case in QCD sector. Considered corrections were preliminarily calculated to be negligible and as the solution for the discrepancy, the change of the value of the coupling $\chi^\text{(r)}$ was offered. A significant difference was found. The results can be used in related problems such as $g$-2 experiment.

\newpage

\subsection{($d,^3$He) reaction for the formation of deeply bound pionic atoms }
\addtocontents{toc}{\hspace{2cm}{\sl N.~Ikeno}\par}

\vspace{5mm}

N.~Ikeno, J.~Yamagata-Sekihara${}^{a}$, H.~Nagahiro, S.~Hirenzaki

\vspace{5mm}

\noindent
Department of Physics, Nara Women's University, Nara 630-8506, Japan\\
${}^{a}$High Energy Accelerator Research Organization (KEK), Ibaraki 305-0801, Japan\\

\vspace{5mm}

An effective Lagrangian calculation of dilepton production in
pion-nucleon collisions is presented \cite{Zetenyi-Wolf}. An important
ingredient of the model is the gauge-invariance-preserving scheme for
the definition of form factors in the Born contributions. For processes
involving real photons this scheme was invented by Davidson and Workman
\cite{Davidson-Workman}, we generalized the scheme for the virtual photon
case.

We discuss some issues related to the consistent treatment of spin$>$3/2
baryon resonances in the effective Lagrangian. It has been shown in
Ref.~\cite{Pascalutsa} that interaction Lagrangians invariant under a
gauge transformation of the higher spin baryon field are consistent in
the sense that they avoid the appearance of contributions from lower
spin degrees of freedom. We demonstrate that these gauge invariant
Lagrangians provide a better description of pion photoproduction cross
sections than the traditional Lagrangians.

We discuss the electromagnetic form factors of baryon resonances in the
vector meson dominance picture, showing that the $\rho^0$ and $\omega$
contributions to the VMD form factor can have both constructive and
destructive interference in different isospin channels (see
Ref.~\cite{Lutz}).

Deeply bound pionic atom is one of the best systems to deduce pion
properties at finite density and to obtain precise information on the
partial restoration of chiral symmetry in nuclei~\cite{Piatom}. 
The deeply bound pionic states have been experimentally produced in the 
($d,^3$He) reactions with Pb and Sn isotope targets 
by following theoretical predictions. 
In Ref.~\cite{KSuzuki}, the energy shifts and widths of the pionic 1$s$ state 
have been measured in three Sn isotopes. 
From these observations,
the changes of the pion decay constant $f_{\pi}$ and 
the chiral order parameter $\langle \bar{q} q \rangle$ in nuclei were concluded.

To develop these studies further, we need to obtain improved and
systematic information on deeply bound pionic states.
The information is, for example, necessary for the unique determination
of the pion-nucleus interaction, which is required to fix the potential
strength related to chiral symmetry~\cite{Ikeno}.
In this paper, we report two recent theoretical studies of the
($d, ^3$He) reactions for the pionic atom formation.

\begin{figure}[!h]
\begin{center}
\includegraphics[width=15cm,height=5cm]{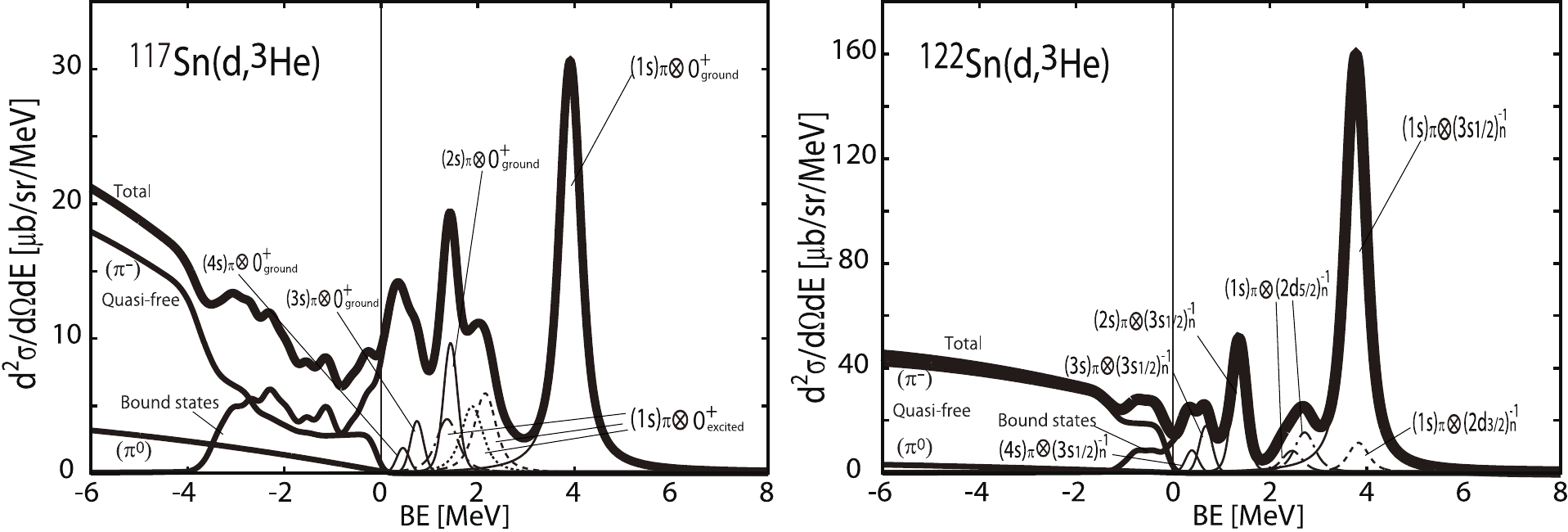}
\vspace{-4.5mm}
\caption{
\small
Calculated spectra for the formation of the pionic states
at $\theta_{d{\rm He}}^{\rm lab}=0^\circ$ 
in the $^{117}$Sn($d, ^{3}$He) (left) and 
the $^{122}$Sn($d, ^{3}$He) (right) reactions
plotted as functions of the pion binding energy~\cite{Ikeno3}.
The dominant subcomponents are also shown in the figures with 
quantum numbers indicated as
$[(n \ell)_{\pi} \otimes J^P]$
in the $^{117}$Sn($d,^3$He) reaction 
and
$[(n \ell)_{\pi} \otimes (n \ell_j)^{-1}_n]$
in the $^{122}$Sn($d,^3$He) reaction, respectively.
}
\label{Cross_0deg}
\end{center}
\end{figure}

The one is the pionic atom formation on the odd-neutron nuclear
target with $J^P =\frac{1}{2}^{+}$such as $^{117}$Sn~\cite{Ikeno3}. 
In Fig.~\ref{Cross_0deg} (left),
we show the calculated results of the $^{117}$Sn($d,^3$He) reaction
spectra at the forward angle
for the formation of the deeply bound pionic states.
We also compare this spectrum with that 
of the even-neutron nuclear target $^{122}$Sn case 
in Fig.~\ref{Cross_0deg}.
For the even-neutron nuclear target cases, 
we may have to take into account the residual interaction effects~\cite{Nose}
to deduce the binding energies of the pionic states
precisely
from the high precision experimental data
since the final pionic states are the 
pion-particle plus neutron-hole states [$\pi \otimes n^{-1}$].
On the other hand,
in the $^{117}$Sn($d,^{3}$He) spectra,
we find that we can see clearly the peak structure 
of the pionic 1$s$ state formation 
with the ground state of the even-even nucleus $^{116}$Sn
as indicated in the figure as [$(1s)_{\pi} \otimes 0^{+}_{\rm ground}$].
This pionic 1$s$ state 
will not have the additional shifts due to the residual
interaction effects. 
Thus,
the formation of the pionic 1$s$ state by the ($d,^3$He) reactions 
on the odd-neutron nuclear target 
is preferable to extract the most accurate information on the
parameter of the QCD symmetry from the observation.
%

\begin{figure}[!h]
\begin{center}
\includegraphics[width=15cm,height=4.8cm]{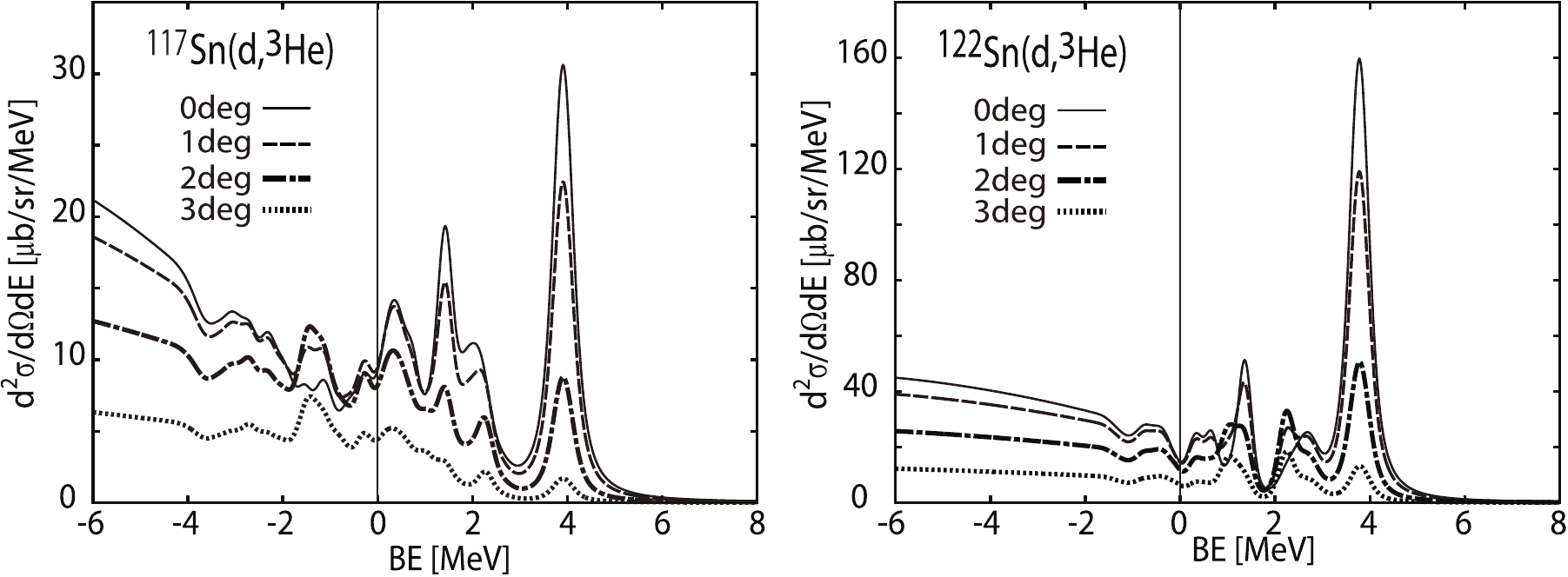}
\vspace{-4.5mm}
\caption{\small
Calculated spectra for the formation of the pionic states 
at $\theta^{\rm lab}_{d{\rm He}}=0^\circ$, 
$1^\circ$, 
$2^\circ$ and $3^\circ$ 
in the $^{117}$Sn($d, ^{3}$He) (left) and 
the $^{122}$Sn($d, ^{3}$He) (right) reactions
plotted as functions of 
the pion binding energy~\cite{Ikeno3}.
}
\label{Cross}
\end{center}
\end{figure}

The other is the pionic atom formation in the ($d, ^3$He) reaction at
finite angles~\cite{Ikeno2}.
In Fig.~\ref{Cross},
we show the calculated spectra
at finite angles for the formation of the pionic states 
in the $^{117}$Sn($d,^{3}$He) and 
$^{122}$Sn($d,^{3}$He) reactions. 
We find that the both spectra have a strong angular dependence 
and the shape of the spectra are much different at finite angles 
from that at the forward angle.
We also find that
we can observe the enhancement of various subcomponents
with different pion angular momentum. 
The observation of several deeply pionic bound states 
in the same nucleus  
will help to deduce precise information on the pion properties 
and the chiral dynamics at finite density~\cite{Ikeno}.

The experiment for the pionic atom formation on the odd-neutron nuclear target
will be performed at RIBF/RIKEN in near future~\cite{RIBF}.
And the $^{122}$Sn($d, ^{3}$He) spectra at finite angles
were obtained from the latest high precision experiment
at RIBF/RIKEN~\cite{Itoh}.
Therefore, we think that 
these studies will provide the systematic information on
the pionic bound states in various nuclei,
%
and
our results will provide a good motivation for further experimental studies.

\vspace{-1.1cm}

\newpage

\subsection{Recipes For The $VP\gamma$ Transition 
Form Factor Analysis}
\addtocontents{toc}{\hspace{2cm}{\sl  S.~Ivashyn}\par}

\vspace{5mm}
S.~Ivashyn$^1$, D.~Babusci$^2$, M.~Mascolo$^3$, D.~Moricciani$^3$,  
\vspace{5mm}

\noindent
$^1$ Akhiezer Institute for Theoretical Physics, NSC ``Kharkiv Institute for Physics and Technology'', Kharkiv UA-61108, Ukraine
\\
$^2$ INFN, Laboratori Nazionali di Frascati, Frascati I-00044, Italy
\\
$^3$ INFN, Sezione Roma ``Tor Vergata'', Roma I-00133, Italy

\vspace{5mm}

In this talk we discuss several issues, which are necessary for conveying a correct and precise 
data analysis aiming at the $VP\gamma$ transition form factor and the $V \to e^+e^-P$ branching ratio measurement.
The consideration is mainly focused on the processes $e^+e^- \to \phi \to e^+e^-\pi^0$ 
and $e^+e^- \to \phi \to e^+e^-\eta$
(these two channels are currently under analysis by KLOE Collaboration~\cite{this:KLOE_vpg}), 
but can also be useful for other similar channels, e.g., for $\omega \to \pi^0 e^+e^-$,
which is important in context of the theoretical difficulties~\cite{this:Carla,this:Sebastian,Ivashyn:2011hb}
of explaining the results of NA60~\cite{Arnaldi:2009aa,Usai:2011zza}
 and Lepton-G~\cite{Dzhelyadin:1980tj} collaborations.

The $V \to P \gamma^\ast$ transition is of an odd-intrinsic-parity type.
It is possible to plug in a model transition form factor, which reflects 
the $e^+e^- \to V \to e^+e^-P$ phenomenology
in the time-like region (e.g.,~\cite{Ivashyn:2011hb,Eidelman:2010ta}), 
into the Monte Carlo generator EKHARA~\cite{Czyz:2010sp,Czyz:2012nq1}.
This allows, for example, to make a quick look at the event distributions.
We compare the polar angle distributions for a final lepton in 
the differential decay width ($V \to e^+e^-P$) 
and in the differential cross section ($e^+e^- \to V \to e^+e^-P$). 
It is stressed that the shapes differ a lot and it is suggested that 
for a signal Monte Carlo in the data analysis one should use
the correct simulation. 

Using the $\phi \to \pi^0 \gamma^\ast$ form factor derived 
from Refs.~\cite{Ivashyn:2011hb,Eidelman:2010ta}
and the numerical results of Ref.~\cite{Schneider:2012ez1}, 
the theory uncertainty in the prediction of the $\phi \to \pi^0 \gamma^\ast$ form factor
is estimated for the timelike photon virtuality $Q^2$ in the range
$4\;m_e^2 \leq Q^2 \leq (\sqrt{s} - m_{\pi^0})^2$ accessible 
in the conversion decay of the $\phi$ meson into the $\pi^0$ and $e^+e^-$ pair. 
An observed significant uncertainty highly motivates the experimental
measurement of this transition form factor.
We stress, that if experimental statistics allows, it is of interest
to study the details of the form factor shape, for example, a possible
kink at the two-pion threshold predicted in Ref.~\cite{Schneider:2012ez1}.
A ``traditional'' way of parametrizing the experimentally studied form factor
in terms of the slope parameter is ambiguous and does not contain
enough information about the shape of the form factor.
Therefore, it is instructive to perform a measurement of the values of the form factor
bin-by-bin in $Q^2$, instead of fitting a differential distribution shape 
with a single-parameter one-pole assumption.
For this purpose we suggest the formulae for the extraction of the transition form factor 
from number of events bin-by-bin and for the extraction of the $V \to e^+e^-P$ branching ratio 
from data.
The form factor and branching ratio extracted this way are consistent with each other,
the normalization condition $F(Q^2=0) \; \equiv \; 1$ holds true by construction
(i.e., no additional fitting of the normalization needed).
The extraction formulae make use of two Monte Carlo bin-by-bin simulations,
one with a ``realistic'' form factor (used for the efficiency estimate)
and the other with constant form factor $F \equiv 1$ (reference value for the form factor shape study).
We argue that the suggested approach allows to considerably 
reduce the ``theoretical'' model dependence of the obtained experimental results
and also to reduce the MC simulation artefacts due to detector simulation,
given the sizes of the $Q^2$ bins are small enough.

It is expected that the new measurements of the $VP\gamma$ transition form factors
in different channels will significantly contribute to understanding the pattern of
the flavor $SU(3)$ symmetry breaking in the light meson sector
and could be helpful for improving the theoretical calculation of the hadronic
light-by-light scattering part of the muon anomalous magnetic moment.

\newpage

\subsection{Towards a Hadron Physics initiative in Horizon 2020}
\addtocontents{toc}{\hspace{2cm}{\sl T.~Johansson}\par}

\vspace{5mm}

T.~Johansson

\vspace{5mm}

\noindent
Uppsala University, Uppsala Sweden\\
\vspace{5mm}

There European Commission has initiated a new framework programme for research and innovation ÒHorizon 2020Ó. The first call for applications is foreseen in 2014, but not many details have been made public yet. The present HadronPhysics3 (HP3) initiative in FP7 will run until 2015. It is therefore time for the European Hadron Physics Community to start to prepare for a new project within Horizon 2020 that can start 2016.

After a public call, Ulrich Wiedner, Bochum University, Germany, has been appointed as the project coordinator for preparing this new proposal for Horizon 2020. The management group of HP3 together with the new coordinator will constitute the Steering Committee. A road map for the continued work has been set up and includes an announcement of an internal call for projects, estimated in August 2013 and with a deadline in November 2013. Afterwards there will be an open meeting in Bochum with presentations of the different projects. The Steering Committee will then start the work on the drafting of the proposal. We expect the publication by the EU of the first calls in Horizon 2020 in January 2014. The ambition is then to be ready to submit is a Hadron Physics proposal in March/April 2014, provided that Integrating  Activities are included in the first call.

\newpage

\subsection{Resonances in the odd-intrinsic sector of QCD}
\addtocontents{toc}{\hspace{2cm}{\sl T.~Kadav\'y}\par}

\vspace{5mm}

T.~Kadav\'y, K.~Kampf and J.~Novotn\'y

\vspace{5mm}

\noindent
Institute of Particle and Nuclear Physics,\\ Faculty of Mathematics and Physics, Charles University, 18000 Prague, Czech Republic

\vspace{5mm}

Using the large $N_C$ approximation we have constructed \cite{Kampf:2011ty} the most general chiral resonance Lagrangian in the odd-intrinsic parity sector that can generate low energy chiral constants (LEC) up to $\mathcal{O}(p^{6})$. For the other works on the odd-intrinsic parity sector see e.g. \cite{Guo:2011ir,Czyz:2012nq,Dumm:2012vb,Terschlusen:2012xw1,RuizFemenia:2003hm}. Integrating out the resonance fields these $\mathcal{O}(p^6)$ constants are expressed in terms of resonance couplings and masses. Using the resonance basis we can also calculate the five possible three-point QCD Green functions of currents of this sector: $\langle VVP\rangle, \langle VAS \rangle, \langle AAP \rangle, \langle VVA \rangle$ and $\langle AAA \rangle$. Employing the high energy constraints we can find additional relations for the resonance couplings. The studies of several phenomenological applications are on progress.\\
\indent Already finished calculations of three-point correlators $\langle VVP\rangle$ and $\langle VAS \rangle$ are published in \cite{Kampf:2011ty}. The first application $VVP$ is the most important example of the
odd-intrinsic sector, both from the theoretical and phenomenological point
of view. We have discussed different aspects of this Green functions. First,
after calculating this three-point correlator within our model and imposing
a QCD high-energy constraint we ended up with the result which depends
only on two parameters. These were further set using new BABAR \cite{Aubert:2009mc} data on $%
\pi\gamma\gamma$ off-shell formfactor and Belle collaboration's \cite{Abe:2006by} limit on $%
\pi^{\prime }\to\gamma\gamma$ decay. After setting these two parameters we
can make further predictions. The outcome of our analysis is for example a
very precise determination of the decay width of a process $\rho\to\pi\gamma$:
$\Gamma_{\rho\to\pi\gamma}=67(2.3)\,\mathrm{keV}$. We have also studied a
relative dependence of the rare decays $\pi^{\prime}\to\gamma\gamma$ and $%
\pi^{\prime }\to\rho\gamma$. Based on the experimental upper limit of the
former one can set the lowest limit of the latter. Prediction of our model
is $30\,\mathrm{keV} \gtrsim\Gamma_{\pi^{\prime}\to\rho\gamma}\gtrsim 4\,\mathrm{keV}$ (based on Belle's $\Gamma_{\pi^{\prime}\to\gamma\gamma} \lesssim 72%
\,\mathrm{eV}$).\\
\indent Next, we have also evaluated the value of $C_7^W$ LEC together
with a short discussion on the $\pi^0$ and $\eta$ two photon decays.
\begin{equation}\nonumber
C_{7}^{W}=\frac{F^{2}}{64M_{V}^{4}}\Bigg(1+2\frac{M_{V}^{2}}{M_{P}^{2}}%
(\delta _{BL}-\delta _{A})\Bigg)\approx (0.35\pm 0.07)\times 10^{-3}\,\mathrm{%
GeV}^{-2}\,.  \label{CW7}
\end{equation}
\indent Last but not
least a very precise determination of the off-shell $\pi^0$-pole
contribution to the muon $g-2$ factor \cite{Jegerlehner:2009ry1} was provided. Our final determination
of this factor is $a_\mu^{\pi^0} = 65.8(1.2) \times 10^{-11}$. The $R\chi T$
approach has thus reduced the error of the similar determination based on
lowest-meson saturation ansatz by factor of ten and is in exact agreement
with the most recent determination based on the AdS/QCD assumptions \cite{Cappiello:2010uy}. Let us note
that the present theoretical error for the complete anomalous magnetic
moment of the muon is around $50\times 10^{-11}$ and the experimental error
around $60\times 10^{-11}$ \cite{Bennett:2006fi} (with the well-know
discrepancy above $3\sigma$). A new proposed experiment at Fermilab E989
\cite{prop989} plans to go down with the precision to the preliminary value $%
16\times 10^{-11}$ and thus the reduction of the error in the theoretical
light-by-light calculation is more than desirable.\\
\indent If $VVP$ represents very important and rich phenomenological
example, the  three-point correlator $\langle VAS \rangle$ is
connected with very rare processes  and represents so far never
studied example of the odd-sector. We have established its OPE
behaviour which enabled us to reduce the dependence of the $VAS$
Green function to one parameter. This opens the possibility of a
future study of these rare but interesting processes.\\
\indent In the last section we have further studied the resonance saturation at low
energies. We have integrated out the resonance fields to establish the
dependence of LECs of odd-sector $C_i^W$ on our parameters. As we are
limited by large $N_C$ we cannot make prediction for $C_3^W$ and $C_{18}^W$
but we have set all other 21 LECs. We have found one relation among $C_{12}^W
$, $C_{14}^W$, $C_{15}^W$ and $C_{22}^W$ free from our parameters.\\
\indent Our main task is now \cite{bakalarka} to study $\langle AAP \rangle$ correlator. We have already finished calculations of the complete $\langle AAP \rangle$ correlator including OPE and now we will study its phenomenology. After that, we keep in mind to expand our attention to the remaining correlators $\langle VVA \rangle$ and $\langle AAA \rangle$ and follow the procedure mentioned earlier.

\newpage

\subsection{Constraints on QCD order parameters from $\eta\to\pi^+\pi^-\pi^0$}
\addtocontents{toc}{\hspace{2cm}{\sl M.~Koles\'ar}\par}

\vspace{5mm}

M.~Koles\'ar, J.~Novotn\'y

\vspace{5mm}

\noindent
Institute of Particle and Nuclear Physics, Charles University, Prague\\

\vspace{5mm}

Quark condensate and pseudoscalar decay constant in the chiral limit are the principal order parameters of spontaneous chiral symmetry breaking in QCD \cite{Gasser:1984gg}. Yet their three flavor values are still only weakly constrained by analyses using experimental data \cite{DescotesGenon:2003cg,DescotesGenon:2007ta}. We try to obtain such constraints by statistical methods from the decay  width of the $\eta$$\,\to\,$$\pi^+\pi^-\pi^0$ decay \cite{Beringer:1900zz,Gasser:1984pr,Bijnens:2007pr} in the framework of resummed chiral perturbation theory \cite{DescotesGenon:2003cg}. We rely on recent estimates of the isospin violating parameter $R$ \cite{Kampf:2011wr}, which is proportional to the difference of the u and d quark masses. Alternatively, by the same methods, we try to extract information on R.

Our calculation closely follows the procedure outlined in \cite{Kolesar:2008jr}. In accord with the method, leading order low energy constants (LECs) are expressed in terms of convenient free parameters

\[
Z = \frac{F_0^2}{F_{\pi}^2}\, ,\ \
X = \frac{2\hat{m}\Sigma}{F_{\pi}^2M_{\pi}^2}\, ,\ \
r = \frac{m_s}{\hat{m}}\, ,\ \
R = \frac{(m_s-\hat{m})}{(m_d-m_u)},
\]
	
\noindent where $F_0$ is the pseudoscalar decay constant in the chiral limit, $\Sigma$ the chiral condensate and $\hat{m}$=$(m_u+m_d)/2$.  
We fix $r\,$=\,25, motivated by lattice results \cite{Colangelo:2010et}. For constraints on $X$ and $Z$ we use the value $R\,$=\,37.8$\pm$3.3 \cite{Kampf:2011wr}. At next-to-leading order, the LECs $L_4$-$L_8$ are algebraically reparametrized using chiral expansions of two point Green functions. For $L_1$-$L_3$ we use the estimate described in \cite{Kolesar:2011wn}. The $O(p^6)$ and higher order LECs, notorious for their abundance, are collected in a relatively smaller number of higher order remainders.  

We use a statistical analysis based on Bayes' theorem \cite{DescotesGenon:2003cg}. The treatment of remainders is based on general arguments about the convergence of the chiral series, leading to

\[G\ =\ G^{(2)}+G^{(4)}+\Delta_G^{(6)},\quad \Delta_G^{(6)}\ \sim \pm 0.1 G,\]

\noindent where $G$ stands for any of our 2- or 4-point Green functions, which generate the remainders. This we statistically implement in two ways, either as a normal or as a uniform probability distribution. We use Monte Carlo sampling with 10000 samples per grid element, the total number of samples being $10^5$-$10^6$.

Our preliminary results have shown that the $\eta\to\pi^+\pi^-\pi^0$ decay width is sensitive to $X$ and $Z$. A large portion of the parameter space can be excluded at 2.0$\,\sigma$ C.L., given information about $R$. It seems $Y\,$=$\,X/Z\,$$\geq$1 is preferred, therefore we have a specific test for $Y$ in preparation. The normal and uniform distributions of the remainders have provided qualitatively similar outcomes.
	
As expected, it's hard to constrain $R$ without information on $X$ and $Z$. Assuming $Z$$\,>\,$0.5 excludes the region $R$$\,>\,$44 at 2.2$\,\sigma$ C.L. and $R$$\,>\,$40 at 1.8$\,\sigma$ C.L.

As an outlook, we work on an in depth statistical stability test of the Monte Carlo sampling and plan to extend the analysis to more parameters and include a wider range of experimental data.

\newpage

\subsection{Status and Recent Results from the CBELSA/TAPS Experiment
at ELSA}
\addtocontents{toc}{\hspace{2cm}{\sl M.~Lang}\par}

\vspace{5mm}

M.~Lang for the CBELSA/TAPS Collaboration

\vspace{5mm}

\noindent
Helmholtz-Institut f\"ur Strahlen- und Kernphysik,\\ 
Rheinische Friedrich-Wilhelms-Universit\"at Bonn, Germany\\
\vspace{5mm}

The CBELSA~/TAPS experiment at the electron stretcher accelerator ELSA in
Bonn~/Germany offers the possibility to carry out precision measurements of
baryon resonances up to a center of mass energy of 2.5~GeV utilising real
photons as electromagnetic probes. It focuses on neutral mesons in the final
state by using a combination of two electromagnetic calorimeters, the Crystal
Barrel and the Mini-TAPS detector, and thus covering almost the entire 4$\pi$ 
solid angular range.

Aside from projects carried out at JLab and MAMI, this experiment strongly
contributes to the international efforts to create a {\em complete data base}
for a {\em complete experiment}. Hence, an interpretation of the data with a
unique partial wave analysis (PWA) solution will be possible. In order to
achieve this goal, it is essential to explore polarization degrees of
freedom \nocite{2007a} aside from precise, unpolarized data already measured 
in Bonn for various final states such as $\mbox{p}\pi^0$,
$\mbox{n}\pi^+$, $\mbox{p}\eta$, $\mbox{K}^+\Lambda$,
$\mbox{p}\pi^+\pi^0$.
In general, 8 carefully chosen polarization observables need to be
measured for the full determination of the CGLN production amplitudes
$F_1(\omega,\Theta)$ to $F_4(\omega,\Theta)$ except for an overall phase ~\cite{1997a}.

At an energy range between the $\pi$ and the $\pi\pi$ photoproduction
thresholds, where the degrees of freedom are reduced, the photoproduction
process is described by a finite number of 4 multipoles. This is given by an 
upper limit of the angular momentum quantum number $l=1$. 
Using the Fermi-Watson theorem and the $\pi N$ scattering phase accessible 
from $\pi N$ scattering
experiments, only two observables are required for a complete
data base and a unique description of the measured values in this case.
It was shown in the analysis of experimental data as presented in \cite{1999a} 
and \cite{2000a}. Thus, a complete set of measured
polarization observables is the key for finding a unique PWA solution for the 
full energy range.

Recent measurements at ELSA
were devoted to collect data for the polarization observables E, G, P,
T and H for various final states. The current focus lies on the reactions
$\gamma\mbox{p}\rightarrow\mbox{p}\pi^0$ and
$\gamma\mbox{p}\rightarrow\mbox{p}\eta$. Recent data measured for the
observable G can be found in~\cite{2012a}.

The Crystal Barrel detector will be modified using avalanche
photodiodes (APDs) and new trigger electronics starting end of 2013 in order
to highly improve its neutral trigger capability and to access entirely 
neutral final states such as $\mbox{n}\pi^0$ with a high detection
efficiency. Exploring neutral channels will clearly improve the precision 
of data currently available representing a further important step towards a
complete experiement.

\newpage

\subsection{Low momentum dielectrons radiated off cold nuclear matter}
\addtocontents{toc}{\hspace{2cm}{\sl M.~Lorenz}\par}

\vspace{5mm}

M.~Lorenz for the HADES Collaboration

\vspace{5mm}

\noindent
Institut f\"ur Kernphysik, Goethe Universt\"at Frankfurt, Germany\\

\vspace{5mm}

The relation between chiral symmetry restoration and hadron properties inside a strongly interacting medium is a much debated topic and has motivated plenty of work, both in theoretical and experimental physics. Spontaneous breaking of chiral symmetry leads to non vanishing values of QCD condensates \cite{Hilger}, most prominently the two-quark condensate, which among others might be related to hadron masses. Various models predict relatively strong changes of particle masses and/or widths already at normal nuclear matter density $\rho_{0}$ for vector mesons  \cite{weise,rho,lee,rapp}.
A consistent picture of in-medium hadron properties has however not yet emerged, asking for more experimental input and yielding the question to what extent a relation between observed hadron masses and chiral symmetry restoration is possible. On the other hand such modifications, are expected to be most pronounced for particles with small relative momenta to the surrounding medium according to hadronic models \cite{leupold}, a region which is challenging to access in experiment and hence has as only been scratched by experiments.

An ideal experimental probe for the study of medium properties is the decay of vector mesons into an e$^{+}$e$^{-}$ pair, since electrons and positrons are not affected by strong final state interactions. 
In order to investigate such modifications one has to distort the QCD vacuum, which can be achieved in heavy-ion collisions or by impinging elementary beams on (heavy) nuclei. Although the effects are expected to be stronger in heavy-ion collisions, compared to induced photon, pion or proton induced reactions on nuclei, measured observables represent an average over the complete space-time evolution in temperature and density evolution of the system and hence are complicated to model. On the other hand, in induced reactions on nuclei the system does not undergo a noticeable density and temperature evolution in time and hence the conditions of the system are well defined. The experimental drawback is that, for a measurement sensitive to the in-medium spectral shape, the decay to an e$^{+}$e$^{-}$ pair has to take place inside the nucleus. Therefore good acceptance for decays of low momentum vector mesons is of crucial importance, in particular, for the relatively long living dilepton sources like $\omega$ and $\phi$ mesons.

As mentioned above, most experiments focusing on the spectral distribution of dielectrons produced off nuclei in photon and proton induced reactions are restricted to relatively high momenta ($P_{ee}>0.8$ GeV/c) and are not conclusive yet. For the $\rho$ meson, the CLAS experiment at JLab \cite{clas} reports a slight broadening and no shift of the $\rho$ pole position in photon induced reactions, while the E325 experiment at KEK \cite{kek} deduced a shift but no broadening in proton induced reactions.

Recently HADES has published data \cite{hades-pNb} on inclusive e$^{+}$e$^{-}$ pair production in p+Nb reactions at E$_{kin} = 3.5$~GeV, representing the first high statistics measurement with small e$^{+}$e$^{-}$ pair momenta relative to the medium ($P_{ee}<0.8$ GeV/c). These results are compared to reference data measured in p+p reactions at the same incident beam energy \cite{hades-pp} in order to extract medium modifications \footnote{Note that, according to the systematics on dielectron emission obtained in p+p and d+p collisions at various beam energies by the DLS collaboration \cite{DLS}, it is safe to conclude that at the kinetic beam energy of 3.5 GeV isospin effects play only a secondary role. Therefore, in order to extract medium effects in p+Nb, the above discussed p+p data at the same kinetic beam energy of 3.5 GeV/c represent a valuable reference.}.

Comparing the shape of the invariant mass spectra separately for pairs with momenta larger and smaller 0.8 GeV/c to pairs from p+p, scaled to the number of participants and the total reaction cross section, we observe a strong e$^+$e$^-$ excess yield below the $\omega$ pole mass at small pair momenta, while for pairs with $P_{ee}>$ 0.8 GeV/c no significant difference in the vector meson mass region within the systematic uncertainties is visible. For pairs with $P_{ee}<$ 0.8 GeV/c the e$^+$e$^-$ yield at the $\omega$ pole mass is not reduced, but as the underlying smooth distribution is enhanced, the yield in the peak is reduced to almost zero within errors. This additional yield we attribute to $\rho$-like channels which are supposed to be the dominating source for radiation from the medium due to the large total width of the $\rho$. For more details, see \cite{hades-pNb}.\newline

\newpage

\subsection{Chiral dynamics with vector mesons }
\addtocontents{toc}{\hspace{2cm}{\sl M.F.M.~Lutz}\par}

\vspace{5mm}

M.F.M.~Lutz$^{1,2}$, I.G.~Danilkin$^{1}$, 
S.~Leupold$^{3}$, C.~Terschl\"usen$^{3}$

\vspace{5mm}

\noindent
$^1$ GSI Helmholtzzentrum f\"ur Schwerionenforschung GmbH,\\
Planckstra\ss e 1, 64291 Darmstadt, Germany\\
$^2$ Institut f\"ur Kernphysik, Technische Universit\"at Darmstadt, 64289 Darmstadt, Germany \\
$^3$ Institutionen f\"or fysik och astronomi, Uppsala Universitet, \\ Box
516,75120 Uppsala, Sweden
\vspace{5mm}

We study the reactions $\gamma\gamma\rightarrow \pi^0\pi^0$,
$\pi^+\pi^-$, $K^0\bar{K}^0$, $K^+K^-$, $\eta\,\eta$ and
$\pi^0\eta$ based on a chiral Lagrangian with dynamical light
vector mesons as formulated within the hadrogenesis conjecture. At
present our chiral Lagrangian contains 5 unknown parameters that
are relevant for the photon fusion reactions. They parameterize
the strength of interaction terms involving two vector meson
fields. These parameters are fitted to photon fusion data
$\gamma\gamma\rightarrow \pi^0\pi^0$, $\pi^+\pi^-, \pi^0\eta$  and
to the decay $\eta\rightarrow\pi^0\gamma\gamma$. In order to
derive gauge invariant reaction amplitudes in the resonance region
constraints from maximal analyticity and exact t coupled-channel
unitarity are used. Our results are in good agreement with the
existing experimental data from threshold up to about 0.9 GeV for
the two-pion final states. The $a_0$ meson in the $\pi^0\eta$
channel is dynamically generated and an accurate reproduction of
the $\gamma\gamma\rightarrow \pi^0\eta$ data is achieved up to 1.2
GeV. Based on our parameter sets we predict the
$\gamma\gamma\rightarrow $ $K^0\bar{K}^0$, $K^+K^-$, $\eta\,\eta$
cross sections.

\newpage

\subsection{Meson assisted baryon-baryon interactions}
\addtocontents{toc}{\hspace{2cm}{\sl H.~Machner}\par}

\vspace{5mm}

H.~Machner

\vspace{5mm}

\noindent
Fakult\"{a}t f\"ur Physik, Universit\"{a}t Duisburg-Essen, Lotharstr. 1, 47048 Duisburg, Germany\\

\vspace{5mm}

We discuss baryon-baryon interaction in three body final states. Starting from experiments dealing with $nd\to pnn$ \cite{Huhn_2001, GonzalezTrotter_2006} we argue that reactions with one meson in the final state are much more favourable, since the meson-two baryon interaction is much weaker than the baryon-baryon interaction. The method to study the interaction is the effective range formalism as discussed by Goldberger and Watson \cite{GW64}.

A first example is the $pp\to pp\pi^0$ reaction, where an Indiana group \cite{Meyer01} questioned the factorisation in the Goldberger Watson approach. An a subsequent study by the GEM collaboration \cite{Betigeri02} it could be shown that differential data as well as the excitation function could be well reproduced with the standard effective range parameters for the $pp$ system in the $s$-wave.

The $pn$ interaction was studied in $pp\to pn\pi^+$ reactions \cite{Boudard96}. The triplet strength was extracted from the deuteron production with the help of a theorem \cite{Faeldt_97}. The remaining cross section was then attributed to the $pn$ singlet scattering. High resolution data from the GEM collaboration \cite{Abdel_Bary05} showed that a huge fraction of the continuum cross section was due to a leaking of the deuteron into the continuum. The new data indicate that there is practically no singlet strength. Also an effect due to the tensor force could be excluded \cite{Budzanowski08}.

The opposite was found by the HIRES collaboration: in $pp\to \Lambda pK^+$ there is no bound deuteron like system and all yield is due to singlet interaction \cite{Budzanowski10}.  Inclusive data allowed to extract the $pp\to \Sigma^+nK^+$ cross section \cite{Budzanowski_2010a}. No indication for a deviation from phase space was found. A search for a theoretical predicted dibaryon with strangeness $S=-1$ yielded negative results \cite{Budzanowski_2011}.

A peak like structure is visible in inclusive as well as exclusive data of the $pp\to \Lambda pK^+$ reaction and also in the $K^-d\to\Lambda p\pi^-$ reaction cross sections close to the $\Sigma N$ threshold \cite{Machner_2013}. No conclusive answer could be given on the nature of this peak: a bound state in the $\Sigma N$ system, a genuine cusp or a resonance in the $\Lambda^+$-system.

The author is grateful to the members of the GEM and HIRES collaborations.

\newpage

\subsection{$\eta$ and $\eta$' - Transition Form Factors from Rational Approximants}
\addtocontents{toc}{\hspace{2cm}{\sl P.~Masjuan}\par}

\vspace{5mm}

P.~Masjuan\footnote{Supported by the Deutsche Forschungsgemeinschaft DFG through the Collaborative Research Center ``The Low-Energy Frontier of the Standard Model" (SFB 1044)}

\vspace{5mm}

\noindent
Institut f\"ur Kernphysik, Johannes Gutenberg Universt\"at Mainz, Germany\\

\vspace{5mm}

$\eta$- and $\eta'$- Transition Form Factors (TFF) in the space-like region have received a lot of attention in the QCD community during last years, triggered by the experimental effort on measuring them~\cite{Behrend:1990sr,Gronberg:1997fj,Acciarri:1997yx,BABAR:2011ad}. Experimentally, one access such space-like form factors using the single-tag method on the $\ee \rightarrow \ee P$ process with one lepton tagged (which emits a highly off-shell photon with momentum transfer $q^2_1=-Q^2$) and the other lepton untagged (emitting then a quasi real photon with $q_2^2 \sim 0$). Theoretically, the limits $Q^2=0$ and $Q^2\rightarrow\infty$ are well known in terms of the axial anomaly in the chiral limit of QCD \cite{Adler:1969gk,Bell:1969ts} and perturbative QCD (pQCD) \cite{Lepage:1980fj}, respectively. The TFF is then calculated as a convolution of a perturbative hard-scattering amplitude and a gauge-invariant meson distribution amplitude (DA)~\cite{Mueller:1994cn} which incorporates the nonperturbative dynamics of the QCD bound-state~\cite{Lepage:1980fj}. Some model needs to be used either for the DA or the TFF itself. The discrepancy among different approaches reflects the model-dependency of that procedure.

We propose~\cite{EscribanoMasjuan}  to use a sequence of rational approximants called Pad\'e Approximants (PA)~\cite{Baker,Masjuan:2007ay,Queralt:2010sv} constructed from the Taylor expansion of the TFF to fit the available experimental data and obtain, in such a way, the derivatives of the TFF at the origin of energies in a simple, systematic and model-independent way~\cite{Masjuan:2008fv,Masjuan:2012wy}. Including the decays of the $\eta^{(')}\rightarrow \gamma \gamma$ in our set of data, we can systematically predict the slope and the curvature of both $\eta^{(')}$-TFFs, and ascribe a systematic error on the procedure~\cite{Masjuan:2008fv,Masjuan:2009wy}. Notice that this method does not allow a determination of the resonance pole properties present on the given channel~\cite{MasjuanP,Masjuan:2013jha}. A matching procedure of our TFF's parameterizations with the pQCD results would provide a description for the whole energy range. The low-energy parameters obtain with this method can be used to constrain the hadronic models used to account for the light-by-light scattering contribution part of the anomalous magnetic moment of the muon~\cite{Masjuan:2012wy,Jegerlehner:2009rym,Masjuan:2012qn}.

The physical $\eta$ and $\eta'$ mesons are an admixture of the $SU(3)$ Lagrangian eignestates~\cite{Leutwyler:1997yr,Feldmann:1998vh,Escribano:2005qq,Escribano:2010wt}. Deriving the parameters governing the mixing is a challenging task. Usually, these are determined through the use of $\eta(')\rightarrow \gamma\gamma$ decays as well as vector radiative decays into $\eta(')$ (see Refs.~\cite{Feldmann:1998vh,Escribano:2005qq}). However, since pQCD predicts 
that the asymptotic limit of the TFF for the $\eta(')$ is essentially given in terms of these mixing 
parameters~\cite{Lepage:1980fj,Feldmann:1998vh}, we use our TFF parametrization to estimate the asymptotic limit and further constrain the mixing parameters with compatible results compared to standard (but more sophisticated) determinations.

Finally, a simple extrapolation of our parameterization of the $\eta$-TFF to the time-like region can be used to confront the available data on the $\eta$ Dalitz decay $\eta \rightarrow \ee \gamma$ (see the contribution of M.~Unverzagt in this mini-proceedings).

\newpage

\subsection{The search for meson-nucleus bound states}
\addtocontents{toc}{\hspace{2cm}{\sl V.~Metag}\par}

\vspace{5mm}

V.~Metag for the CBELSA/TAPS collaboration

\vspace{5mm}

\noindent
II. Physikalisches Institut, Justus Liebig Universit\"at Giessen, Germany\\

\vspace{5mm}

The existence of meson-nucleus bound states has been established in a series of measurements at the fragment separator at GSI \cite{Itahashi,Geissel,Suzuki} by studying recoil-free production of $\pi^-$ mesons in the (d,${}^3\textrm{He})$ reaction on various Pb and Sn isotopes. The widths of these deeply bound pionic states was found to be about 1/10 of the binding energy, giving rise to well resolved peaks in the ${}^{3}\textrm{He}$ kinetic energy spectrum. The origin of these states is due to a pocket-like potential arising from the superposition of the attractive Coulomb potential and the (at low momenta) repulsive $\pi^-$-nucleus potential, leading to a halo-like $\pi^-$ distribution around the nucleus.

The key question is whether mesic states exist with neutral mesons bound to nuclei only by the strong interaction. Theoretical predictions for $\eta-$ \cite{Garcia_Recio,Friedman}, $\omega-$  \cite{Marco_Weise,Nagahiro}, and $ \eta^\prime-$ \cite{Jido} mesic states have been made. In case of the $\eta$ meson the predicted widths are comparable to or even larger than the binding energies. This implies that the strength of the states partially extends into the continuum, allowing for free meson emission. Experimental indications for surprisingly large $\eta$ yields in coherent $\eta$ photo production near threshold have been reported \cite{Pfeiffer,Pheron}. Similar observations have been made by the COSY-ANKE collaboration studying the $pd \rightarrow \eta {}^{3}\textrm{He}$ reaction \cite{Goslawski}. In both cases, the very strong rise of the cross section near threshold is taken to be indicative for a quasi bound state close to threshold. A direct observation of an $\eta$ bound state has been claimed by the COSY-GEM collaboration in the $p+{}^{27}\textrm{Al} \rightarrow {}^{3}\textrm{He} +{}_{\eta}^{25}\textrm{Mg} \rightarrow {}^3\textrm{He} + p +\pi^- +X$ reaction \cite{Budzanowski}. 

The case of $\eta^\prime$ mesic states appears to be promising because of the relatively narrow in-medium width of the $\eta^\prime$ meson of about 20 MeV determined in a transparency ratio measurement \cite{Nanova} (see also contributions by H. Nagahiro and M. Nanova in this mini-proceedings). Corresponding experiments are planned at the GSI fragment separator and later at the SuperFRS at FAIR \cite{Kenta1} as well as at the BGO-OD spectrometer at the electron accelerator ELSA \cite{Volker}.

A search for $\omega$ mesic states has been performed with the CBELSA/TAPS detector, exploiting recoil free kinematics. A carbon target was irradiated with photons of 1250-3100 MeV. Protons detected at forward angles take over the beam momentum, leaving the $\omega$ meson almost at rest so that it can be captured by the nucleus in case of an attractive interaction. The search for $\omega$ bound states is hampered by the large in-medium with of the $\omega$ meson known to be $\Gamma(\rho=\rho_0)=140 $ MeV from a transparency ratio measurement \cite{Kotulla}. No structures were observed in the bound state region \cite{Friedrich}. An attempt is being made to extract information on the sign and strength of the real part of the $\omega {}^{11}\textrm{B}$ potential by analyzing the kinetic energy distribution of the $\omega$ mesons in comparison to a reference measurement on a LH$_2$ target.

\vspace{5mm}
 $^\star$Funded by DFG(SFB/TR-16)

\newpage

\subsection{ Second class $\tau\to\eta\pi\nu$ decay and the $\eta\to 3\pi$ amplitude}
\addtocontents{toc}{\hspace{2cm}{\sl B.~Moussallam}\par}

\vspace{5mm}

B.~Moussallam

\vspace{5mm}

\noindent
IPN, Universit\'e Paris-Sud XI, 91406 Orsay, France

\vspace{5mm}

Semileptonic weak decays which are forbidden in the exact isospin
limit have been labelled as second class by
Weinberg~\cite{Weinberg:1958ut}. It is a curious fact that, while
isospin symmetry is broken both by the $u-d$ quark mass difference and
by QED in the standard model, no  second class decay has been clearly
observed yet. In this talk (based on work in collaboration with
S. Descotes-Genon) we reconsider the isospin violating $\tau$ decay
mode: $\tau\to\eta\pi\nu$. This mode was searched for at the
$B$-factories but not seen, despite the sufficient luminosity, 
because of the difficulty in controlling the background with enough
accuracy~\cite{delAmoSanchez:2010pc}. One may hope that this problem
could be circumvented at tau/charm factories, where the tau's can be
produced practically at rest, such that one could measure not only the
integrated branching fraction but also the energy distribution and
thus gather information on the both the vector and scalar form factors $\fplus$,
$\fzero$  which both encode nontrivial dynamical
information.    

\begin{figure}[h]
\includegraphics[width=0.50\linewidth]{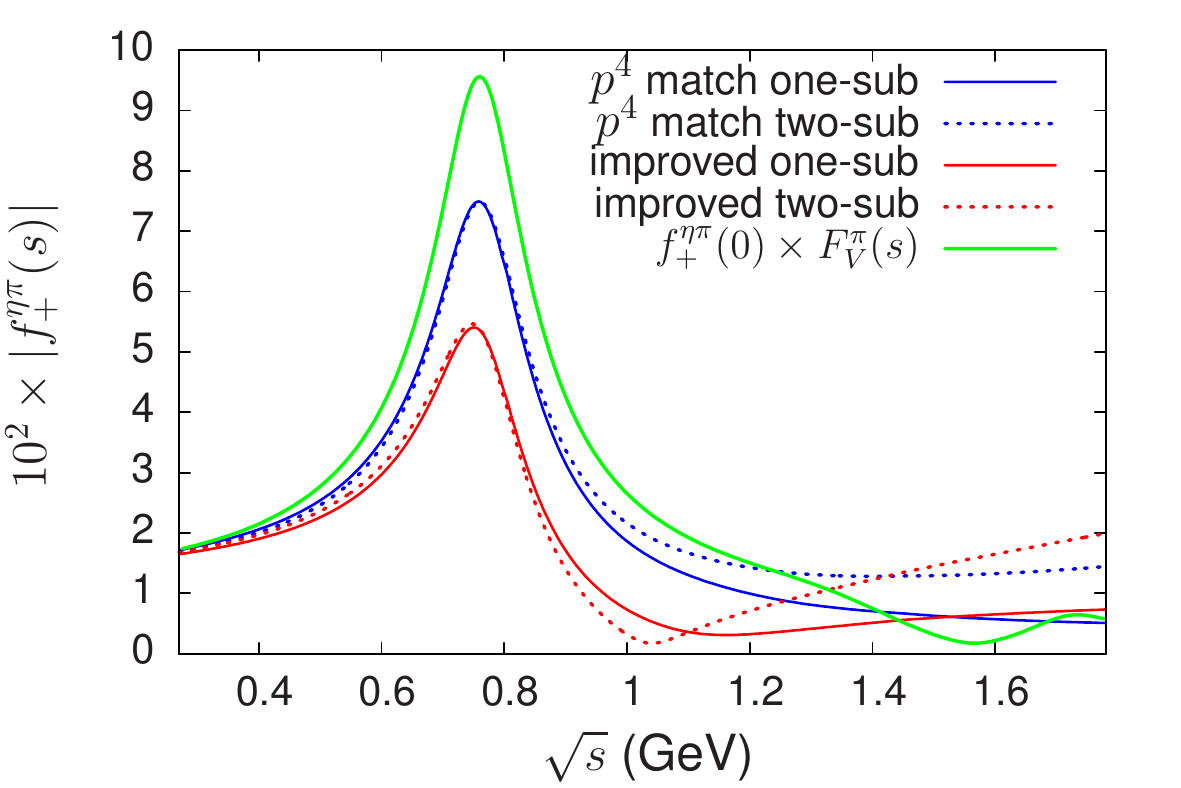}\includegraphics[width=0.50\linewidth]{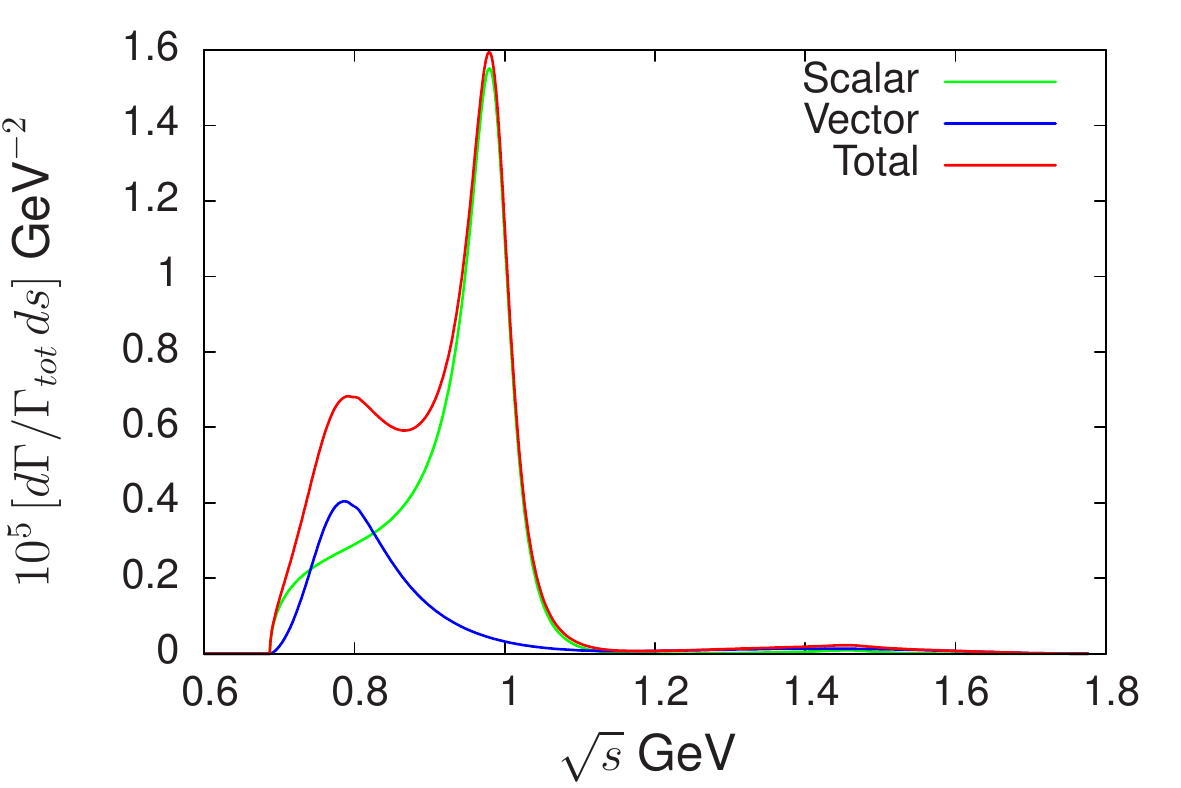}
\caption{{\bf Left:} comparison of  a naive VMD model for the vector form
  factor $|\fplus|$ (green curve) with dispersive calculations using
  Khuri-Treiman solutions. {\bf Right:} spectral function for
  $\tau\to\eta\pi\nu$ decay using central values for the dispersive
  constructions of the vector and scalar form factors.}
\label{fig:m1}
\end{figure}

Theoretical estimates of these two functions can be made based on
their analyticity properties combined with unitarity as well as results
from chiral symmetry which constrain the form factors near $s=0$. In
particular, based on ChPT calculations of the $\eta\pi$ form factors at NLO, a
remarkably simple relation was found in Ref.~\cite{Neufeld:1994eg}
relating the $\eta\pi$ form factor to the $K^+\pi^0$ and $K^0\pi^+$ ones 
\begin{equation}\nonumber
f_+^{\eta\pi}(0)={1\over\sqrt3}\left[\frac{f_+^{\kplus\pizero}(0)}{
  f^{\kzero\piplus}_+(0)}     -1-{3e^2\over4(4\pi)^2}
\log{\mkd\over\mpid}\right]\ ,
\end{equation}  
which gives (see~\cite{Antonelli10}) $f_+^{\eta\pi}(0)=
f_0^{\eta\pi}(0)=(1.49\pm0.23)\,10^{-2}$. In the case of the vector
form factor, the unitarity relation below one GeV involves the
$\pi\pi\to\eta\pi$ amplitude in a partly unphysical kinematical
region. We have evaluated it  based on a four-parameter family of
solutions of the Khuri-Treiman equations (following the approach of
ref.~\cite{Anisovich:1996tx}) constrained from both the  $\eta\to
3\pi$ amplitude computed in ChPT at NLO near the Adler zero and from recent
high statistics experimental data on the Dalitz plot~\cite{kloe08}.
Fig.~\ref{fig:m1} (left) shows that the result of this dispersive construction is
rather different from a naive VMD approach, in particular, the height of
the $\rho$ peak is reduced by as much as a factor of two compared to
VMD when using Khuri-Treiman parameters constrained by experiment. 

In the case of the scalar form factor $\fzero(s)$ now, the unitarity
relation involves (essentially) the elastic $\eta\pi\to\eta\pi$
amplitude below one GeV. It is then convenient to write a phase dispersive
representation. A difficulty, however, is that no detailed measurement of
the $\eta\pi$ scattering phase-shift is available. We will rely here
on the model proposed in ref.~\cite{schechteretapi99}, which
reproduces the properties of the $a_0(980)$ and $a_0(1450)$ resonances
and makes the plausible prediction that the global features of
$\eta\pi$ scattering are rather similar to those of  $\pi\pi$ or $\pi
K$ scattering. Above the leading inelastic threshold, the phase of the
form factor differs from the scattering phase-shift. In the case of
$K\pi$, a modelling of this difference has been
proposed~\cite{Jamin:2001zq} based on a coupled-channel
Muskhelishvili-Omn\`es approach using inputs from elastic as well as
inelastic $K\pi$ scattering. A striking feature of the resulting
$K\pi$ form factor phase is that it undergoes a sharp drop after the
onset of inelasticity. It is plausible that the $\eta\pi$ form factor
phase could display a similar behaviour. A drop in the phase
corresponds to a dip in the modulus of the form factor and thus to a
reduced coupling to the $\bar{u}d$ operator. In this language, the
position of the dip is close to a resonance mass if the resonance
is exotic. Fig.~\ref{fig:m1} (right) shows the spectral function for
$\tau\to\eta\pi\nu$ decay assuming that the  phase drop occurs midway
between the two $a_0$ resonances. Within this scenario, the peak of
the $\rho$ meson, which constitutes a 
background-free signature of isospin violation, is clearly visible. 

\newpage

\subsection{$\eta'(958)$ mesic nuclei formation by (p,d) reaction}
\addtocontents{toc}{\hspace{2cm}{\sl H.~Nagahiro}\par}

\vspace{5mm}

H.~Nagahiro

\vspace{5mm}

\noindent
Department of Physics, Nara Womens' University, Nara, Japan

\vspace{5mm}

 The mass of the $\eta'(958)$ meson
 is much heavier than other octet pseudoscalar mesons,
which is known as the $U_A(1)$ problem.  Because
the $U_A(1)$ symmetry is explicitly broken by quantum anomaly, the
$\eta'$ meson is not necessarily a Nambu-Goldstone boson associated with
spontaneous chiral symmetry breaking. 
However, we have not yet
understood quantitatively generation mechanism of the mass of 
$\eta'$.
Since the chiral symmetry  breaking plays
important role in the generation mechanism of the $\eta'$ mass
\cite{Nagahiro:2006dr,Jido:2011pq}, the study of the in-medium mass of
$\eta'$ gives us 
important information on 
the partial restoration of the chiral symmetry.
The anomaly effect can contribute to the mass of $\eta'$ only with the
presence of the chiral symmetry breaking~\cite{Jido:2011pq}, and a
relatively large mass reduction ($\sim$ 100 MeV) of $\eta'$ at normal
saturation density is expected owing to its partial restoration.
Therefore
we can expect to observe the $\eta'$-nuclear bound states
in appropriate formation reactions.
The formation reaction of the $\eta'$ mesic nuclei was first considered
in \cite{Nagahiro:2004qz} and is considered to be possible at GSI \cite{Itahashi:2012ut}.

\begin{figure}[h]
\center
  \includegraphics[width=0.6\textwidth]{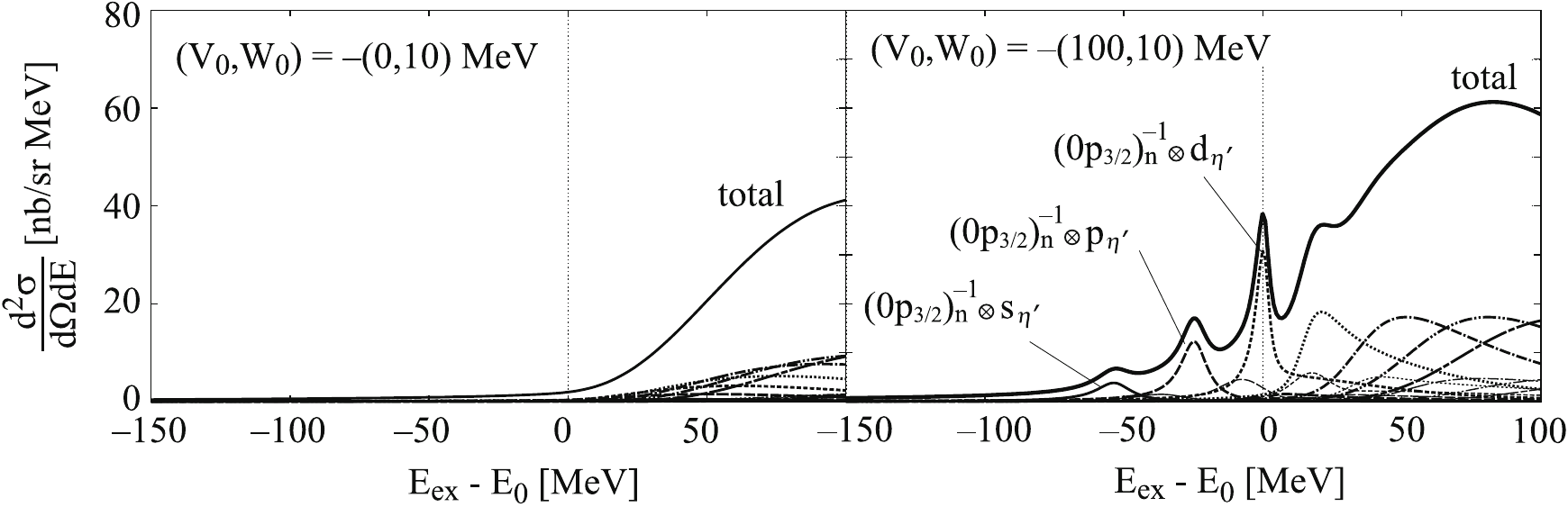}
\caption{Calculated spectra of $^{12}$C$(p,d)$$^{11}$C$\otimes\eta'$ 
 reaction as functions of the excited energy.
}
\label{fig:p_d}       
\end{figure}

In Fig.~\ref{fig:p_d} we show the formation spectra of
$\eta'$-mesic nuclei by the $(p,d)$ reaction with the proton 
kinetic energy $T_p=2.5$ GeV.  
The potential depth is set to 
be $-(0,10)$ MeV and $-(100,10)$ MeV.
A strongly attractive potential admitting
bound states gives rise to clear peak structures and
there is a clear difference between the spectra
obtained using attractive or non-attractive potentials.  In
Refs.~\cite{Itahashi:2012ut,Nagahiro:2011fi} we show formation
spectra for various cases.

\newpage

\subsection{Experimental Determination of the $\eta$' -Nucleus Potential and the Search for $\eta$'-Mesic States}
\addtocontents{toc}{\hspace{2cm}{\sl M.~Nanova}\par}

\vspace{5mm}

M.~Nanova for the CBELSA/TAPS Collaboration

\vspace{5mm}

\noindent
II. Physikalisches Institut, Justus Liebig Universit\"at Giessen, Germany\\

\vspace{5mm}

The origin of the exceptionally large mass of the $\eta$' pseudoscalar meson is
important for our understanding of QCD dynamics as it is linked to the $U_{A}(1) $ axial vector anomaly and the breaking of chiral symmetry. Studying the in-medium properties of the $\eta$'  meson is a first step to search for a mass drop as a sign for a partial restoration of chiral symmetry.
Information on the in-medium properties of the $\eta$'  meson can be gained by determining the optical $\eta$' -nucleus potential:
\begin{equation}\nonumber
U_{\eta^\prime}(r) = V(r) + iW(r),
\end{equation}
where V and W denote the real and imaginary parts of the optical potential, respectively, and r- is the meson-center of nucleus distance. The $\eta^\prime$ in-medium mass shift $\Delta m(\rho_{0})$ at saturation density $\rho_{0}$ can be related to the strength of the real part \cite{Nagahiro1}:
\begin{equation}\nonumber
V(r) = \Delta m(\rho_{0})\cdot \frac{\rho(r)}{\rho_{0}}\, .
\end{equation}
The imaginary part of the potential is responsible for the meson absorption in the medium and is connected with the in-medium width $\Gamma_{0}$ of the meson at normal nuclear matter density by:
\begin{equation}\label{eq.3}
W(r) = -\frac{1}{2}\Gamma_{0}\cdot \frac{\rho(r)}{\rho_{0}}.
\end{equation}
Since the imaginary part of the optical potential represents the meson absorption, the experimental approach to determine it is a measurement of the in-medium width of the $\eta^\prime$ meson. As it has been shown in ~\cite{nanova} the in-medium width of the $\eta^\prime $-meson can be extracted from the attenuation of the $\eta^\prime$-meson flux deduced from a measurement of the transparency ratio for a number of nuclei.
For the $\eta^\prime$ meson an in-medium width of 15-25 MeV at saturation density is obtained at an average recoil momentum $p_{\eta^\prime}$ = 1.05\ GeV/c~\cite{nanova}.
Taking into account Eq.~(\ref{eq.3}) the imaginary part of the optical potential at this density is determined to: $W(\rho_{0})=-(10.0\pm2.5)$ MeV.\\

 Information on the real part of the meson-nucleus potential and on the in-medium mass can be extracted from a measurement of the excitation function and momentum distribution of a meson as discussed in~\cite{Weil}. A downward shift of the meson mass would lower the threshold for meson photo-production. Due to the enlarged phase space, the production cross section for a given incident beam energy will increase as compared to a scenario without mass shift. Furthermore, mesons produced in a nuclear reaction leave the nuclear medium with their free mass. In case of an in-medium mass drop, this mass difference has to be compensated at the expense of their kinetic energy.  As demonstrated in GiBUU  transport-model calculations \cite{Weil}, this leads to a downward shift in the momentum distribution as compared to a scenario without mass shift. A mass shift can thus be indirectly inferred from a measurement of the excitation function and/or the momentum distribution of the meson. 
For the $\eta^\prime$ meson, this idea has independently been pursued on a quantitative level by Paryev \cite{Paryev}.
The data taken with the CB/TAPS detector setup at ELSA accelerator in Bonn on a carbon target have been analised and compared to model calculations  \cite{Paryev} assuming different scenarios for the real part of the potential, related directly to the in-medium mass modification of the  $\eta^\prime$ meson. The preliminary results do not support a very deep but a still sufficiently attractive  $\eta^\prime$ -nucleus potential to allow for the existence of  $\eta^\prime$ -mesic states. The search for such states is encouraged by the relatively small in-medium width of the $\eta^\prime$. There are experiments proposed to search for $\eta^\prime$ bound states via missing mass spectroscopy~\cite{Kenta} at the FRS at GSI (see contribution of H. Nagahiro in this mini-proceedings) and in an exclusive measurement at the BGO-OD setup at the ELSA accelerator in Bonn~\cite{volker}. The observation of an $\eta^\prime$-nucleus bound states would provide important information on the in-medium properties of the $\eta^\prime$ meson.
  \\

 $^\star$Funded by DFG(SFB/TR-16)

\newpage

\subsection{Investigations on $\eta$-mesic nuclei and on the $\eta$-meson mass }
\addtocontents{toc}{\hspace{2cm}{\sl M.~Papenbrock}\par}

\vspace{5mm}

Michael Papenbrock, Christopher Fritzsch, Paul Goslawski, Alfons Khoukaz, Malte Mielke, Marcel Rump, Daniel Schr\"{o}er, Alexander T\"{a}schner\\
for the ANKE Collaboration

\vspace{5mm}

\noindent
Institut f\"{u}r Kernphysik, Westf\"{a}lische Wilhelms-Universit\"{a}t M\"{u}nster\\

\vspace{5mm}

Detailed studies on the reaction $\mathrm{d}+\mathrm{p}\rightarrow{^3{\mathrm{He}}}+\eta$ with unpolarized particles have been performed in former measurements at the ANKE spectrometer \cite{Mers2007,Raus2009}. The rapid rise of the total cross section within the first 0.5 MeV excess energy above threshold implies a very strong final state interaction and might indicate the presence of a quasi-bound $\eta$-mesic state close to threshold \cite{Wilkin2007}. Therefore, systematic investigations on the interaction of the $\eta$ meson with light nuclei have been started. In particular, experiments on the $^3 \mathrm{He} \eta$ final state with a polarized deuteron beam have been performed to investigate possible spin dependent contributions. In addition, very recently measurements on the $\mathrm{d} \eta$ system via quasi-free pn scattering have been taken to study the strength of the final state interaction as function of the mass of the participating nucleus. The reaction $\mathrm{d}+\mathrm{p}\rightarrow{^3{\mathrm{He}}}+\eta$ has also been found to be excellently suited 
for the determination of 
the $\eta$-meson mass by a model independent method based on pure kinematics resulting in the highest precision $\eta$ mass measurement yet \cite{Gosl2012}.
 
The state of the analysis and recent results for these subjects were presented and discussed.

\vspace{5mm}
Supported by the COSY-FFE program.

\newpage

\subsection{N*«s: Electromagnetic form factors}
\addtocontents{toc}{\hspace{2cm}{\sl M.~T.~Pe\~na}\par}

\vspace{5mm}

M.~T.~Pe\~na

\vspace{5mm}

\noindent
CFTP, Instituto Superior T\'ecnico,
Universidade T\'ecnica de Lisboa,\\
Avenida Rovisco Pais, 1049-001 Lisboa, Portugal

\vspace{5mm}

An era of deep physical insight and understanding of the nucleon excitations became recently possible by the experimental studies with electron beam facilities and large acceptance detectors~\cite{Aznauryan09}. Further developments are expected with the forthcoming Jlab upgrade, where the explored space like region will be extended. 
From the theoretical side, several frameworks have been developed, from constituent quark models~\cite{Close72} to Lattice QCD simulations, and from dynamical meson-baryon coupled channel models~\cite{Diaz08} to chiral perturbation theory, QCD in the large $N_c$ limit, and perturbative QCD. The connection between different approaches is sometimes needed to elucidate us on the strong interaction dynamics in the non-perturbative regime.
Alone, each of the approaches, often cannot establish this knowledge.  In our work we connected a constituent quark model to lattice QCD calculations (LQCD). 

Our framework is the Covariant Spectator Theory (CST)~\cite{Gross}, a field-theoretic approach which is the basis of our constituent quark model model~\cite{Nucleon,NDelta,NDeltaD}, and which was proven to be efficient in the description of the $\gamma$ N$\rightarrow$ N* vertices, at least for high momentum transfer.  The connection to LQCD allowed us to constrain our model in a solid way, since it enabled us to disentangle the effects of the
constituent bare quark core structure of the baryons from the meson cloud contributions. The connection to LQCD proceeded by the following realizations~\cite{LatticeD,Lattice}: 1) the pion cloud effects are negligible for large pion masses,
2)  our bare quark core model could be calibrated by the LQCD data for large pion masses, since it uses a electromagnetic quark current inspired by the mechanism of vector meson dominance, and therefore the vector meson mass could be taken as a function of the running pion mass, 3) after that, by taking the limit of the model to the physical pion mass value, the experimental data is well described, at least in the high momentum transfer $Q^2$ region.
Applications of our model include calculations of the form factors for the excitations of the nucleon to the $\Delta$(1232)~\cite{NDeltaD}, N*(1440)~\cite{Roper}, N*(1535)~\cite{S11},
N*(1520), $\Delta$(1600)~\cite{Delta1600}, the baryon octet~\cite{Octet}, the form factors of the $\Delta$~\cite{Delta}, the strangeness sector~\cite{Omega}, Deep Inelastic Scattering~\cite{Nucleon2}, dilepton mass spectrum~\cite{Timelike}, and effects of the nuclear medium~\cite{Medium}.

\newpage

\subsection{Measurements of the  $e^{+}e^{-}\rightarrow\pi^{0}\gamma/\eta\gamma$ cross section at low energies using initial state radiation}
\addtocontents{toc}{\hspace{2cm}{\sl  J.~Pettersson}\par}

\vspace{5mm}

J.~Pettersson

\vspace{5mm}

\noindent
Department of Physics and Astronomy, Uppsala University, Sweden\\

\vspace{5mm}

The anomalous magnetic moment of the muon, $\amu$, is one of the quantities calculable within the Standard Model, known experimentally to the greatest precision \cite{Bennett:2004pv,Jegerlehner:2009ry}. The main uncertainty in theoretical calculations of $\amu$ presently stems from hadronic vacuum polarization and hadronic light-by-light (LbL) scattering contributions \cite{Jegerlehner:2009ry}. The doubly off-shell electromagnetic transition form factors $F_{P\gamma^*\gamma^*}(q^2_1,q^2_2)$ (where P = $\pi^0$,$\eta$) enters the calculations of the LbL amplitude \cite{Bijnens:2007pz, Hayakawa:1997rq}.  Measurements of $\sigma(e^{+}e^{-}\rightarrow\pi^{0}\gamma/\eta\gamma)$ can provide experimental input and constraints for theoretical models from the time-like form factors $F_{P\gamma^*\gamma}(q^2,0)$ \cite{Landsberg:1986fd,Czerwinski:2012ry}, accessible thought the relation, 

\begin{equation}\nonumber
\sigma(e^+e^-\to P\gamma)=4\pi\alpha\Gamma_{\gamma\gamma}\left(\frac{s-m_P^2}{s\, m_P}\right)^3|F_{P\gamma^*\gamma}(s,0)|^2. \label{eq:ee-Pg}
\end{equation}

At present, data available in the low energy region $\sqrt{s} < 1$ GeV is sparse \cite{Akhmetshin:2004gw,Achasov:2003ed,Achasov:2007kw}, and no data currently exists in the region $\sqrt{s} < 0.6$ GeV. Using events with initial state radiation (ISR) offers a possible way to access this energy region. Using ISR data for analysis is desirable since it comes as a by product, thus not requiring any explicit use of beam time. Data from experiments at high luminosity $\ee$ colliders such as KLOE @ DA$\Phi$NE \cite{Bossi:2008aa} and BES III \cite{Li:2011ve} are candidates for analysis. The focus of this work is the issue of the combinatorics problem of ISR and photons in the final state. And subsequently to address the problem of TF extraction in the low energy region approximated by $F_{P\gamma^*\gamma}(s,0) = 1 + as$. 
The simplified "radiator function" Eq.~(\ref{w0}), valid for $4m_e^2 << s$ (full form given in \cite{Druzhinin:2011qd,Actis:2010gg}), describes, to first order, the emission probability of ISR at beam energy fraction $x$ and polar angle $\theta$,

\begin{equation}
w_{0}(\theta,x)=\frac{\alpha}{\pi}\left[\frac{x}{2\tan(\theta)^{2}}+\frac{1-x}{x\sin(\theta)^{2}}\right].
\label{w0}
\end{equation}     
The emission probability calculated from Eq.~(\ref{w0}) can be used to differentiate tagged ISR from final sate photons, to a certain probability, since the distribution of the final state photons is flat in the $P\gamma$ center of mass frame.\\
Other references: \cite{Binner:1999bt,Rodrigo:2001jr,Rodrigo:2001kf, Czyz:2002np}.

\newpage

\subsection{$\gamma\gamma$ interactions at KLOE }
\addtocontents{toc}{\hspace{2cm}{\sl I.~Prado Longhi}\par}

\vspace{5mm}

I.~Prado Longhi on behalf of the KLOE/KLOE-2 Collaboration\footnote{D. Babusci, D. Badoni, I. Balwierz-Pytko, G. Bencivenni, C. Bini, C. Bloise, F. Bossi, P. Branchini, A. Budano, L. Caldeira Balkest\aa hl, G. Capon, F. Ceradini, P. Ciambrone, F. Curciarello, E. Czerwi\'{n}ski, E. Dan\`{e}, V. De Leo, E. De Lucia, G. De Robertis, A. De Santis, A. Di Domenico, C. Di Donato, R. Di Salvo, D. Domenici, O. Erriquez, G. Fanizzi, A. Fantini, G. Felici, S. Fiore, P. Franzini, A. Gajos, P. Gauzzi, G. Giardina, S. Giovannella, F. Gonnella, E. Graziani, F. Happacher, B. H\"{o}istad, L. Iafolla, M. Jacewicz, T. Johansson, K. Kacprzak, A. Kupsc, J. Lee-Franzini, B. Leverington, F. Loddo, S. Loffredo, G. Mandaglio, M. Martemianov, M. Martini, M. Mascolo, R. Messi, S. Miscetti, G. Morello, D. Moricciani, P. Moskal, F. Nguyen, A. Palladino, A. Passeri, V. Patera, I. Prado Longhi, A. Ranieri, C.F. Redmer, P. Santangelo, I. Sarra, M. Schioppa, B. Sciascia, M. Silarski, C. Taccini, L. Tortora, G. Venanzoni, W. Wi\'{s}licki, M. Wolke, J. Zdebik}

\vspace{5mm}

\noindent
Universit\`{a} Roma Tre and INFN Sezione Roma Tre\\

\vspace{5mm}

Hadron production in $\gamma\gamma$ interactions can be studied in $e^+e^-$ colliders using virtual photons radiated by the electron and the positron beams \cite{ref:bro}. Photons are typically radiated at small angles, thus being quasi-real; electron and positron in the final state are themselves scattered by small angles and escape detection. The measurement is said to be performed in no-tag mode.

The KLOE detector at the DA$\Phi$NE $\phi$ factory collected a $\mathcal{L}=$ 242.5 $\mbox{pb}^{-1}$ data sample at $\sqrt{s}=1$ GeV dedicated to $\gamma\gamma$ physics. In the following the studies on $e^+e^-\to e^+e^-\eta$ and $e^+e^-\to e^+e^-\pi^0\pi^0$ processes are presented.
\paragraph{$e^+e^-\to e^+e^-\eta$.}
The $\eta$ meson production in $\gamma\gamma$ interaction has been investigated looking at both the $\eta\to\pi^0\pi^0\pi^0$ and $\eta\to\pi^+\pi^-\pi^0$ decay modes. In both cases the main background process is the annihilation $e^+e^-\to\eta\gamma$ (irreducible when the monochromatic $\gamma$ is lost). Events selections are: no tracks and  6 $\gamma$ from the interaction point (IP) are requested (for $e^+e^-\to e^+e^-\eta$, $\eta\to\pi^0\pi^0\pi^0$); events with 2 $\gamma$ only and  2 tracks with opposite charge coming from the IP are selected for $e^+e^-\to e^+e^-\eta$, $\eta\to\pi^+\pi^-\pi^0$.

In both analyses photons are paired in such a way the two-photons invariant masses reconstruct the $\pi^0$ mass properly, and a kinematic fit is performed asking for the invariant mass of the final state ($6\gamma$ or $\pi^+\pi^- + 2\gamma$) to reconstruct the $\eta$ mass; a cut on a electron/pion likelihood is applied in the $e^+e^-\to e^+e^-\eta$, $\eta\to\pi^+\pi^-\pi^0$ analysis in order to reject $e^+$ and $e^-$ tracks. In both analyses a two-dimensional fit is performed using as fit variables the longitudinal momentum and the squared missing mass (for the $\eta\to\pi^0\pi^0\pi^0$ channel), the transverse momentum and the squared missing mass (for the $\eta\to\pi^+\pi^-\pi^0$ channel). Results on the cross sections obtained within the two analyses are combined in $\sigma(e^+e^-\to e^+e^- \eta, \sqrt{s}=1\,\mbox{GeV}) = (32.7\,\pm 1.3_{\mbox{stat}}\,\pm0.7_{\mbox{syst}})\,\mbox{pb}$. The partial width $\Gamma(\eta\to\gamma\gamma)$ is extracted, $\Gamma(\eta\to\gamma\gamma) = (520\,\pm 20_{\mbox{stat}}\,\pm 13_{\mbox{syst}})\;\mbox{eV}$, which is the most precise measurement to date.  Details of the analysis can be found in \cite{ref:kloe}.

\paragraph{$e^+e^-\to e^+e^-\pi^0\pi^0$.}

$\pi^0\pi^0$ production in $\gamma\gamma$ interaction has been investigated just above the $\pi^0\pi^0$  threshold only by JADE and Crystal Ball experiments \cite{ref:oest,ref:mar}; the process is of great interest as could take place with the production of the $\sigma(500)$ scalar meson as an intermediate resonant state \cite{ref:ach}. The analysis strategy is a combination of a cut-based and a multivariate analysis: the cut-based analysis is meant to reject the main physical background processes ($e^+e^-$ annihilation to $K_SK_L$, $\eta\gamma$, $\omega\pi^0$, $f_0\gamma$,  $a_0\gamma$), while the multivariate approach is useful to handle a copious residual background which is not rejected by the analysis cuts. A longitudinal momentum asymmetry shown by this residual background is a hint for machine background (electro-production of pions on beam gas). Main selections are: 4 $\gamma$ from the IP only not associated with tracks are requested; events with out-of-time clusters are rejected (to reduce the $K_S K_L$ background); photons are paired to reconstruct two $\pi^0$ masses and a cut is applied on a pairing $\chi^2$ variable. 

Residual machine background is studied selecting poorly prompt events, where prompt photons are defined as photons which satisfy the relation $|t-r/c|<5\sigma_t$. These machine background events are selected looking at the tails in the distributions of promptness-sensitive variables (such, for example, the $|t-r/c|$ of the cluster which provides the trigger). Event population selected in such a way shows, beside the $p_L$ asymmetry, high transverse momentum and low $4\gamma$ invariant mass (just above 2$\pi^0$ threshold) distribution. 

The selected machine background sample and MC samples for physical processes (including a generation for $e^+e^-\to e^+e^-\sigma\to e^+e^-\pi^0\pi^0$ events \cite{ref:ngu}) are used to train a multivariate analysis procedure to discriminate the machine background from the signal. The procedure is then applied to the data sample and a likelihood output response is returned for each event, according to whether it has been recognized as a signal event or a machine background event. Cutting on the likelihood response and subtracting from data the remaining contributions from both physical and machine backgrounds one obtains a $\gamma\gamma\to\pi^0\pi^0$  candidate sample of about 2600 events, which shows a distribution in $m_{4\gamma}$ quite in agreement with the MC prediction.

\newpage

\subsection{Study of $\eta$ decays at WASA-at-COSY}
\addtocontents{toc}{\hspace{2cm}{\sl E.~Prencipe}\par}

\vspace{5mm}

E.~Prencipe

\vspace{5mm}

\noindent
Institut f\"ur Kernphysik, Forschugszentrum J\"ulich, Germany\\

\vspace{5mm}

The WASA detector~\cite{wasa::1}, located now in J\"ulich at the COSY facility, started to collect data here since 2007. This experiment allows to perform study of light mesons, such as $\pi^0$, $\eta$ and $\omega$ rare decay processes, in order to perform precise measurements of Branching Ratio (BR), determine Dalitz plot parameters, test symmetry and symmetry breaking, and calculate form factors.

In this experiment a ion beam is scattered against a fix pellet target of hydrogen/deuterium, which allows to analyze the reactions proton-proton ($pp$) or proton-deuteron ($pd$). A high statistics of $\eta$ mesons has been actually collected: in the reaction $pd \rightarrow~^3$He$\eta$, 10$^7$ $\eta$ mesons were tagged at the kinematic energy of the beam in the laboratory system equal to 1.0 GeV, while 5 $\times$ 10$^8$ $\eta$ mesons are produced in the reaction $pp \rightarrow ^3$He$\eta$ at the kinetic energy  of the beam in the laboratory system  equal to 1.4 GeV. This corresponds to the production of 10 $\eta$/s and 100 $\eta$/s, respectively, for the 2 reaction processes. In the $pp$ dataset a higher background level is found compared to the $pd$ dataset; we used to identify the $\eta$ mesons by mean of the missing mass of the event, and the kinematic fit, which rejects main part of the background in our analyses. The advantage in using the $pp$ dataset is indeed that we have a production of $\eta$ mesons almost $\times$10 higher than in the $pd$ dataset: as we plan to measure the BR of very rare processes, high statistics is needed.

An overview of the preliminary results obtained with WASA-at-COSY, using the $pd$ dataset in the sector of the $\eta$ decays, has been given. The processes under study in $pd$ data here presented are:
\begin{itemize}
\item $\eta \rightarrow \pi^+ \pi^- e^+ e^-$ (test of box anomaly contribution)
\item $\eta \rightarrow e^+ e^- e^+ e^-$ (interest in calculating the $\eta$ form factor)
\item $\eta \rightarrow \pi^+ \pi^- \gamma$~\cite{wasa::2} (test the angular distribution and the 2$\pi$ invariant mass distribution; test of box anomaly)
\item $\eta \rightarrow e^+ e^- \gamma$ (interest in the form factor calculation in the time-like region)
\end{itemize} 

A preliminary result on $\eta \rightarrow e^+ e^-$ in $pp$ data was presented as well. This decay is very rare and it is a benchmark channel to study possible effects of new physics beyond the Standard Model (SM) due to the fact that the theoretical predictions quote a BR of about 10$^8$, and the measured value from HADES~\cite{wasa::3} and E799-II~\cite{wasa::4} show an upper limit 2 order of magnitude lower. Only 2 weeks of data runs were used for such a measurement with WASA-at-COSY; we plan to use the full data sample in order to perform the most precise measurement of this upper limit.

Concerning the 4 analyses performed on $pd$ data and here mentioned, all of them were  performed using the full $pd$ dataset and the preliminary results were presented as results of  PhD thesis. BR measurements were introduced, and the analysis strategy was explained. The plan is to publish these preliminary BR results and update those measurements on the $pp$ dataset, where we expect a statistics $\times$10 higher, and eventually combine the two measurements from $pd$ and $pp$ dataset: to run over the full $pp$ dataset will allow us also to cross-check our results on $pd$ data.

\newpage

\subsection{Studies of the neutral kaon regeneration with the KLOE detector}
\addtocontents{toc}{\hspace{2cm}{\sl I.~Balwierz-Pytko}\par}

\vspace{5mm}

I.~Balwierz-Pytko \\ on behalf of the KLOE/KLOE-2 collaboration \footnote{D.~Babusci, D.~Badoni, I.~Balwierz-Pytko, G.~Bencivenni, C.~Bini, C.~Bloise, F.~Bossi, P.~Branchini, A.~Budano, L.~Caldeira~Balkest\aa hl, G.~Capon, F.~Ceradini, P.~Ciambrone, F.~Curciarello, E.~Czerwi\'nski, E.~Dan\`e, V.~De~Leo, E.~De~Lucia, G.~De~Robertis, A.~De~Santis, A.~Di~Domenico, C.~Di~Donato, R.~Di~Salvo, D.~Domenici, O.~Erriquez, G.~Fanizzi, A.~Fantini, G.~Felici, S.~Fiore, P.~Franzini, A.~Gajos, P.~Gauzzi, G.~Giardina, S.~Giovannella, E.~Graziani, F.~Happacher, L.~Heijkenskj\"old, B.~H\"oistad, L.~Iafolla, M.~Jacewicz, T.~Johansson, K.~Kacprzak, A.~Kupsc, J.~Lee-Franzini, B.~Leverington, F.~Loddo, S.~Loffredo, G.~Mandaglio, M.~Martemianov, M.~Martini, M.~Mascolo, R.~Messi, S.~Miscetti, G.~Morello, D.~Moricciani, P.~Moskal, F.~Nguyen, A.~Palladino, A.~Passeri, V.~Patera, I.~Prado~Longhi, A.~Ranieri, C.~F.~Redmer, P.~Santangelo, I.~Sarra, M.~Schioppa, B.~Sciascia, M.~Silarski, C.~Taccini, L.~Tortora, G.~Venanzoni, W.~Wi\'slicki, M.~Wolke, J.~Zdebik}

\vspace{5mm}

\noindent
Institut of Physics, Jagiellonian University, Krakow, Poland\\

\vspace{5mm}

Neutral kaons produced in correlated pairs at a $\phi$-factory offer unique possibilities to perform fundamental tests of CP, CPT invariance, as well as of the basic principles of quantum mechanics. One of the best places to perform such studies is Laboratori Nazionali di Frascati in Italy, where the $K_L$ mesons were produced in the center of the KLOE detector in the collision region of $e^+$ and $e^-$ beams of the DA$\Phi$NE collider that worked at the $\phi$ resonance peak ($\sqrt{s}\approx1020$ MeV/c). 

Parameters that test quantum mechanics at KLOE are, among others, the decoherence and CPT violation parameters \cite{DiDomenico:2007zzb}: $\zeta_{SL}$, $\zeta_{0\bar{0}}$, $\gamma$, $\Re e(\omega)$ and $\Im m(\omega)$. They have been measured at KLOE using interferometric methods by fitting the theoretical function to the distribution of the difference of the decay times ($\Delta t$) between $CP-$violating decays of $K_L\to\pi^+\pi^-$ and $K_S$ decays into two charged pions in the $\phi\to K_L K_S\to \pi^+\pi^- \pi^+ \pi^-$ reaction chain \cite{Ambrosino:2006vr}. The obtained results are \cite{DiDomenico:2009zza}: $\zeta_{SL}=\left(0.3\pm 1.8_{\mathrm{stat}}\pm 0.6_{\mathrm{syst}}\right)\cdot 10^{-2}, \zeta_{0\bar{0}}=\left(1.4\pm 9.5_{\mathrm{stat}}\pm 3.8_{\mathrm{syst}}\right)\cdot 10^{-7}, \gamma=\left(0.7\pm 1.2_{\mathrm{stat}}\pm 0.3_{\mathrm{syst}}\right)\cdot 10^{-21} \mathrm{GeV}, \Re \omega=\left(-1.6^{+3.0}_{-2.1 \ \mathrm{stat}}\pm 0.4_{\mathrm{syst}}\right)\cdot 10^{-4},\Im \omega=\left(-1.7^{+3.3}_{-3.0 \ \mathrm{stat}}\pm 1.2_{\mathrm{syst}}\right)\cdot 10^{-4}$. The uncertainties on these measurements were dominated by the statistical error. At KLOE-2 \cite{AmelinoCamelia:2010me} the statistical error on these parameters can be reduced because of higher luminosity and a new detector close to the interaction point: Cylindrical-GEM Inner Tracker \cite{Archilli:2010xb}.

The main source of these systematic errors is due to the poor knowledge of the incoherent regeneration in the cylindrical beam pipe made of beryllium and located 4.3~cm from the interaction region \cite{Balwierz:2012np}. This is due to the fact that when regeneration occurs, the $K_L$ meson changes into the $K_S$ meson that almost immediately decays into $\pi^+\pi^-$ ($K_L\to K_S^{\mathrm{reg}}\to\pi^+\pi^-$) and which constitutes a background for $K_L K_S\to \pi^+\pi^- \pi^+ \pi^-$ decays.

The data sample composed of $\sim6.6\cdot10^8$ reconstructed neutral kaon pairs was used in this analysis. The $K_L$ mesons were identified based on primary identification of the $K_S$ meson decays into $\pi^+\pi^-$ close to the interaction point. Next, the background normalization was performed by fitting to the data distribution of the regeneration angle (angle between initial $K_L$ direction and the one obtained from two charged particles) the simulated distributions. The background consists mainly of the CP violating events $K_L \to \pi^+\pi^-$ and $K_L$ semileptonic decays. The number of regenerated events $\phi\to K_S K_L\to\pi^+\pi^- K_S^{\mathrm{reg}}\to\pi^+\pi^-\pi^+\pi^-$ will be extracted by fitting to the data the simulated distributions of the vertex position of the $K_L$ meson decays into $\pi^+\pi^-$. In these distributions events corresponding to $K_L\to K_S$ regeneration manifest themselves as peaks at the positions of regenerators. The fit of the simulated distributions of signal to the measured data will be performed and finally the regeneration cross-sections will be determined based on the extracted number of regenerated events and number of the $K_L$ mesons passing through the regenerator.\\ 

This work was supported in part by the European Commission under the 7th Framework Programme through the `Research Infrastructures' action of the `Capacities' Programme, Call: FP7-INFRASTRUCTURES-2008-1, Grant Agreement No. 283286; by the Polish National Science Centre through the Grants No. 0469/B/H03/2009/37, 0309/B/H03/2011/40, DEC-2011/03/N/ST2/02641, 2011/01/D/ST2/00748 and by the Foundation for Polish Science through the MPD programme and the project HOMING PLUS BIS/2011-4/3.

\newpage

\subsection{Production and Dalitz decays of baryon resonances in $p+p$ interactions at $E_{kin}=1.25$ and $3.5~GeV$ beam energy with HADES }
\addtocontents{toc}{\hspace{2cm}{\sl W.~Przygoda}\par}

\vspace{5mm}

W.~Przygoda for the HADES Collaboration

\vspace{5mm}

\noindent
Smoluchowski Institute of Physics, Jagiellonian University, 30-059 Cracow, Poland\\

\vspace{5mm}

HADES is a versatile magnetic spectrometer installed at GSI Darmstadt on SIS18 \cite{L1}. Thanks to its high acceptance, powerful particle ($p/K/\pi/e$) identification and very good mass resolution ($2-3\%$ for dielectrons in the light vector meson mass range) it allows to study both hadron and rare dilepton production in N+N, p+A, A+A collisions an a few AGeV beam energy range.

Nucleon-nucleon reactions play in this context a special role providing a reference for $p+A$ and $A+A$ collisions. Exclusive channels $ppe^{+}e^{-}$ and $ppe^{+}e^{-}\gamma$ as well as $pp\pi^{0}$ and $pn\pi^{+}$ were studied in proton+proton collisions providing stringent constraints on various theoretical models. A main emphasis was placed on understanding of vector mesons ($\omega/\rho$) and baryon resonance production and their dielectron decays.

In $p+p~@~1.25~GeV$ two major sources, $\pi^{0}$ and $\Delta$, are expected to play a dominant role in $e^{+}e^{-}$ production. In particular, for the first time the $\Delta$ resonance was reconstructed exclusively via the $pe^{+}e^{-}$ decay channel. This is an important issue since the branching ratio of electromagnetic $\Delta$ Dalitz decay had not been measured before. The resonance cross section was determined from hadronic channels (one pion production) by means of the PWA (Partial Wave Analysis) \cite{L2}. 

At higher kinetic energy ($3.5~GeV$) dielectron pair production is determined by Dalitz decays of neutral mesons ($\pi^{0}$, $\eta$, $\omega$) and baryonic resonances ($N^{*}$, $\Delta$) and, in the high mass region, by two body decays of $\omega/\rho$. At this beam energy the $e^{+}e^{-}$ invariant mass distribution appears to be sensitive to the structure of the baryon resonance electromagnetic transition form-factors in the time-like region \cite{L3, L4}. The data were compared to various theoretical predictions \cite{L5, L6, L7, L8}. Model deficiencies were unraveled either below the vector mesons pole (missing yield in the $e^{+}e^{-}$ invariant mass spectrum) or at high $pe^{+}e^{-}$ invariant mass with only $\Delta$ form-factor (yield saturated without implementation of e-TFF for higher resonances).

\newpage

\subsection{ $\gamma \gamma$ Physics at BES-III }
\addtocontents{toc}{\hspace{2cm}{\sl C.F.~Redmer}\par}

\vspace{5mm}

C.F.~Redmer

\vspace{5mm}

\noindent
Institut f\"ur Kernphysik, Johannes Gutenberg-Universt\"at Mainz, Germany\\

\vspace{5mm}

The $\gamma \gamma$ physics program~\cite{BES3P} at the BES-III experiment aims at the measurement of the 
electromagnetic transition form factors (TFF) of pseudoscalar mesons in the space-like region. Precise knowledge of the 
TFF is of vital importance to provide experimental input for the calculation of the hadronic light-by-light scattering, 
one of the hadronic contributions to the muon anomaly $a_\mu$, which completely limit the theoretical prediction of 
$a_\mu$.

The BES~III experiment~\cite{BES3H} is located at the BEPC~II $e^+ e^-$ collider, operated at the IHEP in Beijing 
(China). Data can be collected in a center-of-mass energy range from 2.0~GeV to 4.6~GeV. For the determination of the 
TFF, data taken at the $\psi(3770)$ peak are being analyzed. Currently, $2.9\textrm{~fb}^{-1}$ have been collected at 
this energy and it is planned to extend the data set. Additional $2.5\textrm{~fb}^{-1}$ collected at $\sqrt{s} = 4230, 
4260$ and $4360$~GeV are analyzed to benefit from higher cross sections and access the to larger ranges of $Q^2$.

Feasibility studies~\cite{feas}, performed with the Ekhara~\cite{ekhara,2octet} event generator, show that the TFF for 
$\pi^0, \eta$ and $\eta^\prime$ mesons can be extracted at momentum transfers in the range of \mbox{$0.3 \leq Q^2 
[GeV^2] \leq 10$}. Assuming a total integrated luminosity of $10\textrm{~fb}^{-1}$, the statistical accuracy of a TFF 
measurement at BES-III was found to be unprecedented for momentum transfers of $Q^2 \leq 4$~GeV$^2$, a region of special
relevance for the hadronic light-by-light scattering~\cite{theo1,theo2}. At higher momentum transfers the precision is 
compatible with the CLEO~\cite{CLEO} result, allowing for cross checks with previous measurements of TFF's~\cite{OTHER}.

At the moment, data analysis is based on a single-tag technique, where only the produced meson and one of the two 
scattered leptons are reconstructed from detector information. The second lepton is reconstructed from four-momentum 
conservation and required to have a small scattering angle, so that the momentum transfer is small and one of the 
exchanged photons is quasi-real. The ongoing analyses tag the produced pseudoscalar meson in the decay channels 
$\pi^0\rightarrow\gamma\gamma$, $\eta\rightarrow\gamma\gamma$, $\eta\rightarrow\pi^+\pi^-\pi^0$, and 
$\eta^\prime\rightarrow\pi^+\pi^-\eta$. Major sources of background are QED processes such as virtual Compton 
scattering, misidentified hadronic final states, external photon conversion, and on-peak background from two-photon 
processes such as the production of different mesons or and initial state radiation in the signal channel. Conditions 
are being devised to suppress the identified background sources. First results are expected soon.

Future prospects of the $\gamma \gamma$ physics program at BES-III comprise the investigation of multi-meson final 
states to study scalar and tensor meson production, measurement of polarization observables, and double tagged 
measurements of $\gamma \gamma$ processes.

\newpage

\subsection{Study of $\phi \to (\eta/\pi^0) e^+e^-$ at KLOE}
\addtocontents{toc}{\hspace{2cm}{\sl I.~Sarra}\par}

\vspace{5mm}

I.~Sarra  on behalf of the KLOE-2 Collaboration\footnote{D. Babusci, D. Badoni, I. Balwierz-Pytko, G. Bencivenni, C. Bini, C. Bloise, F. Bossi, P. Branchini, A. Budano, L. Caldeira Balkestahl, G. Capon, F. Ceradini, P. Ciambrone, F. Curciarello, E. Czerwinski, E. Dane', V. De Leo, E. De Lucia, G. De Robertis, A. De Santis, A. Di Domenico, C. Di Donato, R. Di Salvo, D. Domenici, O. Erriquez, G. Fanizzi, A. Fantini, G. Felici, S. Fiore, P. Franzini, P. Gauzzi, G. Giardina, S. Giovannella, F. Gonnella, E. Graziani, F. Happacher, B. Hoistad, L. Iafolla, M. Jacewicz, T. Johansson, K. Kacprzak, A. Kupsc, J. Lee-Franzini, B. Leverington, F. Loddo, S. Loffredo, G. Mandaglio, M. Martemianov, M. Martini, M. Mascolo, R. Messi, S. Miscetti, G. Morello, D. Moricciani, P. Moskal, F. Nguyen, A. Palladino, A. Passeri, V. Patera, I. Prado Longhi, A. Ranieri, C.F. Redmer, P. Santangelo, I. Sarra, M. Schioppa, B. Sciascia, M. Silarski, C. Taccini, L. Tortora, G. Venanzoni, W. Wislicki, M. Wolke, J. Zdebik}

\vspace{5mm}

\noindent
INFN - Laboratori Nazionali di Frascati\\

\vspace{5mm}

Conversion decays of vector and pseudoscalar mesons are closely related to corresponding radiative decays ($\rm{V \to P\gamma}$). In conversion decays a radiated photon is virtual and its squared mass $q^2$ (equal to invariant mass of lepton pair $\ell^+\ell^-$) is not equal to zero, therefore studying the lepton-pair invariant-mass spectrum it is possible to learn more about mesons structure and underlying quark dynamics and to measure a so-called transition form factor. The only existing data on $\rm{\phi \to \eta} e^+ e^-$ come from the SND experiment \cite{SND}, which has measured the M$_{ee}$ invariant-mass distribution on the basis of 213 events. At KLOE~\cite{KLOE}, a detailed study of this decay has been performed using both $\rm{\eta \to \pi^+\pi^-\pi^0}$ and $\rm{\eta \to \pi^0\pi^0\pi^0}$ final states. The analyzed sample is based on 1.5 fb$^{-1}$ and 1.7 fb$^{-1}$ data, respectively, collected on $\phi$ peak during the 2004-2005 KLOE data taking.  Simple analysis cuts provide clean signal events (about 14000 and 30000), with a residual background contamination of 2-3\%. 
The fit to the $e^+e^-$ invariant-mass distributions has been performed using the decay parametrization \cite{Landsberg}  folded with the analysis efficiencies and the smearing matrix. The preliminary result for the measurement of the transition form factor, with the $\rm{\eta \to \pi^0\pi^0\pi^0}$ final state, is b$_{\phi\eta}$ = (1.17 $\pm$ 0.11 $\rm{^{+ 0.09}_{-0.08}}$) GeV$^{-2}$. This result is in agreement with the VMD prediction within 1$\sigma$. With the same 29625 candidates we have measured a BR($\rm{\phi \to \eta} e^+ e^-$) = (1.131 $\pm$ 0.031 $\pm$ 0.007 $\rm{^{+ 0.011}_{- 0.006}}$) $\times 10^{-4}$, in agreement with the VMD prediction within 1 $\sigma$ and with the previous measurements of the SND \cite{SND} and CMD-2 \cite{CMD-2} experiments. 

The $e^+e^-$ invariant-mass distributions have been used to set a combined upper limit on the process $\rm{\phi \to \eta} U$, with U $\to e^+e^-$, where U is a vector gauge boson, mediating dark forces. The resulting exclusion plot, obtained by combining both samples ($\rm{\eta \to \pi^+\pi^-\pi^0}$ and $\rm{\eta \to \pi^0\pi^0\pi^0}$), covers the mass range 5 $<$ M$_U$ $<$ 470 MeV and sets an upper limit at 90\% C.L. on the ratio between the U boson coupling constant and the fine structure constant, $\alpha'/\alpha$, of $\le$ 1.5 $\times$ 10$^{-5}$ for 30 $<$ M$_U$ $<$ 420 MeV and $\le$ 5.0 $\times$ 10$^{-6}$ for 60 $<$ M$_U$ $<$ 190 MeV \cite{Uboson}. This result assumes the Vector Meson Dominance expectations for the $\phi \eta \gamma^*$ transition form factor. The dependence of this limit on the transition form factor has also been studied.\\
The $\rm{\phi \to \pi^0} e^+ e^-$ rare decay is being also studied at KLOE using a sample of 1.7 fb$^{-1}$ data collected. The Branching Ratio will be measured, with a consistent reduction of the actual total uncertainty of the world average (25\% \cite{PDG}). The $\pi^0$ Transition Form Factor will also be measured for the first
time in the time-like region, with the momentum transfer given by the $e^+e^-$ invariant mass.

\newpage

\subsection{The WASA-at-COSY pilot study on meson decays in $pp$ reactions above the $\omega$ threshold}
\addtocontents{toc}{\hspace{2cm}{\sl S.~Sawant}\par}

\vspace{5mm}

S.~Sawant, \textit{for the WASA-at-COSY Collaboration}
\let\thefootnote\relax\footnote{ {\Large ${}^{*}$}Supported by JCHP-FFE of Forschungszentrum J\"ulich.}
\vspace{5mm}

\noindent
Indian Institute of Technology Bombay, India\\
Institut f\"ur Kernphysik, Forschungszentrum J\"ulich, Germany
\vspace{5mm}

A pilot study on meson decays in proton-proton collisions above the $\omega$ threshold  (at T$_{p}$=2.06 \textrm{GeV} and 
2.54 \textrm{GeV})  
has been performed with the WASA-at-COSY facility. The challenge of using $pp$ reactions at these energies lies in 
reconstructing fast scattered  protons and the considerable multi-pion background. We study the feasibility of addressing 
rare decays in these reactions, especially conversion decays leading to the determination of transition form factors. 

First we focus on the $\pi^{+}\pi^{-}\pi^{0}$ final state. In particular we study the decay mechanism of the $\omega \to \pi^{+}\pi^{-}\pi^{0}$ decay 
from the density distribution in the Dalitz plot \cite{Leupold, Kubis}.

In the $pp \to pp$ $\pi^{+}\pi^{-}\pi^{0}$ reaction, protons are detected in the forward part 
of the WASA detector and $\pi^{+},$ $\pi^{-}$ 
and $\gamma$ in the central part \cite{WASA_setup}. 
The $\pi^{0}$ is reconstructed from a $\gamma$ pair satisfying the $|m_{\gamma\gamma}-m_{\pi^{0}}| < 35$ 
\textrm{MeV/c$^{2}$} condition.  
A kinematic fit routine is implemented to suppress multi-pion background other than direct 3$\pi$ background and to eliminate wrongly 
reconstructed particles. 
The missing mass of beam, target, and two protons shows clear signals for $\eta$ and $\omega$.

The dynamics of the three body decay is described by two variables $X = \sqrt{3} (T_{\pi^{+}} - T_{\pi^{-}}) / Q_{\omega}$ 
and $Y = 3$ $T_{\pi^{0}} / Q_{\omega} - 1$, where the $T_{\pi}$'s are the pion kinetic energies, 
$Q_{\omega}$ = $T_{\pi^{+}} + T_{\pi^{-}} + T_{\pi^{0}}$. 
For each bin of $X$ and $Y$, the missing mass of beam, target and two protons is compared with the 
corresponding spectrum of a Monte Carlo simulation. For this simulation, the 
$pp \to pp$ $\omega (\to \pi^{+}\pi^{-}\pi^{0})$, 
$pp \to pp$ $\eta (\to \pi^{+}\pi^{-}\pi^{0})$ and $pp \to pp$ $\pi^{+}\pi^{-}\pi^{0}$ reactions are considered. 
From the comparison of the missing mass spectra we extract the number of 
$\omega \to 3\pi$ events for the corresponding bin of $X$ and $Y$. 
 This number of events is plotted in the
corresponding bin of $X$ and $Y$ in the Dalitz plot of $\omega \to 3\pi$.

The analysis is being performed on $pp$ collisions at T$_{p}$=2.06 \textrm{GeV} 
and 2.54 \textrm{GeV}.  
The idea is to combine the data sets to improve statistics.

\newpage

\subsection{Current status and results of the experiments at VEPP-2000}
\addtocontents{toc}{\hspace{2cm}{\sl B.~Shwartz}\par}

\vspace{5mm}

B.~Shwartz

\vspace{5mm}

\noindent
Budker Institute of Nuclear Physics of SB RAS, \\
Novosibirsk State University, \\
630090 Novosibirsk, Russia

\vspace{5mm}

Detail study of $e^+e^-$
annihilation into hadrons at low energies provides important information
about interactions of light quarks and spectroscopy of their bound states. 
One of the important goals of this studies is a measurement of
the total cross section of the hadrons production, characterized by
the ratio R, with high precision. These data are necessary 
for the calculation of the contribution of the hadronic vacuum 
polarization to the muon anomalous magnetic 
moment~\cite{Bouchiat:1961,Gourdin:1969}.
At present the accuracy of the theoretical calculations of the muon $(g-2)$ via 
the Standard Model is dominated  by the precision of the hadronic contribution
while the difference of theoretical and experimental values exceeds three
standard deviations~\cite{eidelman:2009,teubner:2010} .

Recently the VEPP-2000 collider has started operation at BINP (Novosibirsk). 
This machine~\cite{vepp2000:2009}, exploiting the idea of the round beams,
has the energy range extended up to 2 GeV and the project
luminosity up to  $10^{32}$cm$^{-2}$s$^{-1}$.  
Such a high luminosity together with improved 
characteristics of  two modified detectors, CMD-3 and SND~\cite{cmd3-snd:2010}, 
should provide the measurements of the hadronic cross sections with much better
precision than those existing now. 
By now each of the two detectors collected about 60~pb$^{-1}$ of integrated 
luminosity in the CMS energy range from 320~MeV to 2~GeV.
First preliminary results were obtained for many hadronic final 
states~\cite{snd-pre1,snd-pre2,cmd-pre}:
$e^+e^- \to p\overline{p}$, $n\overline{n}$, $\pi^+\pi^-\pi^0$, 
$\pi^+\pi^-\pi^+\pi^-$,
$\pi^+\pi^-\pi^0\pi^0$, $\pi^+\pi^-\pi^+\pi^-\pi^0$, $3(\pi^+\pi^-)$, 
$2(\pi^+\pi^-\pi^0)$, $\pi^+\pi^-4\pi^0$, $\eta \pi^+\pi^-$, $K_S K_L$,
$\omega \pi^0 \to \pi^0\pi^0\gamma $.
QED processes $e^+e^- \to e^+e^-$ and $\gamma\gamma$ were used 
for the luminosity measurement.

Recently CMD-3 published first experimental result concerning the 
interesting process  $e^+e^- \to 6\pi $ \cite{cmd3:2013}. 
Long time ago the DM2 collaboration found an evidence of the sharp deep in 
the energy range near $p\overline{p}$ production threshold \cite{dm2-6pi}. 
Some time ago this was confirmed by the BaBar \cite{bbr-6pi}. 
The CMD-3 results on this process are in a good agreement with \cite{bbr-6pi}. 
It worth noting that the energy resolution in the conventional
$e^+e^-$ experiment is much better than that in the ISR approach.
Thus the sharp deep will be studied in more detail at VEPP-2000.  

The current run will continue until the middle of July. Then a long shutdown
(1-1.5 years) is scheduled to increase the booster energy to 1 GeV and 
to commission the new injection complex to reach the project luminosity.

\newpage

\subsection{Search for eta-mesic helium via the $dd\rightarrow(^{4}\hspace{-0.03cm}\mbox{He}$-$\eta)_{bs}\rightarrow$ $^{3}\hspace{-0.03cm}\mbox{He}\, n \,\pi{}^{0}$ reaction with the WASA-at-COSY facility}
\addtocontents{toc}{\hspace{2cm}{\sl M.~Skurzok}\par}

\vspace{5mm}

M.~Skurzok, W.~Krzemien, P.~Moskal

\vspace{5mm}

\noindent
Faculty of Physics, Astronomy and Applied Computer Science, Jagiellonian~University~in~Cracow, Poland\\

\vspace{5mm}

The existence of $\eta$-mesic nuclei in which the $\eta$ meson is bound to a nucleus by means of the strong interaction was postulated already in 1986~\cite{HaiderLiu1} but it has not been confirmed experimentally yet. The discovery of this new kind of exotic nuclear matter would be very important as it might allow for
a better understanding of the $\eta$ meson structure and its interaction with nucleons~\cite{InoueOset,BassThomas,BassTom}. The search for $\eta$-mesic helium ($^{4}\hspace{-0.03cm}\mbox{He}$-$\eta$) has been carried out with high statistics and high acceptance with the WASA detector, installed at the cooler synchrotron COSY of the Forschungszentrum J\"ulich.

\indent The search is conducted via the measurement of the excitation function for selected decay channels of the $^{4}\hspace{-0.03cm}\mbox{He}$-$\eta$ system. The deuteron beam - deuteron target collisions lead to the creation of the $^{4}\hspace{-0.03cm}\mbox{He}$ nucleus bound with the $\eta$ meson via strong interaction. The $\eta$ meson can be absorbed by one of the nucleons inside helium and may propagate in the nucleus via consecutive excitation of nucleons to the $N^{\star}$~(1525) state until the resonance decays into the pion-nucleon (n,$\pi^{0}$ or p,$\pi^{-}$) pair outgoing from the nucleus~\cite{KrzeMosSmy}. The relative nucleon-pion angle is equal to 180$^{\circ}$ in the $N^{\star}$ reference frame and it is smeared by about 30$^{\circ}$ in the center-of-mass frame due to the Fermi motion of the nucleons inside the~helium~nucleus.\\
\indent In June 2008 a search for the $^{4}\hspace{-0.03cm}\mbox{He}$-$\eta$ bound state was performed by measuring the excitation function of the $dd\rightarrow$ $^{3}\hspace{-0.03cm}\mbox{He} p \pi{}^{-}$ reaction near the $\eta$ production threshold.~During the experiment the deuteron beam momentum was varied continuously from 2.185~GeV/c to 2.400~GeV/c corresponding to an excess energy variation from -51.4 MeV to 22 MeV. 
The excitation function was determined after applying cuts on the $p$ and $\pi^{-}$ kinetic energy distribution and the $p - \pi^{-}$ opening angle in the CM system~\cite{Krzemien_PhD}. The relative normalization of the $dd\rightarrow$ $^{3}\hspace{-0.03cm}\mbox{He} p \pi{}^{-}$ excitation function was based on quasi-elastic proton-proton scattering.
In the excitation function there is no structure which could be interpreted as a resonance originating from decay of the $\eta$-mesic $^{4}\hspace{-0.03cm}\mbox{He}$~\cite{Krzemien_MESON2012,Wojtek_publ}.

During the second experiment in November 2010 two possible channels of the $\eta$-mesic helium decay were measured:  $dd\rightarrow(^{4}\hspace{-0.03cm}\mbox{He}$-$\eta)_{bs}\rightarrow$ $^{3}\hspace{-0.03cm}\mbox{He} p \pi{}^{-}$ and  $dd\rightarrow(^{4}\hspace{-0.03cm}\mbox{He}$-$\eta)_{bs}\rightarrow$ $^{3}\hspace{-0.03cm}\mbox{He} n \pi{}^{0} \rightarrow$ $^{3}\hspace{-0.03cm}\mbox{He} n \gamma \gamma$~\cite{SkuMosKrze}. The measurement was performed with the beam momentum ramping from 2.127 GeV/c to 2.422 GeV/c, corresponding to the range of the excess energy \mbox{$Q\in$(-70,30)~MeV}. Until now the \mbox{$dd\rightarrow(^{4}\hspace{-0.03cm}\mbox{He}$-$\eta)_{bound}\rightarrow$ $^{3}\hspace{-0.03cm}\mbox{He} n \pi{}^{0} \rightarrow$ $^{3}\hspace{-0.03cm}\mbox{He} n \gamma \gamma$} reaction is analysed. The $^{3}\hspace{-0.03cm}\mbox{He}$ is identified in the Forward Detector based on the $\Delta E$-$E$ method. The neutral pion $\pi^{0}$ is reconstructed in the Central Detector from the invariant mass of two gamma quanta originating from its decay while the neutron four-momentum is calculated using the missing mass technique.
The excitation function for the $dd\rightarrow$ $^{3}\hspace{-0.03cm}\mbox{He} n \pi{}^{0} \rightarrow$ $^{3}\hspace{-0.03cm}\mbox{He} n \gamma \gamma$ reaction is studyied for the "signal-rich" region corresponding to $^{3}\hspace{-0.03cm}\mbox{He}$ momenta in the CM system below 0.3 GeV/c and the "signal-poor" region for the $^{3}\hspace{-0.03cm}\mbox{He}$ CM momenta above 0.3 GeV/c. The contributions from different background reactions is under investigation. 

We acknowledge support by the Foundation for Polish Science - MPD program, co-financed by the European Union within the European Regional Development Fund, by the Polish National Science Center through grant No. 2011/01/B/ST2/00431 and by the FFE grants of the Forschungszentrum J\"ulich.

\newpage

\subsection{ Experimental search for $\eta '$ mesic nuclei with  ($p$,$d$) reaction at GSI and FAIR }
\addtocontents{toc}{\hspace{2cm}{\sl Y.K.~Tanaka}\par}

\vspace{5mm}

Y.K.~Tanaka for the $\eta$-PRiME collaboration

\vspace{5mm}

\noindent
Department of Physics, University of Tokyo, Japan\\

\vspace{5mm}

The large mass of  the $\eta '$ meson is theoretically understood as the $U_A$(1) axial anomaly effect. 
Since the strength of this effect is considered to be related to the chiral condensate $\langle \bar{q} q\rangle$ \cite{Ref1,Ref1_2},  
the $\eta '$ mass may be reduced in the nuclear medium, where the chiral symmetry is partially restored. 
The Nambu-Jona-Lasinio model calculation shows a mass reduction of about 150 MeV/$c^2$ at nuclear saturation density \cite{Ref2,Ref2_2}.
Such a large mass reduction suggests a strong attractive potential between $\eta '$ and a nucleus and the existence of 
$\eta '$ meson-nucleus bound states ($\eta '$ mesic nuclei) \cite{Ref2_2,Ref3,Ref4}. 

As for the decay width, the CBELSA/TAPS collaboration reported that the absorption width of the $\eta '$ meson at the nuclear saturation density 
is around 15 - 25 MeV at the average $\eta '$ momentum of 1050~MeV/c \cite{Ref6}. 
This implies that the decay width of $\eta '$ mesic nuclei could be small as well, 
and they may be observed as narrow peaks experimentally.

We plan an experimental search for $\eta '$ mesic nuclei at GSI and FAIR. 
We will perform a missing-mass spectroscopy of {the} $^{12}$C($p$, $d$)$\eta ' \otimes ^{11}$C reaction near the 
$\eta '$ emission threshold. 
A 2.5~GeV proton beam from SIS-18 at GSI or SIS-100 at FAIR will be injected to a $^{12}$C target. 
The missing mass of the reaction will be analyzed by the momentum measurement of the ejectile deuterons using FRS or Super-FRS.

\begin{figure}[t]
\centering \includegraphics[scale=0.25]{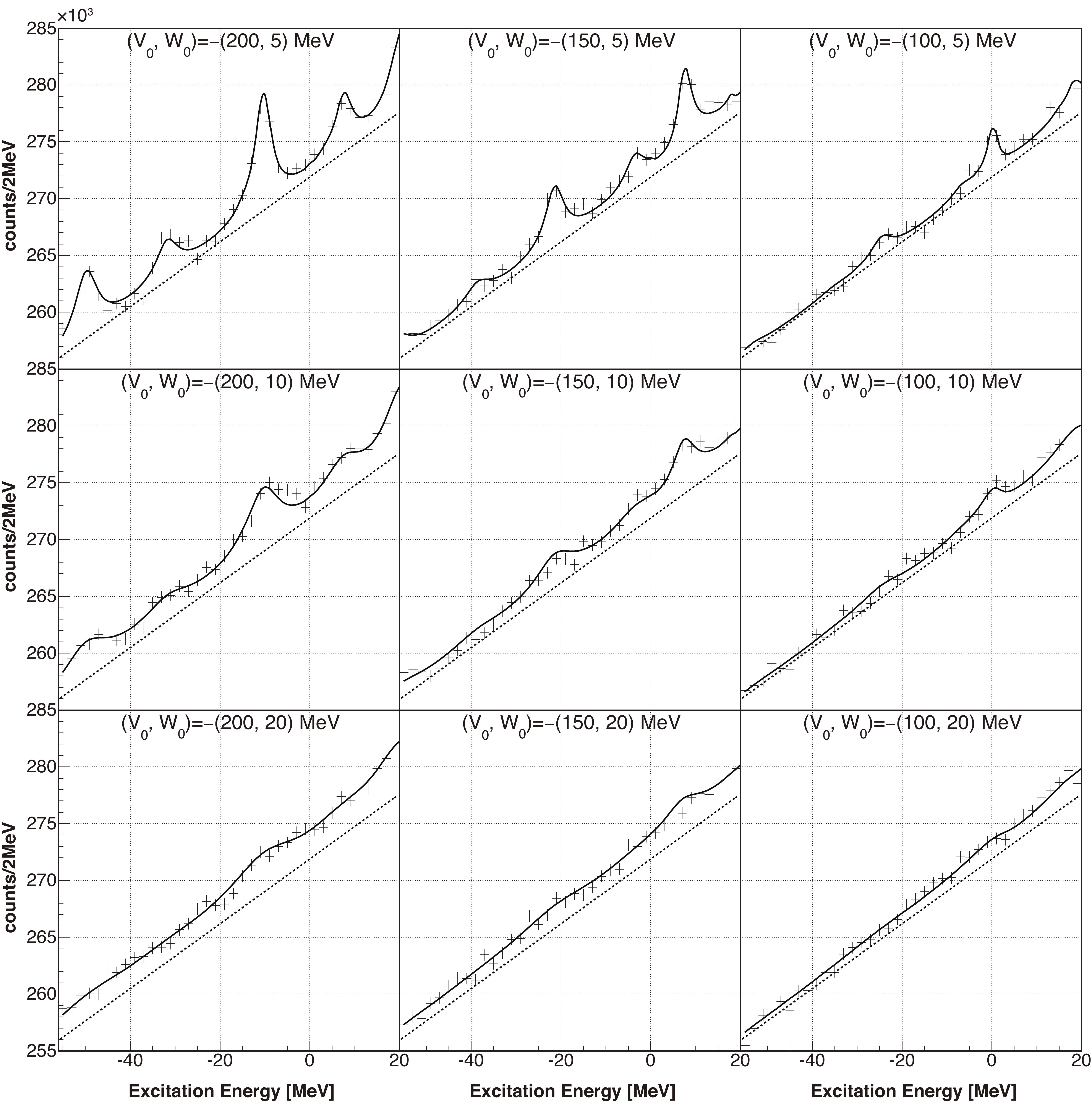}
  \caption{Simulated inclusive spectra expected in a 4.5 day data acquisition 
  with $3.24\times 10^{14}$ protons on a 4~g/cm${}^2$-thick $^{12}$C target.
  This figure is taken from K. Itahashi {\it{et al.}}~\cite{Ref8}. 
$V_0$ is the real part and $W_0$ is the imaginary part of the optical potential at normal nuclear density.
The in-medium mass reduction and width correspond to $|V_0|$ and $2|W_0|$, respectively. 
The amount of the background processes is shown by the dashed line.}
\label{fig:t1}
\end{figure}

At GSI, as a first step, we will perform an inclusive measurement, where only the ejectile deuterons are measured. 
In this measurement, the signal-to-noise ratio is expected to be very small due to background processes dominated 
by quasi-free multi-pion production ($p+N \rightarrow d + \pi$'s).
Thus, a high-statistics measurement is essential using an intense proton beam ($\sim 10^{10}$~/spill) and a thick production target ($\sim4$\,g/cm$^{2}$).
Figure~\ref{fig:t1} shows a simulated result of the inclusive spectra for several cases of mass reductions and decay widths assuming a 4.5 day data acquisition. 
If the mass reduction in 
the medium is sufficiently large and the decay width is small around 20~MeV,
peak structures may be observed even in the inclusive spectra \cite{Ref7,Ref8}. 

Moreover, we are planning a semi-exclusive measurement of the ($p$,$dp$) reaction at FAIR. 
We will measure protons from the decay of the $\eta '$ mesic nuclei, in addition to the missing-mass measurement with Super-FRS. 
Tagging high-energy protons from the absorption of $\eta '$ meson at a backward angle
could drastically improve the signal-to-noise ratio. Then, $\eta '$ mesic nuclei may be observed even in case of a smaller mass reduction. 

\vspace{5mm}
\small{This work is partly supported by Grants-in-Aid for Scientific Research (Nos. 24105705 and 25707018)
and a Grant-in-Aid for JSPS Fellows 
(No. 13J08155)
in Japan.}

\newpage

\subsection{ Electromagnetic transition form factors between vector and pseudoscalar mesons}
\addtocontents{toc}{\hspace{2cm}{\sl  C.~Terschl\"usen}\par}

\vspace{5mm}

 C.~Terschl\"usen$^1$, S.~Leupold$^1$, M.F.M.~Lutz$^2$

\vspace{5mm}

\noindent
$^1$ Department of Physics and Astronomy, Uppsala University, Sweden\\
$^2$ GSI, Darmstadt, Germany\\ 

\vspace{5mm}

Recently, a chiral Lagrangian has been suggested which is inspired by the framework of effective field theories and contains the low-mass pseudoscalar and vector mesons as dynamical degrees of freedom \cite{LL, TLL} (see contribution of M.F.M.~Lutz in this mini-proceedings). While the pseudoscalar mesons are special because of their character as Goldstone bosons of chiral symmetry breaking, vector mesons are important for any process which couples electromagnetism to hadrons (vector-meson dominance). 

Some interactions between hadrons and electromagnetism can be successfully described by the standard vector-meson dominance model, however, it fails to describe the $\omega$-$\pi^0$ transition form factor ($\omega$ coupled to a neutral pion and a virtual photon) \cite{NA60}. Using our leading-order chiral Lagrangian, this transition form factor can be described much better \cite{TLL, TL}. Furthermore, the partial decay widths for the decays into dimuon and dielectron agree very well with the experimental data. The same Lagrangian has been used in \cite{TLL} to predict several other transition form factors including, e.g., the transition of eta-prime to omega and the corresponding decay width.

\newpage

\subsection{Status Report Institute for Nuclear Physics in Mainz}
\addtocontents{toc}{\hspace{2cm}{\sl M.~Unverzagt}\par}

\vspace{5mm}

M.~Unverzagt

\vspace{5mm}

\noindent
Institut f\"ur Kernphysik, Johannes Gutenberg-Universt\"at Mainz, Germany
\vspace{5mm}

In the last few years the Institute for Nuclear Physics in Mainz, Germany, has initiated two major structural measures to strengthen its position in the research field of hadron physics and to guarantee long term funding. These measures are the successful applications for a Sonderforschungsbereich (CRC1044)~\cite{crc} funded by the German Science Foundation (DFG), and the PRISMA~\cite{prisma} cluster in the excellence initiative in Germany.

The CRC 1044 \textquotedblleft The Low-Energy Frontier of the Standard Model: From Quarks and Gluons to Hadrons and Nuclei\textquotedblright comprises a 12 years physics programme including precision test of the Standard Model and physics beyond, time-like and space-like observables in hadron structure, structure and dynamics of light mesons, and nuclear few-body systems and baryon-baryon interactions. Many aspects will be studied in complementary measurements at MAMI using electron scattering and photoproduction, and with the BES-III detector at the $e^+e^-$-collider BEPC-II in Beijing, China. For instance, electromagnetic transition form factors of light mesons will be studied in the space-like momentum transfer region with BES-III and in the time-like region in decays with the Crystal Ball at MAMI. Furthermore, many decays of light mesons and their production mechanisms will be investigated.

Three institutes of the University in Mainz have formed the excellence cluster \textquotedblleft Precision Physics, Fundamental Interactions and Structure of Matter\textquotedblright (PRISMA). It is based on the existing local research infrastructure (MAMI accelerator, TRIGA research reactor, WILSON high-performance PC cluster) but also includes new developments, such as the construction of the high-intensity electon accelerator MESA at the Institute for Nuclear Physics, and the foundation of the Mainz Institute for Theoretical Physics (MITP)~\cite{mitp}. Key experiments with MESA are a high-precision measurement of the weak mixing angle in the low energy region, and the search for the Dark Photon.

Furthermore, the Helmholtz-Institute Mainz (HIM)~\cite{him} has been established in 2009 as a new research institution of the German Helmholtz-Gemeinschaft on the campus of the University in Mainz. Its aim is to strengthen the co-operation between several institutes of the university and the GSI in Darmstadt, Germany, around the FAIR project. One of the main research areas is hadron physics, which mainly will be carried out with the PANDA-detector currently under construction.

\newpage

\subsection{Meson Physics with the Crystal Ball at MAMI  }
\addtocontents{toc}{\hspace{2cm}{\sl  M.~Unverzagt}\par}

\vspace{5mm}

 M.~Unverzagt

\vspace{5mm}

\noindent
Institut f\"ur Kernphysik, Johannes Gutenberg Universt\"at Mainz, Germany\\

\vspace{5mm}

The A2 collaboration at the Institute for Nuclear Physics in Mainz, Germany, studies photoproduction mechanisms and decays of light mesons. The photon beam is produced through Bremsstrahlung of electrons in a thin radiator. The electrons are delivered by the MAMI accelerator~\cite{Jan06,Kai08} with a maximum energy of $E_e = 1604$~MeV. Post-radiating electrons are momentum-analised in the Glasgow-tagging spectrometer~\cite{McG08}. In 2012, a new tagging spectrometer was successfully installed to cover the accessible $\eta'$-photoproduction range. The photon-beam hits a target (e.g. liquid hydrogen or polarised Butanol) in the centre of the Crystal Ball-spectrometer~\cite{Sta01}. Forward angles are covered by a spectrometer-wall consisting of TAPS-crystals~\cite{Nov91}. Here, some of the latest preliminary results will be briefly described.

The first $\eta'$ measurements of the A2 collaboration were used to determine preliminary total and differential photoproduction cross-sections near threshold from $\eta' \to \eta \pi^0 \pi^0$ decays where the $\eta$ decayed into two photons. These results are currently the most accurate existing determinations. First studies of the $\eta' \to \eta \pi^0 \pi^0$ Dalitz-plot and the $\eta' \to e^+ e^- \gamma$ Dalitz-decay will also be performed. After an upgrade of the data-acquisition system a factor 4 in speed will be gained which is important to fulfill the goals of the A2 collaboration to measure 400,000 $\eta' \to \eta \pi^0 \pi^0$, 200-300 $\eta' \to e^+ e^- \gamma$ decays.

In 2011, the A2 collaboration had published a determination of the $\eta$ transition form factor based on $\sim$1350 $\eta \to e^+ e^- \gamma$ events~\cite{Ber11}. An independent analysis of 3 times more data than in the publication mentioned above using a kinematic fitting technique allowed for the extraction of roughly 20,000 $\eta \to e^+ e^- \gamma$ events. The resulting transition form factor agrees very well with all earlier measurements. Comparisons to theoretical calculations show very good agreement, though the statistical accuracy does not allow one to rule out any prediction. Thus, the need for higher precision measurements is evident. The new result of the A2 collaboration will be published soon.

Recently, a long series of measurements with a transversally polarized target was concluded. Using a circularly polarized photon-beam it is possible to measure the polarisation observable $F$, and also in parallel $T$ which does not require any beam polarisation. In $\eta$-photoproduction $T$ is proportional to an interference term between $s$- and $d$-waves. Though all isobar models predict it to be zero, the PHOENICS experiment in Bonn has seen a behaviour that indicates the contrary~\cite{Boc98}. In Mainz $T$ was determined with improved statistics but the analysis is still ongoing, and thus, no conclusion can be drawn yet. The observable $F$ has been measured for the first time with the Crystal Ball-setup in Mainz. 
The analysis of $F$ is also still preliminary but both results will be published soon.

\newpage

\subsection{Formation spectrum for ${\bar K}NN$ system via kaon/pion induced reactions}
\addtocontents{toc}{\hspace{2cm}{\sl J.~Yamagata-Sekihara}\par}

\vspace{5mm}

J.~Yamagata-Sekihara, T. Sekihara, D. Jido$^{a}$, S. Hirenzaki$^{b}$

\vspace{5mm}

\noindent
 Institute of Particle and Nuclear Studies, High Energy Accelerator Research Organization (KEK), Ibaraki 305-0801, Japan\\
$^{a}$Department of Physics, Tokyo Metropolitan University, Tokyo 192-0397, Japan\\
$^{b}$Department of Physics, Nara Women's University, Nara 630-8506, Japan\\
\vspace{5mm}

Meson-nucleus systems such as mesic atoms and mesic nuclei are very important and useful objects to extract the meson properties at finite nuclear density,
 which may have the close connections to the symmetry breaking pattern of QCD and its partial restoration in the nucleus.
Especially, we are very interested in the ${\bar K}$-nucleus systems, which are expected to provide information on the ${\bar K}$ meson properties at finite density. 
The study of the ${\bar K}$-nucleus system is important because the system gives information on the controversial ${\bar K}$-nucleus interaction and on the basic properties of high density matter such as neutron stars.
In this paper, we report the calculated formation spectra of ${\bar K}NN$ system via the kaon/pion induced reactions.

\begin{figure}[!h]
\centerline{
\includegraphics[width=16cm]{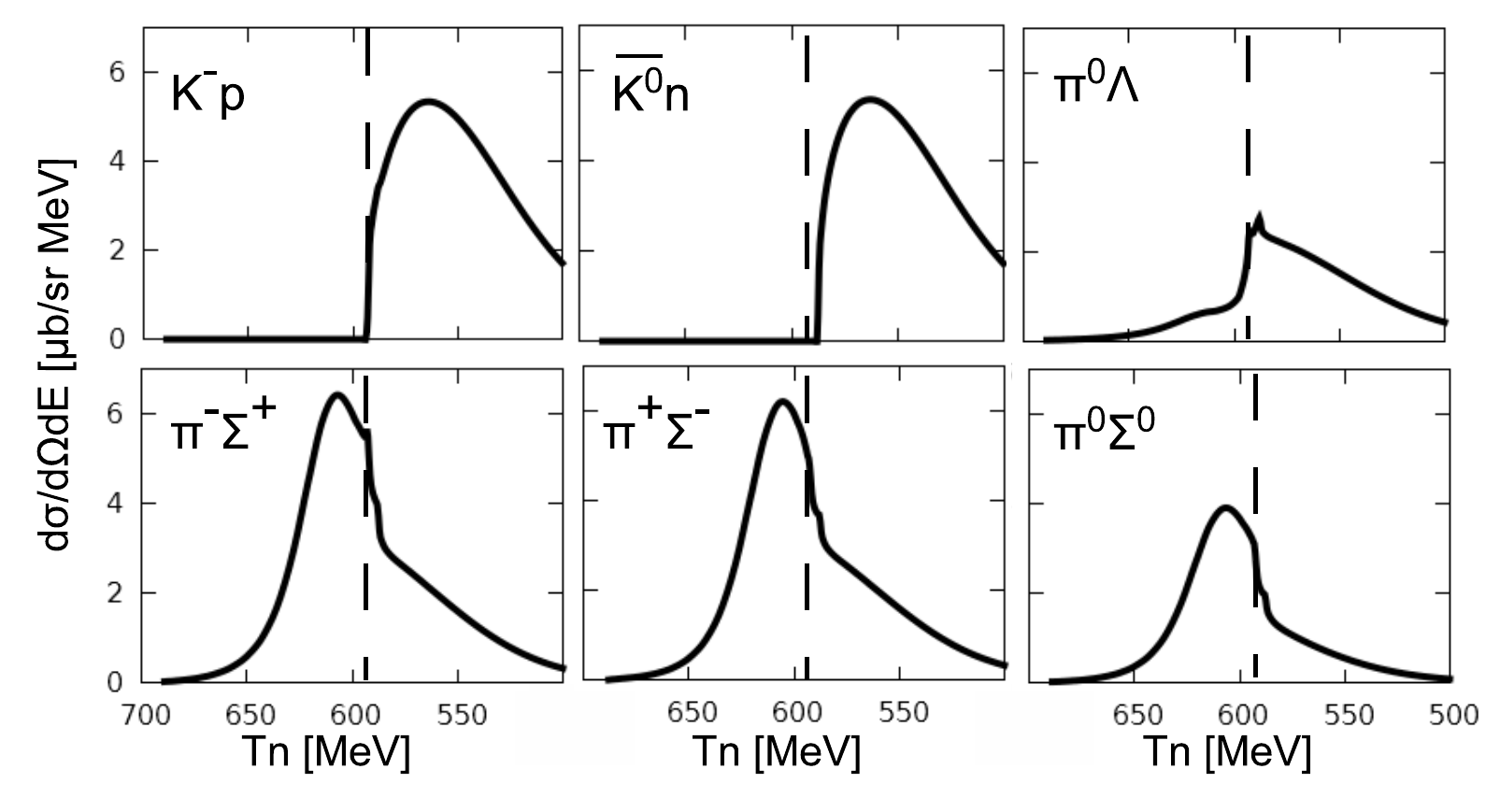}}
\vspace*{-0.5cm}
\caption{Mesonic decay exclusive spectra of the ${\bar K}NN$ formation in the $^3$He($K^-,n$) reactions at $T_{K^-}=600$~MeV at $\theta^{\rm Lab}_n=0$(deg.) calculated based on the chiral unitary model.
The vertical dashed line indicates the kaon production threshold.}
\label{Fig1y} 
\end{figure}

\begin{figure}[!h]
\centerline{
\includegraphics[width=16cm]{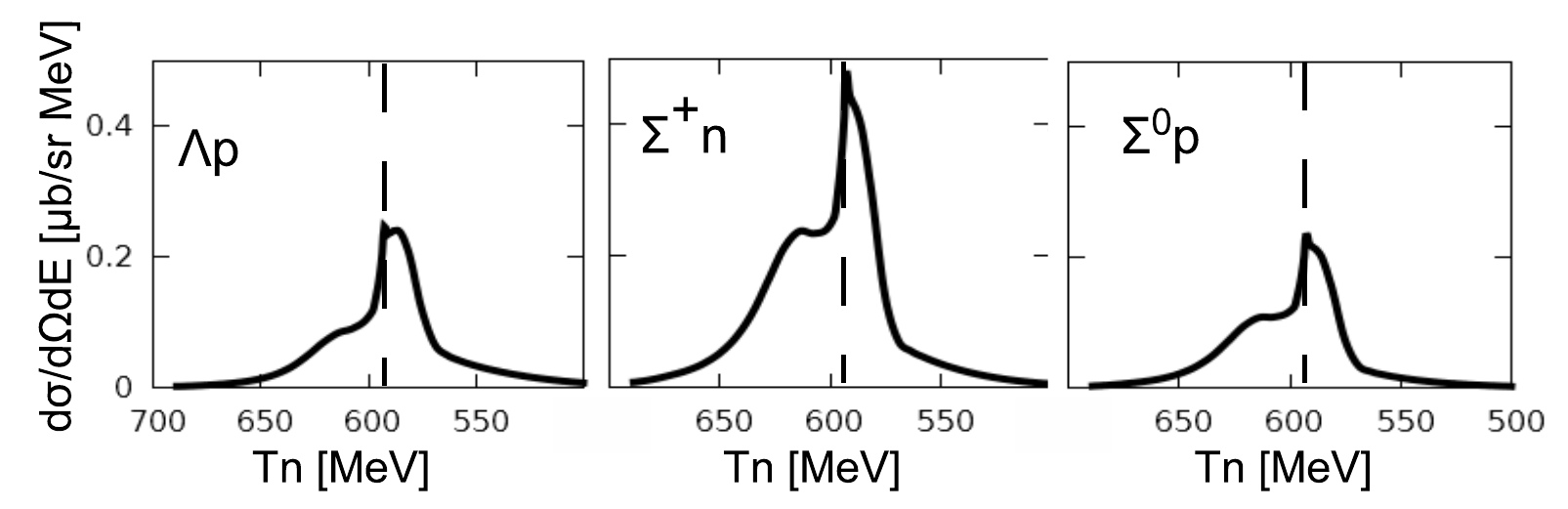}}
\vspace*{-0.5cm} 
\caption{Non-mesonic decay exclusive spectra of the ${\bar K}NN$ formation. Same as described in the caption of Fig.~\ref{Fig1y}.}
\label{Fig2y}
\end{figure}

For the kaon induced reaction, we considered the ${\bar K}NN$ system produced by the ($K^-,n$) reaction, which is found to be dominated by the ${\bar K}NN$ system with $I=1/2$.
Using the ${\bar K}$-nucleus interaction obtained from the chiral unitary model~\cite{Oset:2001cn,Hyodo:2002pk,Sekihara:2012wj}, we calculated the (semi-) exclusive spectra in coincident with the particle pair emissions due to kaon absorption in ${\bar K}NN\to MBN$ (mesonic decay process) and ${\bar K}NN\to YN$ (non-mesonic decay process),
where $M$, $B$ and $Y$ indicate meson, baryon and hyperon, respectively, with the same theoretical approach as Ref.~\cite{YamagataSekihara:2008ji}.
The calculated results are shown in Figs. \ref{Fig1y} and \ref{Fig2y}.
As we can see from the figures, by measuring the exclusive spectra, we have the strength in the bound region.
Especially, for $\pi\Sigma$ channel, we expect to obtain the information on the ${\bar K}NN$ bound state around $T_n=610$~MeV.

For the pion induced reaction, we calculated the formation spectra of the ${\bar K}NN$ system as a $\Lambda(1405)p$ bound state produced by the $(\pi^+,K^+)$ reaction.
In this work, we considered combination of two elementary reactions.
The first step is the production of a free $\Lambda(1405)$ state by the $\pi^++n\to K^++\Lambda(1405)$ reaction, and the second is the $\Lambda(1405)+p\to Y+p$ reaction, which contains the $\Lambda(1405)p$ interaction.
For the $\Lambda(1405)p$ interaction, we assumed that the $\Lambda(1405)p$ state makes a bound state, which is completely composite, and we took the binding energy of the $\Lambda(1405)p$ system and its width to $Yp$ as parameters.
The details of this study will be reported in Ref.~\cite{gata}.
With a certain parameter set, we could see the strength of the contribution from the $\Lambda(1405)p$ bound state.
As future work, we will search the observable case by using several parameters.
In addition, we will improve our calculation by using more realistic amplitudes.


\newpage

\subsection{Determination of the light quark masses}
\addtocontents{toc}{\hspace{2cm}{\sl M.~Zdr\'{a}hal}\par}

\vspace{5mm}

M.~Zdr\'{a}hal

\vspace{5mm}

\noindent
Institute of Particle and Nuclear Physics, Faculty of Mathematics and Physics, \\
Charles University in Prague, Czech Republic\\

\vspace{5mm}

Despite a significant progress in lattice QCD and in the field of sum rules of QCD, for the precise determination of the individual current quark masses of the $u$ and $d$ quarks
the combination of results of these methods together with some isospin breaking study performed in ChPT is required. A suitable process for the latter is $\eta\to3\pi$ decay since its decay rate happens to be proportional to $m_d-m_u$, represented in our work by the isospin breaking parameter $R$. (For more details see either contribution of G.~Colangelo or \cite{Zdrahal:2011ta}.)

There already exists a two-loop ChPT computation \cite{BG}, which should gives us a value of this parameter with a good precision. However, this precision is questioned by the observed problems of the computed amplitude: large chiral corrections in the first three orders; big discrepancies between the values of Dalitz parameters obtained from this computation and those measured in experiments; and finally, the computation depends on numerical values of some low-energy constants $C_i$, which are poorly known.
This has inspired various studies of the origin of the discrepancies and its effect on the determination of $R$ \cite{Schneider:2010hs,Kolesar:2011wn,Nehme:2011ez}. The most advanced with respect to the second aspect are the dispersive methods \cite{Colangelo:2009db,Kampf:2011wr}. 

In my talk I have also discussed the similarities and the main differences between them. If one wanted to emphasize the differences just in an appropriate naming of the methods, I would distinguish between Numerical fully dispersive approach \cite{Colangelo:2009db} and our Analytical perturbatively dispersive approach \cite{Kampf:2011wr}.
Both of them can be used for obtaining ``corrections" of the chiral amplitudes that incorporate various effects that were proposed for explanation of the discrepancies mentioned above. The approach \cite{Colangelo:2009db} is more suitable for incorporation of higher $\pi\pi$ rescattering effects, whereas our approach \cite{Kampf:2011wr} is of use if one wants to study the effect of the change of low-energy constants (note the existence of $C_i$-independent combination of Dalitz parameters).
However, in the construction of a model-independent parametrization of the $\eta\to3\pi$ amplitude that goes beyond the usual $O(p^6)$ chiral counting and using this parametrization for fitting the data, the only principal difference could appear in the procedure used for fixing the normalization of the amplitude from ChPT that is needed for extraction of $R$. The discussion of this procedure is also presented below.

Our parametrization of $\eta\to3\pi$ amplitude is in a form of six linear terms each time composed of one parameter multiplied by some simple function of Mandelstam variables and $\pi\pi$ scattering parameters and thus the parametrization of any theoretical model comprises a simple linear fit of those six independent parameters. The situation when one fits an experiment is slightly more complicated since one has to fit a square of the amplitude on a rather small physical region and the fit cease to be linear.
Fortunately, the fit to our parametrization worked so far well for two data sets constructed from KLOE 2008 results \cite{Ambrosino:2008ht}, the optimistic and the more realistic ones with 2500 and 154 data points, respectively. Real data sets will be available soon
(see contribution of L.~Caldeira Balkest\aa{}hl in this mini-proceedings).

For both of the data sets we have performed an analysis studying the possibility that the Dalitz parameters discrepancy is due to the incorrectly determined values of $C_i$s. Both of them were compatible with this assumption, the optimistic distribution led to a quite stable value of $R$ \cite{Kampf:2011wr}, however, the errors of the parameters of the fit for the realistic distribution were much larger, which reflected itself in an extensive range of the possible values of $R$.

Our more conclusive analysis has used a direct fit of the data to our parametrization. As I have stated above, in order to obtain any information about the quark masses, the fixing of the correct normalization from ChPT is necessary. In connection to that there is usually mentioned Adler theorem. However, for a determination of the normalization the theorem offers no help since it does not tell anything about the chiral convergence of the series in the Adler region and the ratio between two corrections in that region is in no way guaranteed to be smaller than the ratios in any other points. Moreover, the existence of zeros in real or imaginary parts in various orders means that at that point the correction of the respective order is of the same size as the complete amplitude in the lower order (otherwise they would not cancel). Our requirements for the matching point were therefore different. We have demanded: the matching dependent on the values of $C_i$s as less as possible; within 
that region the chiral expansion should work well; also in higher orders; the physical amplitude should have the similar behavior as the chiral amplitude inside the region. Using our parametrization of the physical data and of the chiral amplitude \cite{BG}, we have found a suitable region and also for the realistic data set obtained quite narrow interval for the values of $R$. Taking into account various data sets reproducing the values of Dalitz parameters obtained by KLOE \cite{Ambrosino:2008ht} and other sources of errors, our result is $R=39.6^{+2.5}_{-5.1}$.

In reaction to the observations of \cite{Lanz:2013ku,Leutwyler:2013wna} about the behavior of extrapolation of our original physical parametrization, we have found that the parametrization at the physical region and all conclusions of our analyses (including the values for $R$) are only mildly dependent on the value of one of our parameter. There is therefore no problem to set it to the value that provides a reasonable behavior of the amplitude in the Adler region for $s=t$ and fit just the other parameters to the physical amplitude.

\newpage

\subsection{Baryon Resonances in Effective Lagrangian Models }
\addtocontents{toc}{\hspace{2cm}{\sl M.~Z\'et\'enyi}\par}

\vspace{5mm}

M.~Z\'et\'enyi

\vspace{5mm}

\noindent
Wigner Research Centre for Physics, Budapest, Hungary\\

\vspace{5mm}

An effective Lagrangian calculation of dilepton production in
pion-nucleon collisions is presented \cite{Zetenyi-Wolf}. An important
ingredient of the model is the gauge-invariance-preserving scheme for
the definition of form factors in the Born contributions. For processes
involving real photons this scheme was invented by Davidson and Workman
\cite{Davidson-Workman}, we generalized the scheme for the virtual photon
case.

We discuss some issues related to the consistent treatment of spin$>$3/2
baryon resonances in the effective Lagrangian. It has been shown in
Ref.~\cite{Pascalutsa} that interaction Lagrangians invariant under a
gauge transformation of the higher spin baryon field are consistent in
the sense that they avoid the appearance of contributions from lower
spin degrees of freedom. We demonstrate that these gauge invariant
Lagrangians provide a better description of pion photoproduction cross
sections than the traditional Lagrangians.

We discuss the electromagnetic form factors of baryon resonances in the
vector meson dominance picture, showing that the $\rho^0$ and $\omega$
contributions to the VMD form factor can have both constructive and
destructive interference in different isospin channels (see
Ref.~\cite{Lutz}).

\newpage

\newpage

\section{List of participants}

\begin{flushleft}
\begin{itemize}
\item Moskov Amaryan, Old Dominion University, {\tt mamaryan@odu.edu}
\item Mikhail Bashkanov, Universit\"at T\"ubingen, {\tt bashkano@pit.physik.uni-tuebingen.de}
\item Maurice Benayoun, LPNHE des Universites Paris 6 et 7, {\tt benayoun@in2p3.fr}
\item Florian Bergmann, Westf\"alische Wilhelms-Universit\"at M\"unster, {\tt  florianbergmann@uni-muenster.de}
\item Johan Bijnens, Lund University, {\tt  bijnens@thep.lu.se}
\item Petr Bydzovsky, Nuclear Physics Institute, Czech Republic  {\tt bydz@ujf.cas.cz}
\item Li Caldeira Balkestahl, Uppsala University, {\tt li.caldeira\_balkestahl@physics.uu.se}
\item Ales Cieply, Nuclear Physics Institute, Czech Republic,  {\tt cieply@ujf.cas.cz}
\item Heinz Clement, Universit\"at T\"ubingen, {\tt clement@pit.physik.uni-tuebingen.de}
\item Gilberto Colangelo,  University of Bern, {\tt gilberto@itp.unibe.ch}
\item Eryk Czerwinski, Jagiellonian University, {\tt  eryk.czerwinski@uj.edu.pl}
\item Johanna Daub, Universit\"at Bonn, {\tt daub@hiskp.uni-bonn.de}
\item Jiri Dolejsi, Charles University, {\tt jiri.dolejsi@mff.cuni.cz}
\item Simon Eidelman, Novosibirsk State University, {\tt eidelman@mail.cern.ch }
\item Bossi Fabio, Laboratori Nazionali di Frascati, {\tt fabio.bossi@lnf.infn.it}
\item Shuangshi Fang, IHEP Beijing, {\tt fangss@ihep.ac.cn}
\item Kjell Fransson, Uppsala University, {\tt kjell.fransson@fysast.uu.se}
\item Hiroyuki Fujioka, Kyoto University, {\tt fujioka@scphys.kyoto-u.ac.jp}
\item Aleksander Gajos,  Jagiellonian University, {\tt  aleksander.gajos@uj.edu.pl}
\item Paolo Gauzzi, Sapienza Universit\'a di Roma e INFN, {\tt paolo.gauzzi@roma1.infn.it}
\item Simona Giovannella, Laboratori Nazionali di Frascati, {\tt  simona.giovannella@lnf.infn.it}
\item Frank Goldenbaum, Forschungszentrum Juelich, {\tt  f.goldenbaum@fz-juelich.de}
\item Evgueni Goudzovski, University of Birmingham, {\tt  eg@hep.ph.bham.ac.uk}
\item Dieter Grzonka, Forschungszentrum Juelich, {\tt  d.grzonka@fz-juelich.de}
\item Carl-Oscar Gullstršm, Uppsala University,  {\tt carl-oscar.gullstrom@physics.uu.se}
\item Malgorzata Gumberidze, TU Darmstadt, {\tt m.gumberidze@gsi.de}
\item Lena Heijkenskjšld, Uppsala University, {\tt lena.heijkenskjold@physics.uu.se}
\item Volker Hejny, Forschungszentrum Juelich, {\tt V.Hejny@fz-juelich.de}
\item Satoru Hirenzaki, Nara Women's University, {\tt zaki@cc.nara-wu.ac.jp}
\item Malgorzata Hodana, Jagiellonian University, {\tt m.hodana@gmail.com}
\item Martin Hoferichter, University of Bern, {\tt hoferichter@itp.unibe.ch}
\item Tom\'a\v s Husek, Charles University, {\tt Tomas.Husek@cern.ch}
\item Natsumi Ikeno, Nara Women's University, {\tt jan\_ikeno@cc.nara-wu.ac.jp}
\item Kenta Itahashi, RIKEN, {\tt itahashi@riken.jp}
\item Sergiy Ivashyn, Institute for Theoretical Phyisics, Kharkiv, {\tt s.ivashyn@gmail.com}
\item Tord Johansson, Uppsala University, {\tt tord.johansson@physics.uu.se}
\item Tom\'a\v s Kadav\'y, Charles University, {\tt tomas.kadavy@gmail.com}
\item Karol Kampf, Charles University, {\tt karol.kampf@mff.cuni.cz}
\item Farha Khan, Forschungszentrum Juelich, {\tt f.khan@fz-juelich.de}
\item Mari\'an Koles\'ar, Charles University, {\tt  kolesar@ipnp.troja.mff.cuni.cz}
\item Bastian Kubis, University of Bonn, {\tt kubis@hiskp.uni-bonn.de}
\item Andrzej Kupsc, Uppsala University, {\tt Andrzej.Kupsc@physics.uu.se }
\item Michael Lang, Helmholtz-Institut f\"ur Strahlen- und Kernphysik, {\tt mlang@hiskp.uni-bonn.de}
\item Daniel Lersch, Forschungszentrum Juelich, {\tt d.lersch@fz-juelich.de}
\item Stefan Leupold, Uppsala University, {\tt stefan.leupold@physics.uu.se}
\item Manuel Lorenz, Goethe University Frankfurt, {\tt  m.lorenz@gsi.de}
\item Matthias F.M. Lutz, GSI Darmstadt, {\tt  m.lutz@gsi.de}
\item Hartmut Machner, University Duisburg-Essen, {\tt hartmut.machner@uni-due.de}
\item Jiri Mares, Nuclear Physics Institute, Czech Republic,  {\tt mares@ujf.cas.cz}
\item Pere Masjuan, Universt\"at Mainz, {\tt masjuan@kph.uni-mainz.de}
\item Volker Metag, Giessen University, {\tt volker.metag@exp2.physik.uni-giessen.de}
\item Pawel Moskal, Jagiellonian University, {\tt p.moskal@uj.edu.pl}
\item Bachir  Moussallam, IPN, Universit\'e Paris-Sud XI, {\tt moussall@ipno.in2p3.fr}
\item Hideko Nagahiro, Nara Women's University, {\tt  nagahiro@cc.nara-wu.ac.jp}
\item Mariana Nanova, Giessen University, {\tt  Mariana.Nanova@exp2.physik.uni-giessen.de}
\item Franz Niecknig, Universit\"at Bonn, {\tt Niecknig@hiskp.uni-bonn.de}
\item Jiri Novotny, Charles University, {\tt  jiri.novotny@mff.cuni.cz}
\item Iryna Ozerianska, UJ Krakow, {\tt i.ozerianska@gmail.com}
\item Michael Papenbrock, Westf\"alische Wilhelms-Universit\"at M\"unster, {\tt  michaelp@uni-muenster.de}
\item Teresa Pena, IST, Lisboa, {\tt teresa.pena@ist.utl.pt}
\item Joachim Pettersson,Uppsala University, {\tt Joachim.Pettersson.@physics.uu.se}
\item Ivan Prado Longhi, Roma Tre University, {\tt plonghi@roma3.infn.it}
\item Elisabetta Prencipe, Forschungszentrum Juelich, {\tt  e.prencipe@fz-juelich.de}
\item Witold Przygoda, Jagiellonian University,  {\tt przygoda@if.uj.edu.pl}
\item Damian Pszczel, NCNR, Poland, {\tt damian.pszczel@fuw.edu.pl}
\item Izabela Pytko, Jagiellonian University, {\tt iza.pytko@gmail.com}
\item Christoph Florian Redmer, Universit\"at Mainz, {\tt redmer@uni-mainz.de }
\item Piotr Salabura, Jagiellonian University,  {\tt piotr.salabura@uj.edu.pl}
\item Ivano sarra,  Laboratori Nazionali di Frascati, {\tt  ivano.sarra@lnf.infn.it}
\item Siddhesh Sawant, Indian Institute of Technology Bombai, {\tt sawantsiddhesh08@gmail.com}
\item Sebastian Schneider, Universit\"at Bonn, {\tt schneider@hiskp.uni-bonn.de}
\item Boris Shwartz,  Budker Institute, {\tt shwartz@inp.nsk.su}
\item Magdalena Skurzok, Jagiellonian University,  {\tt mskurzok@gmail.com}
\item Jaroslav Smejkal, UTEF Prague, {\tt jaroslav.smejkal@utef.cvut.cz}
\item Joanna Stepaniak, NCBJ, Warsaw, {\tt Joanna.stepaniak@fuw.edu.pl}
\item Yoshiki K. Tanaka, University of Tokyo, {\tt tanaka@nucl.phys.s.u-tokyo.ac.jp}
\item Carla Terschluesen, Uppsala University, {\tt carla.terschluesen@physics.uu.se}
\item M.~Unverzagt, Universit\"at Mainz, {\tt unvemarc@kph.uni-mainz.de }
\item Magnus Wolke, Uppsala University, {\tt magnus.wolke@fysast.uu.se}
\item Junko Yamagata-Sekihara, KEK, {\tt yamajun@post.kek.jp}
\item Martin Zdr\'ahal, Charles University, {\tt  martin.zdrahal@mff.cuni.cz}
\item Mikl\'os Z\'et\'enyi, Wigner Research Centre for Physics, Budapest, {\tt  zetenyi.miklos@wigner.mta.hu}
\item Jozef Zlomanczuk, Uppsala University, {\tt jozef-zlomanczuk@physics.uu.se}
\end{itemize}
\end{flushleft}

\end{document}